%% file: PDF4LHC_Run2_recommendations-arXiv.tex
\documentclass[prd,aps,preprint,tightenlines,superscriptaddress,nofootinbib,showpacs,showkeys]{revtex4}
\usepackage{mathrsfs}
\usepackage{amsfonts}
\usepackage{array}
\usepackage{graphics}
\usepackage{rotating}
\usepackage{multirow}
\usepackage{color}

\usepackage{amsmath}    
\allowdisplaybreaks
\usepackage{hyperref}   

\DeclareGraphicsExtensions{.eps,.pdf}

\usepackage{pslatex}
\usepackage{txfonts}
\usepackage[latin1]{inputenc}
\usepackage[T1]{fontenc}

\newcommand{\GeV}{\ensuremath{\,\mathrm{GeV}}}
\newcommand{\TeV}{\ensuremath{\,\mathrm{TeV}}}
\newcommand{\msbar}{$\overline{\mathrm{MS}}\, $}

\newcommand{\gsim}{\raisebox{-0.07cm}{$\:\:\stackrel{>}{{\scriptstyle
 \sim}}\:\: $} }
\newcommand{\lsim}{\raisebox{-0.07cm}{$\:\:\stackrel{<}{{\scriptstyle
 \sim}}\:\: $} }

\newcommand{\tablefootnote}{\footnote}

\begin{document}
\thispagestyle{empty}
\begin{flushright}
  DESY 16-041 \\
  DO-TH 16/05 \\
  LTH 1081\\
  JLAB-THY-16-2231\\
\end{flushright}

\vspace{1.0cm}
\begin{center}
{\Large\bf 
    A Critical Appraisal and Evaluation of Modern PDFs
}
\vspace{1.25cm}

   A.~Accardi$^{\, a,b}$,	
   S.~Alekhin$^{\, c,d}$,
   J.~Bl\"umlein$^{\, e}$,
   M.V.~Garzelli$^{\, c}$,
   K.~Lipka$^{\, f}$, \\[0.5ex]
   W. Melnitchouk$^{\, b}$, 
   S.~Moch$^{\, c}$,
   J.F.~Owens$^{\, g}$,
   R.~Pla\v cakyt\. e$^{\, f}$,
   E.~Reya$^{\, h}$,
   N.~Sato$^{\, b}$,
   A.~Vogt$^{\, i}$ \\[0.5ex]
   and
   O. Zenaiev$^{\, f}$
   \\

\vspace{1.0cm}
{\it \small
   $^a$ Hampton University, Hampton, VA 23668, USA \\
   \vspace{0.1cm}
   $^b$ Jefferson Lab, Newport News, VA 23606, USA \\
   \vspace{0.1cm}
   $^c$ II. Institut f\"ur Theoretische Physik, Universit\"at Hamburg \\
   Luruper Chaussee 149, D--22761 Hamburg, Germany \\
   \vspace{0.1cm}
   $^d$Institute for High Energy Physics \\
   142281 Protvino, Moscow region, Russia \\
   \vspace{0.1cm}
   $^e$Deutsches Elektronensynchrotron DESY \\
   Platanenallee 6, D--15738 Zeuthen, Germany \\
   \vspace{0.1cm}
   $^f$Deutsches Elektronensynchrotron DESY \\
   Notkestra{\ss}e 85, D--22607 Hamburg, Germany \\
   \vspace{0.1cm}
   $^g$ Florida State University, Tallahassee, FL 32306, USA \\
   \vspace{0.1cm}
   $^h$ Institut f\"ur Physik, Technische Universit\"at Dortmund \\ 
   D--44221 Dortmund, Germany \\
   \vspace{0.1cm}
   $^i$ Department of Mathematical Sciences, University of Liverpool \\
   Liverpool L69 3BX, United Kingdom
}
\end{center}

\vspace{0.5cm}

\begin{center}
{\bf \large Abstract:}
\end{center}
\noindent
We review the present status of the determination of parton distribution functions (PDFs) 
in the light of the precision requirements for the LHC in Run 2 and other future hadron colliders.
We provide brief reviews of all currently available PDF sets and use them 
to compute cross sections for a number of benchmark processes,
including Higgs boson production in gluon-gluon fusion at the LHC.
We show that the differences in the predictions obtained with the various PDFs
are due to particular theory assumptions made in the fits of those PDFs. 
We discuss PDF uncertainties in the kinematic region covered by the LHC and 
on averaging procedures for PDFs, such as advocated by the PDF4LHC15 sets,
and provide recommendations for the usage of PDF sets
for theory predictions at the LHC.
\setcounter{page}{0}
\clearpage

\tableofcontents
\newpage

\section{Introduction}
\label{sec:introduction}

In Run 2 of the Large Hadron Collider (LHC), 
the very details of the Standard Model (SM), including cross sections of different processes 
and Higgs bosons properties, are being measured with very high precision.
At the same time, the new data at the highest center-of-mass collision energies
ever achieved ($\sqrt{s}=13$~TeV) are used to search for physics phenomena beyond the SM (BSM).
The experimental data used to perform those measurements are generally expected
to have percent-level accuracy, depending on details such as the final states 
and the acceptance and efficiency of the detectors in particular kinematics ranges.

To further test the SM and to identify signals for new physics,
measurements need to be compared to precise theoretical predictions, 
which need to incorporate higher order radiative corrections in 
Quantum Chromodynamics (QCD) and, possibly, the electroweak sector of the SM.
In order to reach the benchmark precision set by the accuracy of the
experimental data, next-to-next-to-leading order (NNLO) corrections in QCD are often required.
At next-to-leading order (NLO) in QCD, the residual theoretical uncertainty 
from truncating the perturbative expansion commonly estimated by variations of
the renormalization and factorization scales $\mu_r$ and $\mu_f$  are often
too large compared to the experimental accuracy.
Nonetheless, for observables with complex final states, 
and indeed for many BSM signals, one must still contend with NLO calculations, 
which will continue to require corresponding NLO fits.

Parton distribution functions (PDFs) in the proton serve as an essential input for 
any cross section prediction at hadron colliders and have been measured with
increasing precision over the last three decades.
Likewise, the strong coupling constant $\alpha_s(M_Z)$ at the $Z$ boson mass scale $M_Z$
and the masses $m_h$ of the heavy quarks $h=c, b, t$
are well constrained by existing data and their determination is accurate at least to NNLO.
However, despite steady improvements in the accuracy of PDF determinations over the years, 
the uncertainties associated with PDFs, the strong coupling
$\alpha_s(M_Z)$, and quark masses still dominate many calculations
of cross sections for SM processes at the LHC.
A particularly prominent example is the cross section for the production of a
SM Higgs boson in the gluon-gluon fusion channel.

The currently available PDF sets are
CJ15~\cite{Accardi:2016qay}, 
accurate to NLO in QCD, 
as well as 
ABM12~\cite{Alekhin:2013nda},
CT14~\cite{Dulat:2015mca},
HERAPDF2.0~\cite{Abramowicz:2015mha},
JR14~\cite{Jimenez-Delgado:2014twa},
MMHT14~\cite{Harland-Lang:2014zoa},
and 
NNPDF3.0~\cite{Ball:2014uwa}
to NNLO in QCD.
These provide a detailed description of the parton content of the proton, which
depends on the chosen sets of experimental data as well as on the theory
assumptions and the underlying physics models used in the analyses.
Both theoretical and experimental inputs have direct impact on the obtained nonperturbative parameters,
namely, the fitted PDFs, the value of $\alpha_s(M_Z)$ and the quark masses. 
Moreover, they can lead to large systematic shifts compared to the
uncertainties of the experimental data used in the fit.
For precision predictions in Run 2 of the LHC it is therefore very
important to quantify those effects in detailed validations of the 
individual PDF sets in order to reduce the uncertainties in those
nonperturbative input parameters.
Moreover, this will allow one to pinpoint problems with the determination of certain PDFs.
Any approach to determine the parton luminosities at the LHC which implies mixing or averaging 
of various PDFs or of their respective uncertainties, such as that
advocated in the recent PDF4LHC recommendations~\cite{Butterworth:2015oua}, is therefore potentially 
dangerous in the context of precision measurements, in particular, 
or when studying processes at kinematic edges such as at large values of
Bjorken $x$ or small scales $Q^2$. 
The precision measurements of the LHC experiments themselves help 
to constrain the different sets of PDFs and may even indicate deviations from SM processes, 
cf.~\cite{Ridder:2016nkl} for an example. 
It is thus of central importance that comparisons for all available PDF
sets are performed in a quantitative manner and with the best available accuracy.

In this paper we briefly discuss the available world data used to constrain PDFs in Sec.~\ref{sec:datapdfs} 
and stress the need to include only compatible data sets in any analysis.
The data analysis relies on comparison with precise theoretical predictions, 
with many of these implemented in software tools.
In this respect, we underline in Sec.~\ref{sec:th4pdfs} the importance of open-source code to provide 
benchmarks and to facilitate theory improvements through 
indication and reduction of possible errors. 
In addition, Sec.~\ref{sec:th4pdfs} is devoted to a discussion of a number of crucial theory aspects in PDF fits.
These include the treatment of heavy quarks and their masses,
QCD corrections for $W^\pm$- and $Z$-boson production applied 
in the fit of light-flavor PDFs, and the importance of nuclear corrections 
in scattering data off nuclei.
The strong coupling constant is correlated with the PDFs and is therefore an
important parameter to be determined simultaneously with the PDFs.
The state of the art is reviewed in Sec.~\ref{sec:alphas}.
The need to address PDF uncertainties for cross section predictions is illustrated in Sec.~\ref{sec:crs-at-lhc}, 
with the Higgs boson cross section in the gluon-gluon fusion channel being the most prominent case.
Other examples include the production of heavy quarks at the LHC in different kinematical regimes.
Our observations illustrate important shortcomings of the recent PDF4LHC
recommendations~\cite{Butterworth:2015oua} which are addressed in Sec.~\ref{sec:recommendations}, 
where alternative recommendations for the usage of sets of PDFs for theory predictions at the LHC are provided.
Finally, we conclude in Sec.~\ref{sec:conclusion}.

\section{Data sets and results for PDF fits}
\label{sec:datapdfs}

We begin with an overview of the currently available data which can be used to
determine PDFs and present the fit results of the various groups.

%
\input{table-data}

\subsection{Data sets used in PDF fits}

The data used in the various PDF fits overlap to a large extent, as indicated in Tab.~\ref{tab:chi2}. 
However, there are also substantial differences which are related to  
the accuracy required in the analysis, 
the feasibility of efficiently implementing the corresponding theoretical computations,
or the subjective evaluation of the data quality, to name a few.

The core of all PDF fits comprises the deep-inelastic scattering (DIS) data 
obtained at the HERA electron-proton ($ep$) collider and in fixed-target experiments. 
While the former has used only a proton target, the latter have collected large amounts of data
for the deuteron and heavier targets as well. 
The analysis of nuclear-target data requires an accurate account of nuclear effects. 
This is challenging already in the case of the loosely-bound deuteron (cf.~Sec.~\ref{sec:th4pdfs}),
and even more so for heavier targets. Therefore, in general, data sets
for DIS on targets heavier than deuteron are not used. 
Nonetheless, different combinations of data sets for the neutrino-induced DIS 
off iron and lead targets obtained by the CCFR/NuTeV, CDHSW, and CHORUS experiments  
are included in the CT14, MMHT14, and NNPDF3.0 analyses, 
but are not used by other groups 
to avoid any influence of nuclear correction uncertainties. 
One can also point out the abnormal dependence of the DIS structure functions on the 
beam energy in the NuTeV experiment~\cite{Paukkunen:2013grz} and the poor agreement 
of the CDHSW data with the QCD predictions on the $Q^2$ 
slope of structure functions~\cite{Altarelli:1992ei,Blumlein:1989pd,Berge:1989hr} 
as an additional motivations to exclude these data sets. 

The kinematic cuts applied to the commonly used DIS data also differ in
various analyses in order to minimize the influence of higher twist contributions.
Another important feature of the DIS data analyses in PDF fits concerns the 
use of data for the DIS structure function $F_2$ instead of the data for the measured cross sections. 
These aspects will be discussed in Sec.~\ref{sec:th4pdfs}. 

The inclusive DIS data are often supplemented by the semi-inclusive data on the  
neutral-current and charged-current DIS charm-quark production. 
The neutral-current sample collected by the HERA experiments provides a valuable tool 
to study the heavy-quark production mechanism. 
This is vital for pinning down PDFs, in particular the gluon PDF at small $x$, 
relevant for important phenomenological applications at the LHC (cf.~Sec.~\ref{sec:crs-at-lhc}).
The charged-current charm production data help to constrain the strange sea PDF, 
which is strongly mixed with contributions from non-strange PDFs in other observables (cf.~Sec.~\ref{sec:th4pdfs}). 

The Drell-Yan (DY) data are also a necessary ingredient of any PDF analysis 
since DIS data alone do not allow for a comprehensive disentangling of the quark 
and anti-quark distributions.  
Historically, for a long time only fixed-target DY data were available for PDF fits.
In particular, this did not allow for a model-independent separation of the valence and sea quarks at small-$x$. 
The high precision DY data obtained in proton-proton ($pp$) and proton--anti-proton ($p\bar{p}$) collisions
from the LHC and the Tevatron open new possibilities to study the PDFs at small and large $x$. 
The LHC experiments are quickly accumulating statistics and are currently
providing data samples at $\sqrt{s}=$7 and 8 TeV
for $W$- and $Z$-boson production with typical luminosities of $\sim 1~{\rm fb}^{-1}$. 
The rapid progress in experimental measurements 
causes a greatly non-uniform coverage of the recent DY data in various PDF fits 
(cf.~Tabs.~\ref{tab:dydata} and ~\ref{tab:dydata2})
and leads to corresponding differences in the accuracy of the extracted PDFs.
Another issue here is the theoretical accuracy achieved for the description 
of the DY data. This varies substantially and will be discussed in Sec.~\ref{sec:th4pdfs}.

Often, jet production in $pp$ and $p\bar{p}$ collisions is used 
as an additional process to constrain the large-$x$ gluon PDF.
Here, the QCD corrections are known to NLO and the calculation 
of the NNLO ones is in progress~\cite{Ridder:2013mf}.
The incomplete knowledge of the latter is problematic in view of 
a consistent PDF analysis at NNLO when including those jet data.
This will be discussed in Sec.~\ref{sec:alphas} in connection with 
the determination of the value of the strong coupling constant $\alpha_s$.

In addition to these major categories of data commonly used to constrain PDFs, 
some complementary processes are also employed in some cases, as indicated in Tab.~\ref{tab:chi2}.
These comprise the hadro-production of top-quark pairs from $pp$ and $p\bar{p}$ collisions 
and the associated production $W$ bosons with charm quarks in $pp$ collisions. 
Sometimes, also jet production in $ep$ collisions 
and prompt photon ($\gamma$+jet) production from $pp$ and $p\bar{p}$ collisions is considered. 
Except for $t\bar{t}$ production the necessary QCD corrections are known to
NLO only, so that the same arguments as in the case of jet hadro-production
data apply, if those data are included in a fit at NNLO accuracy.
For $t\bar{t}$ production, only the inclusive cross section is considered at
the moment in the available PDFs and there is a significant correlation with PDFs, especially of the gluon
PDF with the top-quark mass.

Taken together, the set of these data has a number of data points (NDP) 
of the order of few thousand, and provides sufficient information to describe the PDFs 
with an ansatz of about ${\cal O}(30)$ free parameters.
The parameters can include the strong coupling constant $\alpha_s(M_Z)$ and the heavy-quark
masses $m_c$, $m_b$ and $m_t$, which are correlated with the PDFs, as will 
be discussed in Secs.~\ref{sec:th4pdfs} and \ref{sec:alphas}.
This provides sufficient flexibility for all PDF groups 
and it is routinely checked 
that no additional terms are required to improve the quality of fit.
The exception is the NNPDF group, which typically uses ${\cal O}(250)$ free parameters 
in the neural network.

Apart from those considerations there is the general problem of the quality of the experimental data, 
that is to say whether or not the PDFs are extracted from a consistent data set. 
The various groups have different approaches, which roughly fall into two classes 
according to the different confidence level (c.l.) criteria for the
value of $\chi^2$ in the goodness-of-fit test.
One approach is to fit to a very wide (or even the widest possible) set of
data, while the other one rejects inconsistent data sets.
In the former case, a tolerance criterion for $\Delta \chi^2$ is introduced 
(e.g. $\Delta \chi^2=100$), while the latter approach maintains that $\Delta \chi^2=1$.
For the various PDF groups this information 
is listed in Tab.~\ref{tab:chi2}.

  For further reference, we quote here the definition of $\chi^2$ used in data
  comparisons (Tabs.~\ref{tab:mcmass}, \ref{tab:mcmass-ctd}, 
  \ref{tab:higgs-mc-mstw}--\ref{tab:higgs-mc-nnpdf}, 
  \ref{tab:ttbar-mc-mmht}--\ref{tab:lhcbbeauty}).
  It follows the definition described 
  in Refs.~\cite{Aaron:2012qi,Adloff:2003uh,Chekanov:2002pv} and is expressed as follows:
\begin{equation} 
  \chi^2 =  
  \sum_i \frac{\left[ {\mu_i} - m_i \left( 1 - \sum_j \gamma^i_j b_j \right) \right]^2}
    { \textstyle \delta^2_{i,{\rm unc}}m_i^2 + \delta^2_{i,{\rm stat}}\, {\mu_i} m_i  }
 + \sum_j b^2_j
 + \sum_i \ln \frac{ \textstyle \delta^2_{i,{\rm unc}}m_i^2 + \delta^2_{i,{\rm stat}}\, {\mu_i} m_i }
 { \textstyle \delta^2_{i,{\rm unc}}\mu_i^2 + \delta^2_{i,{\rm stat}}\mu_i^2 },
 \label{eq:chi2formula}
\end{equation}
where $\mu_i$ represents the measurement at the point $i$, $m_i$ is the corresponding 
theoretical prediction and $\delta_{i,{\rm stat}}$, $\delta_{i,{\rm unc}}$ are the 
relative statistical and uncorrelated systematic uncertainties, respectively.
$\gamma^i_j$ denotes the sensitivity of the measurement to the correlated systematic 
source $j$ and $b_j$ their shifts, with a penalty term $\sum_j b_j^2$ added.
In addition, a logarithmic term is introduced arising from the likelihood transition to $\chi^2$ 
when scaling of the errors is applied~\cite{Aaron:2012qi}.

It is important to note that the $\chi^2$ values obtained with
  Eq.~(\ref{eq:chi2formula}) 
  will not necessarily correspond to numbers
  quoted by PDF groups due to different $\chi^2$ definitions, data treatment
  and other parameters, see also Tab.~\ref{tab:chi2}.

%
\input{table-WandZ-data}

\begin{figure}[tH!]
\begin{center}
\vspace*{-5mm}
\includegraphics[width=0.95\textwidth]{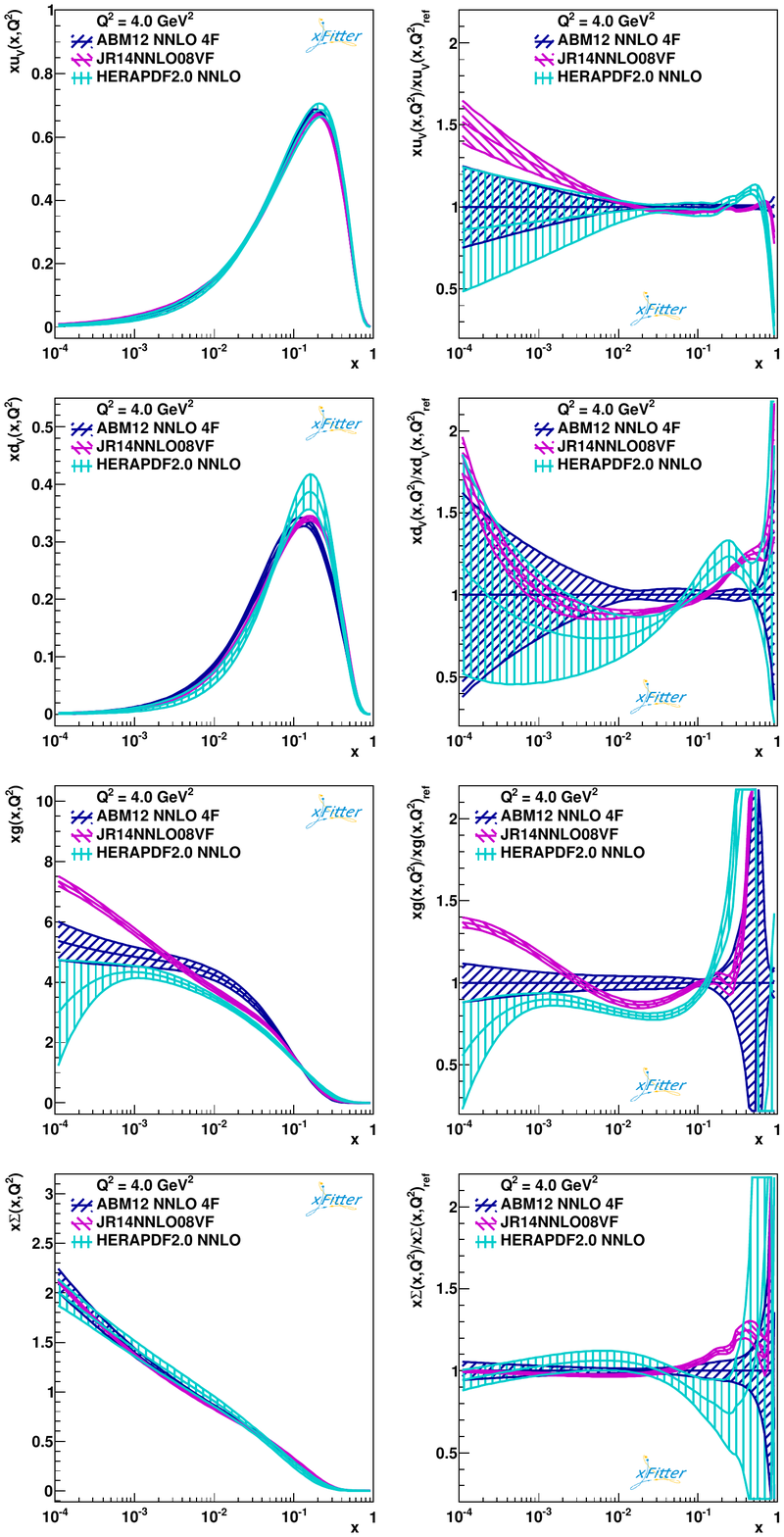}
\vspace*{-5mm}
\caption{\small 
\label{fig:pdfs-abm-4} 
The $u$-valence, $d$-valence, gluon and sea quark ($x\Sigma = 2x(\bar{u}+\bar{c}+\bar{d}+\bar{s})$) 
PDFs with their 1~$\sigma$ uncertainty bands of 
ABM12~\cite{Alekhin:2013nda},  
HERAPDF2.0~\cite{Abramowicz:2015mha} and 
JR14 (set {\tt JR14NNLO08VF})~\cite{Jimenez-Delgado:2014twa} 
at NNLO at the scale $Q^2=4~\GeV^2$;  
absolute results (left) and ratio with respect to ABM12 (right).
}
\end{center}
\end{figure}

\begin{figure}[tH!]
\begin{center}
\vspace*{-5mm}
\includegraphics[width=0.95\textwidth]{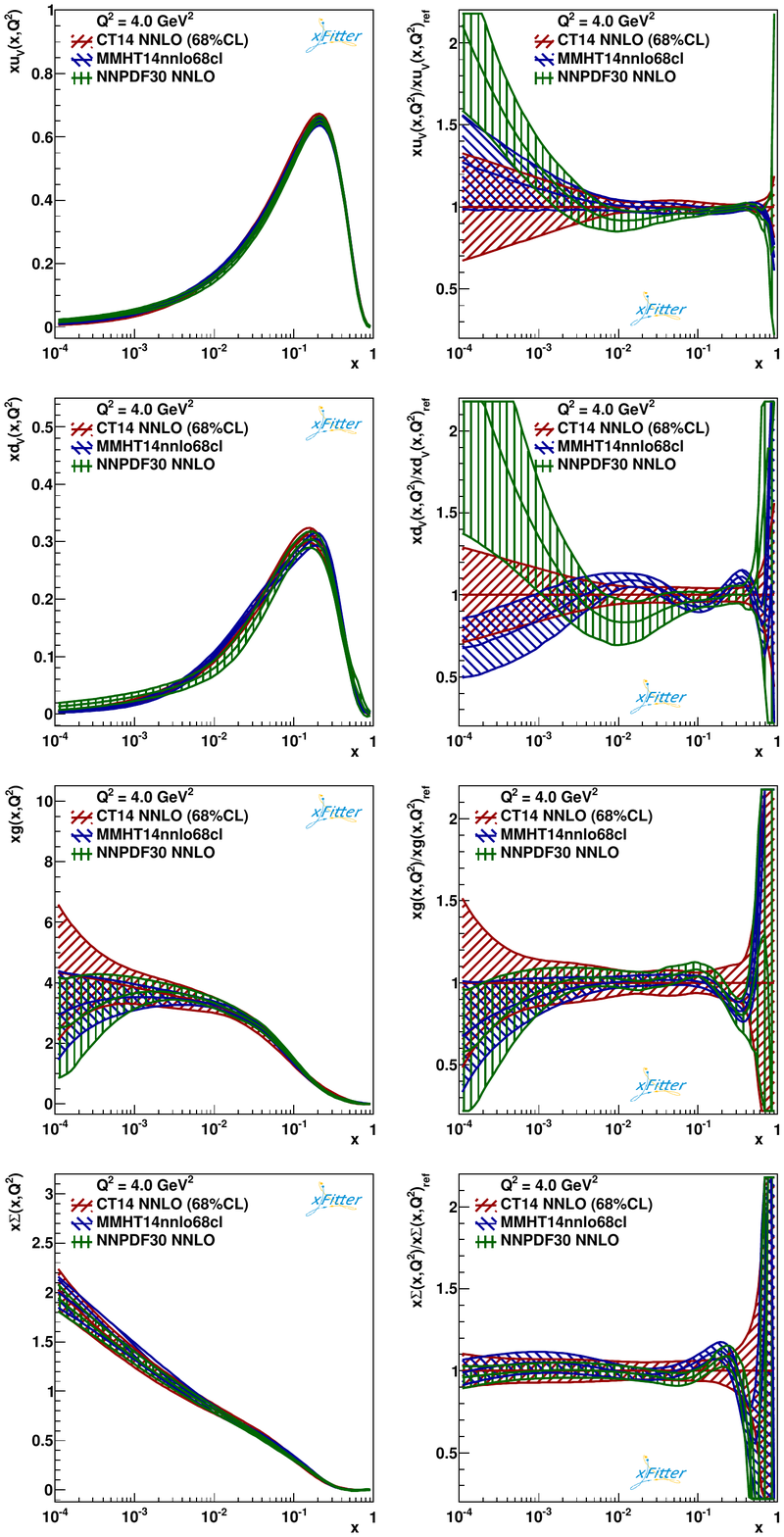}
\vspace*{-5mm}
\caption{\small 
\label{fig:pdfs-ct14-4} 
Same as Fig.~\ref{fig:pdfs-abm-4} for the 
CT14~\cite{Dulat:2015mca},
MMHT14~\cite{Harland-Lang:2014zoa}
and 
NNPDF3.0~\cite{Ball:2014uwa} PDF sets 
with their 1~$\sigma$ uncertainty bands 
at NNLO; absolute results (left) and ratio with respect to CT14 (right).
}
\end{center}
\end{figure}

\begin{figure}[tH!]
\begin{center}
\vspace*{-5mm}
\includegraphics[width=0.95\textwidth]{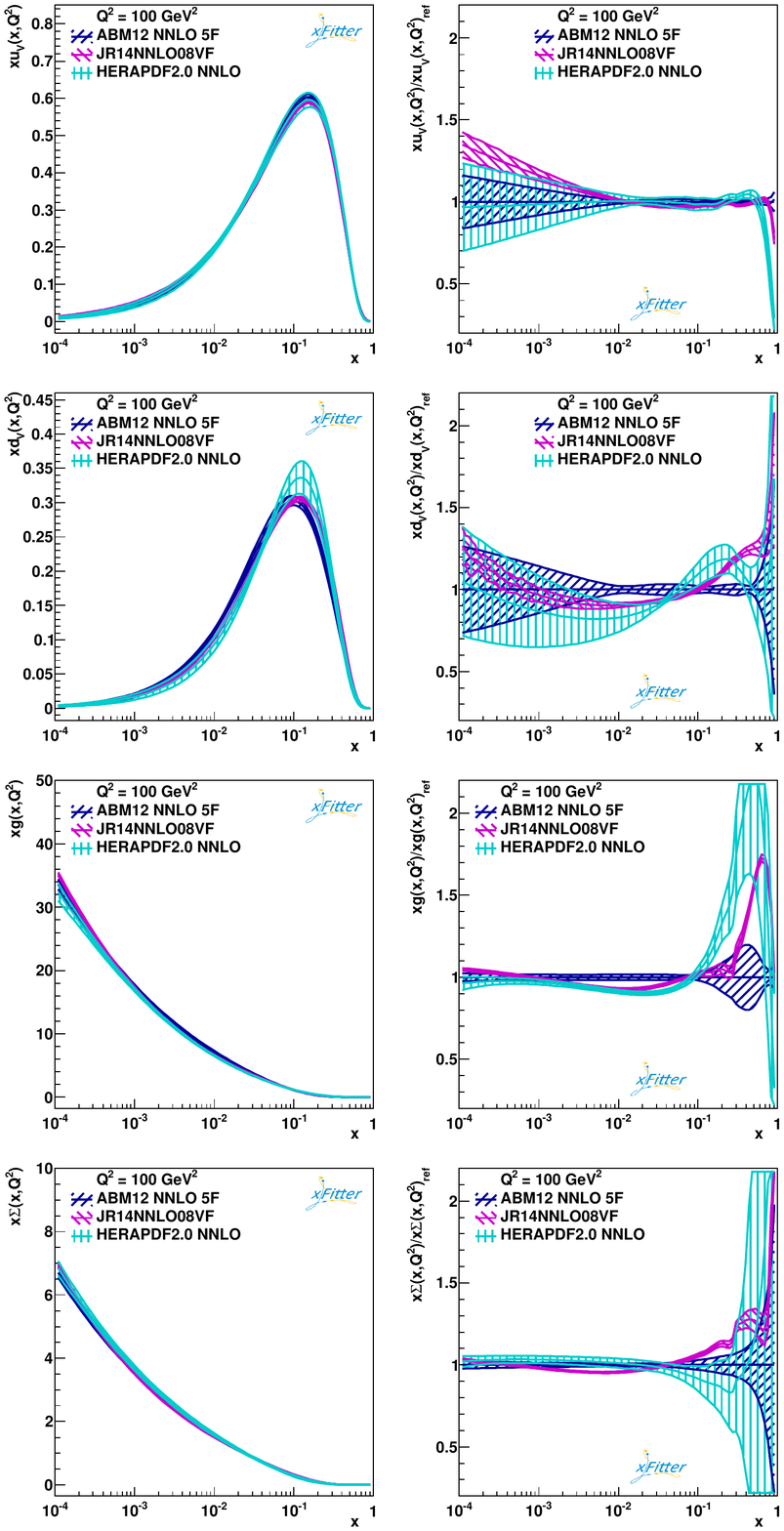}
\vspace*{-5mm}
\caption{\small 
\label{fig:pdfs-abm-100} 
Same as Fig.~\ref{fig:pdfs-abm-4} at the scale $Q^2=100~\GeV^2$ 
with the sea $ x\Sigma = 2x(\bar{u}+\bar{c}+\bar{d}+\bar{s}+\bar{b})$.
}
\end{center}
\end{figure}

\begin{figure}[tH!]
\begin{center}
\vspace*{-5mm}
\includegraphics[width=0.95\textwidth]{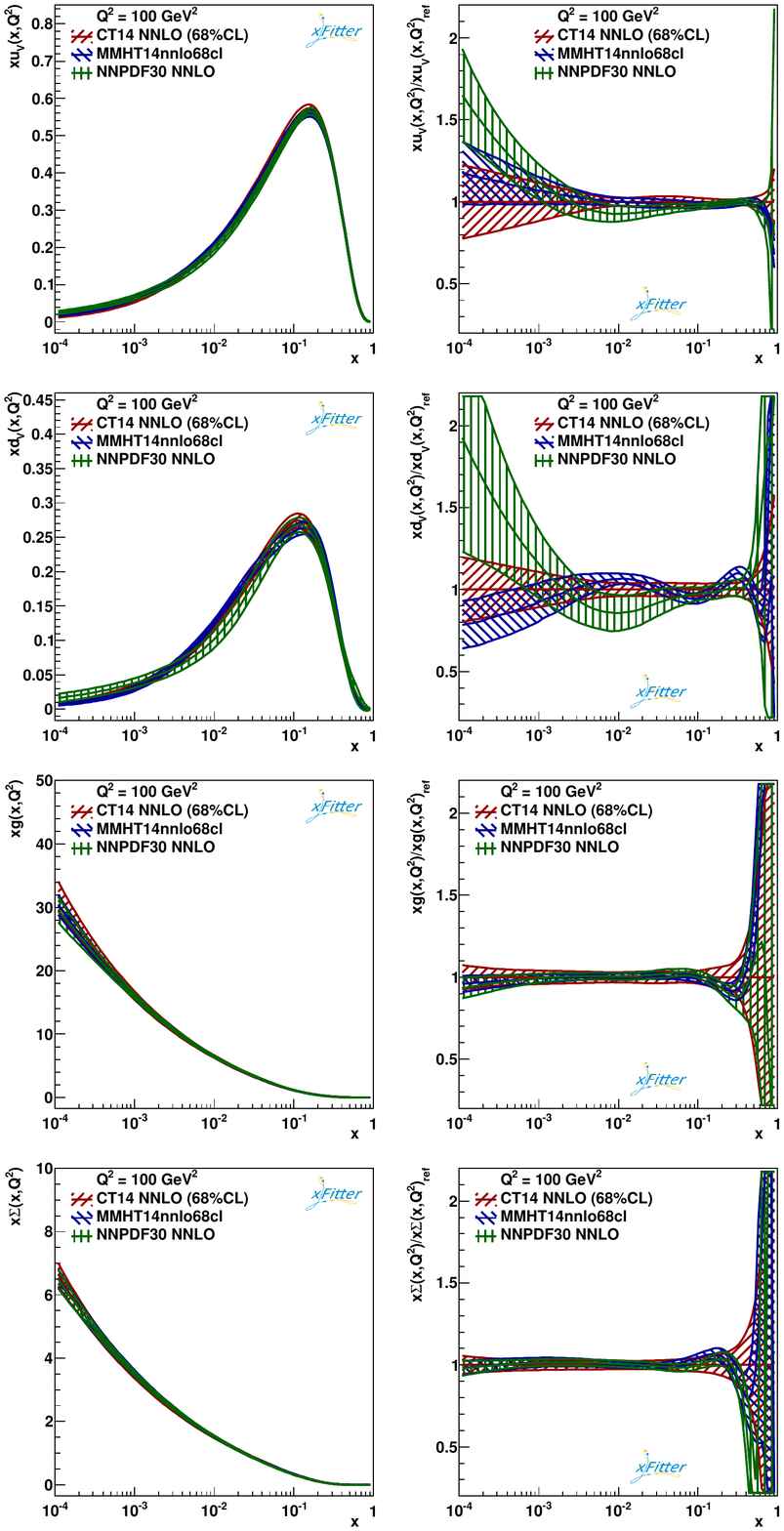}
\vspace*{-5mm}
\caption{\small 
\label{fig:pdfs-ct14-100} 
Same as Fig.~\ref{fig:pdfs-ct14-4} at the scale $Q^2=100~\GeV^2$
with the sea $ x\Sigma = 2x(\bar{u}+\bar{c}+\bar{d}+\bar{s}+\bar{b})$.
}
\end{center}
\end{figure}

\begin{figure}[tH!]
\begin{center}
\vspace*{-5mm}
\includegraphics[width=0.95\textwidth]{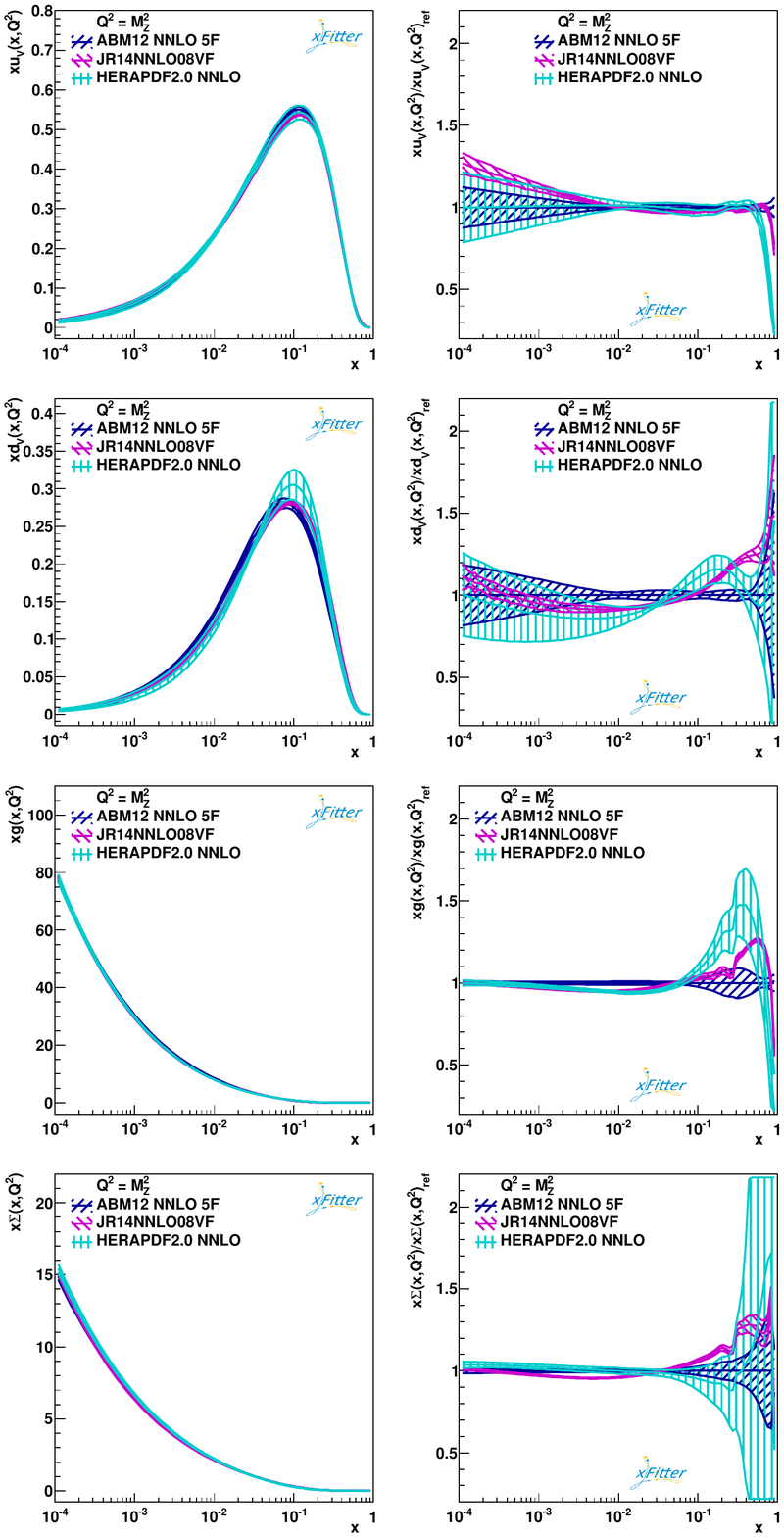}
\vspace*{-5mm}
\caption{\small 
\label{fig:pdfs-abm-MZ} 
Same as Fig.~\ref{fig:pdfs-abm-100} at the scale $Q^2=M_Z^2$.
}
\end{center}
\end{figure}

\begin{figure}[tH!]
\begin{center}
\vspace*{-5mm}
\includegraphics[width=0.95\textwidth]{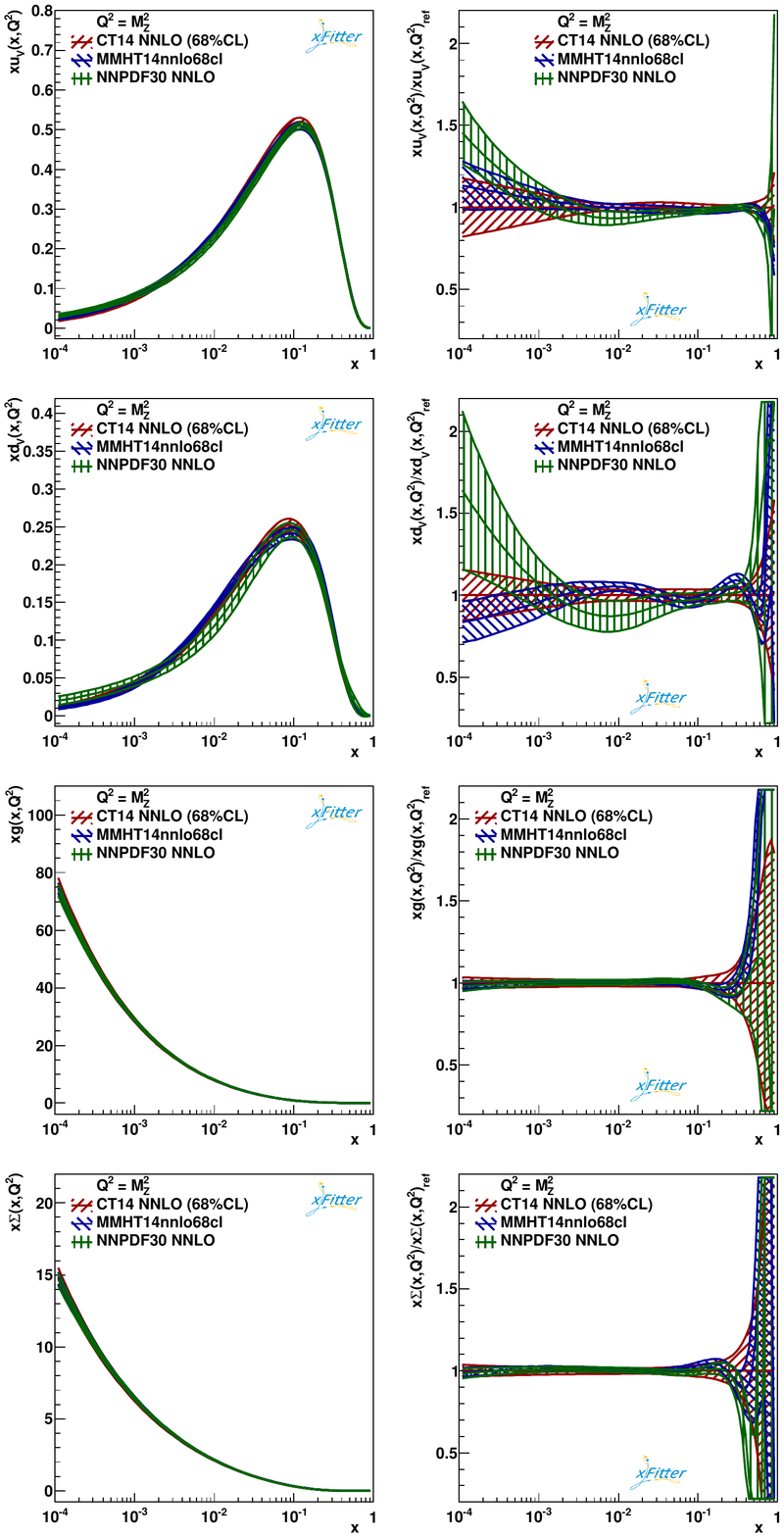}
\vspace*{-5mm}
\caption{\small 
\label{fig:pdfs-ct14-MZ} 
Same as Fig.~\ref{fig:pdfs-ct14-100} at the scale $Q^2=M_Z^2$.
}
\end{center}
\end{figure}

\subsection{Results for PDFs}

Before we start a detailed discussion of the theoretical aspects of the PDF
determinations we would like to illustrate the present status of PDF sets at
NNLO in QCD and discuss briefly some differences, which are clearly visible.
The currently available sets at NNLO in QCD are shown in Figs.~\ref{fig:pdfs-abm-4}--\ref{fig:pdfs-ct14-MZ}. 
The light-quark ($u$, $d$) valence PDFs together with the gluon and the quark sea distributions 
($x\Sigma = 2x(\bar{u}+\bar{c}+\bar{d}+\bar{s})$ for four active flavors) 
with the respective uncertainty bands are 
displayed in Figs.~\ref{fig:pdfs-abm-4}, \ref{fig:pdfs-abm-100} and \ref{fig:pdfs-abm-MZ} 
at the scales $Q^2=4~\GeV^2$, $100~\GeV^2$ and $M_Z^2$ in the range $10^{-4} \le x \le 1$
for the sets ABM12~\cite{Alekhin:2013nda}, HERAPDF2.0~\cite{Abramowicz:2015mha} and JR14~\cite{Jimenez-Delgado:2014twa}. 
Likewise,  Figs.~\ref{fig:pdfs-ct14-4}, \ref{fig:pdfs-ct14-100} and \ref{fig:pdfs-ct14-MZ} 
show the sets CT14~\cite{Dulat:2015mca}, MMHT14~\cite{Harland-Lang:2014zoa} and NNPDF3.0~\cite{Ball:2014uwa}.

The main features of the present NNLO PDFs in Figs.~\ref{fig:pdfs-abm-4}--\ref{fig:pdfs-ct14-MZ} 
in the main kinematic region of $x$ and $Q^2$ relevant for hard scattering events at Tevatron and the LHC 
can be characterized as follows.
The agreement in the distributions $xu_v$, and to a slightly lesser extent $\Sigma$,
is very good for ABM12, JR14 and HERAPDF2.0, as shown in Fig.~\ref{fig:pdfs-abm-4}. 
For the valence PDF $xd_v$ there is also an overall reasonable agreement, 
but the distribution deviates by more than  $1 \sigma$ at $x \gtrsim 0.1$ in the case of HERAPDF2.0. 
One should note that $xd_v$ is more difficult to measure in $e^{\pm}p$ DIS at HERA than $xu_v$ 
and additional constraints from deuteron data are important to fix the details of this PDF, 
as discussed in Sec.~\ref{sec:th4pdfs} below.

The results on the gluon momentum distribution $xg$ are clearly different at low values of $x$. 
Here, JR14 obtains the largest values, followed by ABM12 and HERAPDF2.0, 
with the latter displaying a valence-like shape below $x = 10^{-3}$. 
For CT14, MMHT14 and NNPDF3.0 there is very good agreement for $xu_v$, cf.~Fig.~\ref{fig:pdfs-ct14-4}. 
Some differences are visible in case of $xd_v$, where CT14 reports larger
values than NNPDF3.0 at $x \gsim 5 \cdot 10^{-3}$ and vice versa for smaller $x$. 
The spread in $\Sigma$ for the sets in Fig.~\ref{fig:pdfs-ct14-4} is much greater
than those by ABM12, JR14 and HERAPDF2.0. 
This is true as well for the gluon PDF $xg$ with the CT14 uncertainty band for the gluon PDF also covering the predictions
  for the distributions by ABM12, and HERAPDF2.0.
Note that the error bands for CT14 in Figs.~\ref{fig:pdfs-ct14-4}, \ref{fig:pdfs-ct14-100} and \ref{fig:pdfs-ct14-MZ} 
correspond to the c.l. of 68\%. 

The disagreement in $xd_v$ between HERAPDF2.0 and ABM12 or JR14 persists through the evolution from $Q^2 = 4~\GeV^2$ to 
$Q^2 = M_Z^2$, cf.~Fig.~\ref{fig:pdfs-abm-100} and \ref{fig:pdfs-abm-MZ}. 
Likewise, the spread in $xd_v$ between CT14, MMHT14 and NNPDF3.0 becomes more pronounced,
as shown in Fig.~\ref{fig:pdfs-ct14-100} and \ref{fig:pdfs-ct14-MZ}.  
On the other hand, differences in the singlet PDFs $\Sigma$ and $xg$, while still somewhat visible at $Q^2 = 100~\GeV^2$, 
largely wash out at scales $Q^2 = M_Z^2$ which govern the physics of central rapidity events at the LHC.
Those remaining differences persist at large scales (as in the case of the gluon PDFs at large $x >
0.1$) and will have a significant impact. 
The crucial test for all PDF sets comes through a detailed comparison of cross section predictions to data. 
This will be discussed in the remainder of the
paper, in particular in Secs.~\ref{sec:th4pdfs} and \ref{sec:crs-at-lhc}.

\newpage
\cleardoublepage

\section{Theory for PDF fits}
\label{sec:th4pdfs}

In the following we describe the basic theoretical issues for a consistent
determination of the twist-two PDFs from DIS and other hard scattering data, on
the basis of perturbative QCD at NNLO using the \msbar\ scheme for renormalization 
and factorization.

\subsection{Theory for analyses of DIS data}

The world DIS data are provided in terms of reduced cross sections by the different experiments.
QED and electroweak radiative corrections \cite{Kwiatkowski:1990es,Arbuzov:1995id} are applied,  
which requires careful study of different kinematic variables 
\cite{Blumlein:1990wz,Blumlein:1994ii,Blumlein:2002fy,Arbuzov:1995id}. 
In this way also the contributions from the exchange of more than one gauge boson 
to the partonic twist-2 terms are taken care of. 
In part, also the very small QED corrections to the hadronic tensor are already accounted for. 
These have a flat kinematic behavior and amount to ${\cal O}(1\%)$ 
or less \cite{Kripfganz:1988bd,Blumlein:1989gk,Spiesberger:1994dm,Roth:2004ti}.

The reduced cross sections are differential in either two of the kinematic variables in the set $\{x,y,Q^2\}$. 
The virtuality $Q^2 = -q^2$ of the process is given by the 4-momentum transfer $q$ to the hadronic system.
The Bjorken variable is defined as $x = Q^2/(s y)$, with $y = 2 p \cdot q/s$, and 
$s = (p+l)^2$ the squared center-of-mass energy, where $p$ and $l$ denote the 4-momenta of the nucleon and the lepton.
At energies much greater than the nucleon mass $M$, in the nucleon rest frame $y$ is the fractional energy of 
the lepton transferred to the nucleon. 
The double differential cross sections used in the QCD analyses are given by \cite{Arbuzov:1995id,Blumlein:2012bf,Agashe:2014kda}
\begin{eqnarray}
\label{eq:ncB}
\frac{d^2 \sigma_{\rm NC}^{l^\pm N}}{dx dy} &=& 
\frac{2 \pi \alpha^2 s}{Q^4} \Biggl\{ \left[ 2(1-y) - 2xy \frac{M^2}{s}
\right] F^{NC}_2(x,Q^2) 
+ Y_- xF^{NC}_3(x,Q^2) 
\nonumber\\ &&
+\, y^2\left(1 - \frac{2m_l^2}{Q^2}\right)
2xF^{NC}_1(x,Q^2)
\Biggr\}\, ,
\\
\label{eq:ncNU}
\frac{d^2 \sigma_{\rm NC}^{\nu(\bar{\nu})N}}{dx dy} &=& \frac{G_F^2 s}{16 \pi} \left[ \frac{M_Z^2}{Q^2 +M_Z^2}\right]^2
\left\{Y_+ W_2^{\rm NC}(x,Q^2) \pm Y_- xW_3^{\rm NC}(x,Q^2) - y^2 W_L^{\rm NC}(x,Q^2)\right\}\, ,
\\
\label{eq:ccB}
\frac{d^2 \sigma_{\rm CC}}{dx dy} &=& \frac{G_F^2 s}{4 \pi} \left[ \frac{M_W^2}{Q^2 + M_W^2}\right]^2
\left\{Y_+ W_2^{\rm CC}(x,Q^2) \pm Y_- xW_3^{\rm CC}(x,Q^2) - y^2 W_L^{\rm CC}(x,Q^2)\right\}\, ,
\end{eqnarray}
where $\alpha$ and $G_F$ denote the fine-structure and Fermi constants, $Y_\pm = 1 \pm (1-y)^2$ 
and we keep the dependence on the masses of the nucleon ($M$), the $W$ and $Z$ boson ($M_{W}$, $M_{Z}$)
and the lepton ($m_l$).

The structure functions $F^{NC}_i$ and $W_i$ are nonperturbative quantities defining the hadronic tensor.
They can be measured by varying $y$ at fixed $Q^2$ and $x$ and form the input to the subsequent analysis.
Note that in some previous experiments, assumptions were made about the longitudinal
structure functions $F^{NC}_L$ and $W_L$, where (in the massless limit)
\begin{eqnarray}
  \label{eq:CallanGross}
  F^{NC}_L(x,Q^2) = F^{NC}_2(x,Q^2) - 2x F^{NC}_1(x,Q^2)\, ,
\end{eqnarray}
since at the time of the data analysis 
the corresponding QCD corrections were still missing.
Therefore, it is important to use the differential cross sections in Eqs.~(\ref{eq:ncB})--(\ref{eq:ccB}) 
and to add the correct longitudinal structure functions \cite{Moch:2004xu,Vermaseren:2005qc}, 
cf.~also \cite{Blumlein:2006be,Alekhin:2011ey}.
The structure functions are measured for DIS off massive proton and deuteron targets and are, therefore, subject 
to target mass corrections, which play an important role in the region of lower values of $Q^2$ and larger values of $x$. 
They are available in Refs.~\cite{Georgi:1976ve,Blumlein:2012bf,Steffens:2012jx}.

The neutral- and charged-current structure functions $F^{NC}_i$, $W^{NC}_i$ and $W^{CC}_i$ 
consist of a sum of several terms, each weighted by powers of the QED and electroweak couplings, 
and $F^{NC}_i$ also include the $\gamma-Z$ mixing, which has to be accounted for, cf.~\cite{Arbuzov:1995id,Blumlein:2012bf,Agashe:2014kda}. 
Then, considering one specific gauge boson exchange, 
one arrives at a representation for the individual structure functions $F_i$, 
which are only subject to QCD corrections. 
For example, for pure photon exchange, they are given by
\begin{eqnarray}
F_i(x,Q^2) =  F_i^{\tau =2}(x,Q^2) + \sum_{k=2}^\infty \frac{C^{\tau=2k}_i(x,Q^2)}{Q^{2(k-1)}}~,
\end{eqnarray}
where $F_i^{\tau =2}$ denotes the leading-twist term and the coefficients $C^\tau_i$ parametrize
the higher twist contributions. 
The latter terms are of relevance for many DIS data sets, see Sec.~\ref{sec:datapdfs}.

Present day QCD analyses are aimed at determining the leading-twist contributions to the structure functions.
There are two ways to account for the higher twist terms:
\begin{itemize}

\item[{\bf (i)}] 
  One is fitting the higher twist terms in $F_i$. 
  A rigorous approach requires the knowledge of their scaling violations
  (term by term) and of the various Wilson coefficients to higher orders in $\alpha_s$, 
  see e.g. Sec.~16 in Ref.~\cite{Blumlein:2012bf}. Since at present this is practically out of reach, 
  such fits remains rather phenomenological. 
  Moreover, the size of the (non-singlet) higher twist contributions to the structure function $F_2$ 
  vary strongly with the correction applied to the leading-twist term up to 
  next-to-next-to-next-to-leading order (N$^3$LO),
  as shown in Ref.~\cite{Blumlein:2006be,Blumlein:2008kz}.
  Also, the non-singlet and singlet higher twist contributions are different~\cite{Alekhin:2000ch,Alekhin:2012ig}.

\item[{\bf (ii)}] 
  One has to find appropriate cuts to sufficiently reduce the higher twist terms.
  For instance, in the flavor non-singlet analysis of Ref.~\cite{Blumlein:2006be} 
  the cuts are taken to be $Q^2 \geq 4~\GeV^2$, $W^2 = M^2 + Q^2(1-x)/x \geq 12.5~\GeV^2$.
  In the combined singlet and non-singlet analysis of Ref.~\cite{Alekhin:2012ig},
  $Q^2 \geq 10~\GeV^2$, $W^2 \geq 12.5~\GeV^2$ have been used.
  These bounds are found empirically by cutting on $W^2$ and/or $Q^2$ starting
  from larger values. 
  Applying these cuts severely limits the amount of large-$x$ DIS data
  to be fitted, and usually leads to an increase of the errors of $\alpha_s(M_Z)$ 
  and other fitted fundamental parameters and distributions.
\end{itemize} 
Both methods {\bf (i)} and {\bf (ii)} allow to access the leading-twist
contributions to the DIS structure functions, with some qualifications, however.

The cuts suggested in {\bf (ii)} remove the large-$x$ region potentially
sensitive to the higher twist terms. 
However, they do not affect the data at $x \lesssim 0.1$, 
where higher twist terms still play an important role~\cite{Alekhin:2012ig,Harland-Lang:2016yfn}. 
To some extent, the influence of higher twist can be dampened by using the DIS 
data for the structure function $F_2$ instead of the cross section, 
since in this case the contribution to the structure function $F_L$ need not
be considered.
It should be kept in mind, though, that the experimental separation of the structure functions $F_{2}$ and $F_{L}$ in 
the full phase space of common DIS experiments is
very difficult without dedicated longitudinal--transverse cross section separations.
Therefore, the data on $F_{2}$ and $F_{3}$ are typically extracted from the cross section 
once a certain model for the structure function $F_L$ is taken. 
This approach is justified only at large $x$, however, where the contribution of $F_L$ 
is small and even large uncertainties in the modeling of $F_L$ cannot affect the extracted values of $F_{2}$ and $F_{3}$.
The procedure is not applicable for HERA kinematics, on the other hand, and introduces a 
bias into the analysis of the data taken by the New Muon Collaboration (NMC), 
in particular, a shift in the value of $\alpha_s$ preferred by the fit~\cite{Alekhin:2011ey,Gao:2013xoa}, cf.~Sec.~\ref{sec:alphas}. 
Nonetheless, the MMHT14 analysis~\cite{Harland-Lang:2014zoa} is still based on the DIS structure 
function data, as are the CJ15 and CT14 analyses~\cite{Accardi:2016qay,Dulat:2015mca}. 
The latter two use cross section data for HERA, and for HERA and NMC, respectively, 
and structure function data elsewhere. 
While CT14 performed this important change for the HERA and NMC data, the authors of 
Ref.~\cite{Thorne:2011kq} report that the change has little impact. 
Refs.~\cite{Alekhin:2011ey,Alekhin:2012ig}, on the other hand, disagree with this claim.

The deep-inelastic structure functions are inclusive quantities and contain massless parton and heavy-quark 
contributions,
\begin{eqnarray}
F_i^{\tau = 2}(x,Q^2) =  F_i^{\rm massless}(x,Q^2) + F_i^{\rm massive}(x,Q^2)\, . 
\label{EQ:Fi}
\end{eqnarray}
Here the massless terms are given by
\begin{eqnarray}
F_i^{\rm massless}(x,Q^2) = \sum_{j}\, C_{i,j}\left(x,\frac{Q^2}{\mu^2}\right) \otimes f_j(x,\mu^2)\, ,
\label{eq:MASSL}
\end{eqnarray}
where $C_{i,j}$ denote the massless Wilson coefficients, $f_j$ the massless PDFs and 
$\mu^2$ is the factorization scale. 
The Mellin convolution is abbreviated by $\otimes$ and 
the sum over $j$ is over all contributing partons. 
The renormalization group equation for $F_i^{\rm massless}$ allows one to eliminate the 
dependence on $\mu^2$ order-by-order in perturbation theory. 
This also applies to $F_i^{\tau = 2}$. 
Through the massive contributions $F_i^{\rm massive}$ there is a dependence on the heavy-quark masses $m_c$ and $m_b$ in the 
present world DIS data. 
Note that $F_i^{\rm massive}$ is {\it not} the structure function of a
tagged heavy-flavor sample, which would be infrared sensitive~\cite{Chuvakin:1999nx}. 
Rather, $F_i^{\rm massive}$ is just given as the difference of the complete structure function 
$F_i^{\tau = 2}$ and the massless one in Eq.~(\ref{eq:MASSL}).

\subsubsection{Massless PDFs}

For all QCD calculations we use perturbation theory. The factorized
representation in terms of Wilson coefficients and PDFs is obtained using the light-cone
expansion~\cite{Wilson:1969zs,Zimmermann:1970,Brandt:1970kg,Frishman:1971qn}. 
For a proper definition of the Wilson coefficients and the PDFs 
one has to use the LSZ formalism and refer to asymptotic states at large times $t \rightarrow \pm \infty$,  
given by massless partons. 
We first describe the massless contributions in Eqs.~(\ref{EQ:Fi}) and (\ref{eq:MASSL}), and then
discuss the contribution of heavy quarks.
The Wilson coefficients in Eq.~(\ref{eq:MASSL}) have a perturbative expansion in the strong coupling constant.
At one- \cite{Furmanski:1981cw}, two- 
\cite{Zijlstra:1991qc,vanNeerven:1991nn,Zijlstra:1992qd,Kazakov:1987jk,Kazakov:1990fu,SanchezGuillen:1990iq,Zijlstra:1992kj,
Moch:1999eb}, and three-loop order \cite{Moch:2004xu,Vermaseren:2005qc,Moch:2007gx,Moch:2007rq,Moch:2008fj} 
they have been calculated for the neutral-current structure functions $F_{i}$, with $i=1,2,3$, 
except for the $\gamma-Z$ mixing contribution at three loops.

The structure functions in general depend on the following three non-singlet and singlet combinations of parton 
densities: 
\begin{eqnarray}
q_{jk}^{\pm} \,=\, f_j \pm \bar{f}_j - (f_k \pm \bar{f}_k) \, ,
\qquad\qquad
q^{v} \,=\, \sum_{l=1}^{n_f} (f_l - \bar{f}_l) \, ,
\qquad\qquad
q^{s} \,=\, \sum_{l=1}^{n_f} (f_l + \bar{f}_l)\, ,
\end{eqnarray}
with the light-quark distributions $f_i$ of flavor $i$ and $n_f$ the number of massless flavors.
These combinations evolve in $\mu^2$ from an initial scale $\mu_0^2$ by the QCD evolution equations, 
where the singlet distribution $q^{s}(x,\mu^2)$ mixes with the gluon
distribution $g(x,\mu^2)$,
\begin{eqnarray}
  \label{eq:evo1}
  \frac{d}{d \ln(\mu^2)} q^{i}(x,\mu^2) &=& 
  P^{i}(x) \otimes q^{i}(x,\mu^2)\, , \qquad\qquad i = \pm, v\, ,
  \\
  \label{eq:evo2}
  \frac{d}{d \ln(\mu^2)} 
  \left(\begin{array}{c} q^{s}(x,\mu^2)\\
      g(x,\mu^2) \end{array} \right)
  &=& 
  \left(\begin{array}{cc} P_{qq}(x) &  P_{qg}(x) \\
      P_{gq}(x) &  P_{gg}(x) \end{array} \right) \otimes
  \left(\begin{array}{c} q^{s}(x,\mu^2)\\
      g(x,\mu^2) \end{array} \right)
  \, .
\end{eqnarray}
The non-singlet splitting functions are given by
\begin{eqnarray}
P^{\pm} (x) &=& P_{qq}(x) \pm P_{q\bar{q}}(x)\, ,
\\
P^{v} (x) &=& P_{qq}(x) - P_{q\bar{q}}(x) + {n_f}\left(P_{qq}^s(x) - P_{q\bar{q}}^s(x)\right)
\, ,
\end{eqnarray}
while the anomalous dimensions $\gamma_{ij}$ corresponding to the splitting
functions $P_{ij}$ are obtained by a Mellin transform,
\begin{eqnarray}
\label{eq:Mellindef}
\gamma_{ij}(N) = - \int_0^1 dx\, x^N\, P_{ij}(x)
\, ,
\end{eqnarray}
where we suppress for brevity the dependence of $P_{ij}$ and $\gamma_{ij}$ on 
the strong coupling $a_s(\mu^2) = \alpha_s(\mu^2)/(4\pi)$. 
The $P_{ij}$ are known as well 
at one- 
\cite{Gross:1973ju,Gross:1974cs,Georgi:1951sr,Parisi:1976qj,Kim:1977hp,Altarelli:1977zs}, 
two- 
\cite{Floratos:1977au,Floratos:1978ny,GonzalezArroyo:1979df,GonzalezArroyo:1979ng,Curci:1980uw,Furmanski:1980cm,
Floratos:1980hk,Floratos:1981hs,GonzalezArroyo:1979he,Floratos:1980hm,Hamberg:1991qt,Ellis:1996nn,Moch:1999eb}
and at three-loop order \cite{Moch:2004pa,Vogt:2004mw} 
(see also \cite{Ablinger:2014nga,Charalampos:2016xyz} for checks of $P_{ps}$ and $P_{gg}$ at that order).
The scale evolution of the strong coupling constant in the $\overline{\rm MS}$ scheme is given by
\begin{eqnarray}
\frac{d a_s(\mu^2)}{d \ln(\mu^2)} = - \sum_{k=0}^\infty \beta_k\, a_s^{k+2}(\mu^2)\, ,
\end{eqnarray}
where $\beta_k$ denote the expansion coefficients of the QCD $\beta$-function
\cite{Khriplovich:1969aa,tHooft:unpub,Politzer:1973fx,Gross:1973id,Caswell:1974gg,
Jones:1974mm,Tarasov:1980au,Larin:1993tp,vanRitbergen:1997va,Czakon:2004bu}.

The evolution equations (\ref{eq:evo1}), (\ref{eq:evo2}) can be either solved in $x$- or Mellin (or moment) $N$-space. 
In Mellin-space, defined by the transform Eq.~(\ref{eq:Mellindef}), 
an analytic solution is possible~\cite{Diemoz:1987xu,Gluck:1991ee,Ellis:1993rb,Blumlein:1997em}
by arranging the solution systematically in powers of the coupling constants $a_s(\mu^2)$ and $a_s(\mu_0^2)$,
and even forming factorization-scheme invariant expressions.
In case of the $x$-space solutions this is usually not done due to the necessary iterative solution. 
In the small-$x$ region the iterative solution usually leads to a pile-up of a few per cent~\cite{Blumlein:1996gv}. 
This can be corrected for in $x$-space solutions by applying the method given in~\cite{Rossi:1983xz}. 
Likewise, the iterated solution can be obtained in Mellin $N$-space 
and is a standard option of the evolution program {\tt QCD-Pegasus}~\cite{Vogt:2004ns}.

\subsubsection{Heavy-quark structure functions}

Disregarding contributions from charm at the input scale (``intrinsic charm''),
cf.~\cite{Brodsky:1980pb,Jimenez-Delgado:2014zga,Blumlein:2015qcn}, 
the heavy-flavor corrections to the DIS functions are described by Wilson coefficients.
The leading order results are of ${\cal O}(a_s)$. 
Higher order corrections in the perturbative expansions are, therefore, 
of ${\cal O}(a_s^2)$ at NLO, and of ${\cal O}(a_s^3)$ at NNLO, similar to the case
of the longitudinal structure function \cite{Moch:2004xu,Vermaseren:2005qc}.
The corrections in the neutral- and charged-current cases are available in 
one- \cite{Witten:1975bh,Babcock:1977fi,Novikov:1977yc,Leveille:1978px,Gluck:1980cp,Gottschalk:1980rv,
Gluck:1996ve,Blumlein:2011zu} and two-loop
order \cite{Laenen:1992zk,Laenen:1992xs,Riemersma:1994hv,Bierenbaum:2009zt}, 
where the latter corrections were given in semi-analytic form. 

For the neutral-current exchange the heavy-flavor contributions to the structure functions
$F_{i}$ with $i=2,L$ are \cite{Behring:2014eya,Bierenbaum:2009mv}:
\begin{eqnarray}
  \label{eqF2}
  F_{i}^{\rm massive}(x,n_f\!\!\!&+&\!\!\!2,Q^2,m^2_c,m^2_b) = 
  \nonumber\\
  && 
  \sum_{i=c,b} x \Biggl\{\sum_{k=1}^{n_f}e_k^2\Biggl\{
  L_{i,q}^{ns}\left(x,n_f+1,\frac{Q^2}{\mu^2},\frac{m^2_i}{\mu^2}\right)
  \otimes
  \Bigl[f_k(x,\mu^2,n_f)+f_{\bar k}(x,\mu^2,n_f)\Bigr]
  \nonumber\\ &&\hspace{14mm}
  +\frac{1}{n_f}L_{i,q}^{ps}\left(x,n_f+1,\frac{Q^2}{\mu^2},\frac{m^2_i}{\mu^2}\right)
  \otimes q^{s}(x,\mu^2,n_f)
  \nonumber\\ &&\hspace{14mm}
  +\frac{1}{n_f}L_{i,g}^{s}\left(x,n_f+1,\frac{Q^2}{\mu^2},\frac{m^2_i}{\mu^2}\right)
  \otimes g(x,\mu^2,n_f)
  \Biggr\}
  \nonumber\\
  &+&e_i^2\Biggl[
  H_{i,q}^{ps}\left(x,n_f+1,\frac{Q^2}{\mu^2},\frac{m^2_i}{\mu^2}\right)
  \otimes q^{s}(x,\mu^2,n_f)
  \nonumber\\ &&\hspace{7mm}
  +H_{i,g}^{s}\left(x,n_f+1,\frac{Q^2}{\mu^2},\frac{m^2_i}{\mu^2}\right) 
  \otimes g(x,\mu^2,n_f)
  \Biggr]\Biggr\} \nonumber\\ &&
  + \delta_{i,2} F_{2}^{{\rm massive},\{c,b\}}(x,n_f + 2,Q^2,m^2_c,m_b^2)
  \, . 
\end{eqnarray}
They are determined by five massive Wilson coefficients, $L_{i,k}^{\{ns,ps,s\}}$ and $H_{i,k}^{\{ps,s\}}$, where the electroweak current 
couples either to a massless ($L_{i,k}$) or the massive ($H_{i,k}$) quark line. 
From three-loop order onwards there are contributions containing both heavy flavors $c$ and $b$ 
in a non-separable form, denoted by $F_{2}^{{\rm massive},\{c,b\}}$, in Eq.~(\ref{eqF2}).
The PDFs and the coupling constant in Eq.~(\ref{eqF2}) are defined in the \msbar scheme, 
while the heavy quark masses are taken either in the on-shell or \msbar schemes \cite{Bierenbaum:2009mv,Alekhin:2010sv}.
The relations of the heavy quark masses between the pole mass (on-shell scheme) 
and the $\overline{\rm MS}$ scheme are available to four-loop order~\cite{Marquard:2015qpa}. 
Due to its better perturbative stability, the $\overline{\rm MS}$ scheme for
the definition of the heavy-quark mass is preferred.

For $Q^2 \gg m_i^2$ the asymptotic corrections to $F_L$ are available at three-loop order~\cite{Blumlein:2006mh,Behring:2014eya}.
For $F_2$, four out of the five massive Wilson coefficients, $L_{2,q}^{ns}$, $L_{2,q}^{ps}$, $L_{2,g}^{s}$ and $H_{2,q}^{ps}$ are known as
well~\cite{Ablinger:2010ty,Blumlein:2012vq,Behring:2014eya,Ablinger:2014vwa,Ablinger:2014nga} 
at large scales $Q^2$.
For the remaining coefficient, $H_{2,g}^{s}$, an estimate has been made in Ref.~\cite{Kawamura:2012cr} 
based on the anticipated small-$x$ behavior \cite{Catani:1990eg}, a series of moments 
calculated in \cite{Bierenbaum:2009mv}, and two-loop operator matrix elements from 
Refs.~\cite{Bierenbaum:2007qe,Bierenbaum:2008yu}. 
This provides a good approximation of the NNLO corrections.

\subsection{Heavy-flavor PDFs}

An important issue in PDF fits concerns the number of active quark flavors 
and the theoretical description of heavy quarks such as charm and bottom.
Due to the large range of hard scales $Q$ for the scattering processes considered,
different effective theories may be applied.
At low scales, when $Q \simeq {\cal O}({\rm few})~\GeV$,  
one typically works with $n_f = 3$ massless quark flavors,  
setting $n_f = 3$ in the hard scattering cross section, the evolution kernels and the anomalous dimensions.
In this case, only the light-quark PDFs for up, down and strange are taken into account.
At higher scales, e.g., for hadro-production of jets at high transverse
momentum $p_t$ or top quarks, 
additional dynamical degrees of freedom lead to theories with $n_f > 3$.
By means of the renormalization group and matching these are related to the case with $n_f = 3$ massless quarks.
Technically, one has to apply decoupling relations~\cite{Appelquist:1974tg} at some matching scale $\mu$, 
for instance in the transition of $\alpha_s^{(n_f)} \to \alpha_s^{(n_f+1)}$. This
introduces some logarithmic dependence 
on the masses of the heavy quarks $m_c$, $m_b$ and $m_t$ for charm, bottom and top.
One should also note that the matching of the effective theories for $n_f \to n_f+1$ does not need to be smooth.
In fact, it introduces discontinuities, such as for the running coupling as a solution
of the QCD $\beta$-function at higher order in the perturbative expansion, where
$\alpha_s^{(n_f)}(\mu) \neq \alpha_s^{(n_f+1)}(\mu)$ in the \msbar scheme at the matching scale $\mu$, 
see e.g., \cite{Chetyrkin:2000yt}.

In a similar manner, PDFs in theories with a fixed number $n_f > 3$ of quark
flavors are related to those for $n_f = 3$ 
with the help of heavy-quark operator matrix elements (OMEs) $A_{ij}$ at a chosen matching scale $\mu$.
Potential non-universal non-logarithmic heavy-flavor effects are taken care of
by the Wilson coefficients.
Starting with the PDFs in a so-called fixed-flavor number scheme (FFNS) with 
$n_f$ fixed, one has $f_i^{(n_f)} \to f_i^{(n_f+1)}$ for the light-quark distributions $f_i$ 
and $(q^{s,\, (n_f)}, g^{(n_f)}) \to (q^{s,\, (n_f+1)}, g^{(n_f+1)})$ 
for the gluon and the singlet quark distributions 
with operator mixing in the singlet sector.
In particular, one has~\cite{Buza:1996wv,Bierenbaum:2009mv}
\begin{eqnarray}
  \label{eq:VFNS1}
  f_k(n_f+1, \mu^2) + f_{\bar k}(n_f+1, \mu^2)
  &=& A_{qq,h}^{ns}\Big(n_f, \frac{\mu^2}{m^2}\Big) \otimes \left[f_k(n_f, \mu^2) + f_{\bar k}(n_f, \mu^2)\right]
  \nonumber\\
  &&
  + \frac{1}{n_f} A_{qq,h}^{ps}\Big(n_f, \frac{\mu^2}{m^2}\Big) \otimes {q^{s}(n_f, \mu^2)}
  \nonumber\\
  && + \frac{1}{n_f} A_{qg,h}^{s}\Big(n_f, \frac{\mu^2}{m^2}\Big) \otimes {g(n_f, \mu^2)}
  \, ,
  \\
  g(n_f+1, \mu^2)
  &=& A_{gq,h}^{s}\Bigl(n_f,\frac{\mu^2}{m^2}\Bigr) \otimes q^{s}(n_f,\mu^2)
  + A_{gg,h}^{s}\Bigl(n_f,\frac{\mu^2}{m^2}\Bigr) \otimes g(n_f,\mu^2)
  \, ,
  \\
  q^{s}(n_f+1,\mu^2)
  &=& \left[ A_{qq,h}^{ns}\left(n_f, \frac{\mu^2}{m^2}\right) + 
    A_{qq,h}^{ps} \left(n_f, \frac{\mu^2}{m^2}\right) +
    A_{hq}^{ps} \left(n_f, \frac{\mu^2}{m^2}\right) \right]
  \otimes q^{s}(n_f,\mu^2) \nonumber\\ &&
  +\left[A^{s}_{qg,h}\left(n_f, \frac{\mu^2}{m^2}\right)
    + A^{s}_{hg}\left(n_f, \frac{\mu^2}{m^2}\right) \right]
  \otimes g(n_f,\mu^2)
  \, .
  \label{eq:VFNS2}
\end{eqnarray}
PDFs for charm and bottom ($h=c,b$) are then constructed as
\begin{eqnarray}
  \label{eq:VFNS3}
  f_{h+\bar h}(n_f+1, \mu^2)
  &=& {A_{hq}^{ps}\Big(n_f, \frac{\mu^2}{m^2}\Big)} \otimes {q^{s}(n_f, \mu^2)}
  + {A_{hg}^{s}\Big(n_f, \frac{\mu^2}{m^2}\Big)} \otimes {g(n_f, \mu^2)}
\end{eqnarray}
at the matching scale $\mu$ from the quark singlet and gluon PDFs with $h = {\bar h}$.

The matching conditions are typically imposed at the scale $\mu = m_h$,
and $f_{h + {\bar h}} = 0$ is assumed for scales $\mu \le m_h$.
The necessary heavy-quark OMEs $A_{ij}$ depend logarithmically on the heavy-quark masses 
as $\alpha_s^l \ln^k(\mu^2/m_h^2)$ with $0 \le k \le l$ in the perturbative expansion.
As discussed above, the OMEs are known to NLO analytically~\cite{Buza:1995ie,Bierenbaum:2007qe} 
and at NNLO either exactly or to a good approximation~\cite{Bierenbaum:2009mv,Ablinger:2010ty,Kawamura:2012cr,Ablinger:2014lka,Ablinger:2014nga}. 
Thus, charm and bottom PDFs can be consistently extracted in QCD with a fixed number $n_f =3,4$ or 5. 

It should be stressed, however, that the decoupling relations for PDFs in Eqs.~(\ref{eq:VFNS1})--(\ref{eq:VFNS3})
assume the presence of one heavy quark at a time 
upon moving from lower scales to higher ones. 
Beginning at three-loop order, however, there are graphs containing both charm- and bottom-quark 
lines, and charm quarks cannot be treated as massless at the scale of the bottom-quark due to $(m_c/m_b)^2 \approx 1/10$. 
Such terms cannot be attributed to either the charm- or bottom-quark PDFs,
but rather one has to decouple charm and bottom quarks together at some large scale. 
The simultaneous decoupling of bottom and charm quarks in the strong coupling constant $\alpha_s$
is discussed, for instance, in Ref.~\cite{Grozin:2011nk}.

%
\input{table-mc}

\newpage
\cleardoublepage

\subsection{Heavy-quarks schemes}

\subsubsection{Variable-flavor number schemes}

The hard scattering cross sections also depend on the number of flavors $n_f$ and additional parton channels may open up,
which have to be included as well.
In addition, processes involving massive quarks depend logarithmically
on the ratio $Q^2/m_h^2$, where $Q$ is some hard scale associated with the scattering.
For the heavy-flavor Wilson coefficients in Eq.~(\ref{eqF2}) these logarithms 
are of the type $\alpha_s^l \ln^k(Q^2/m_h^2)$ with $1 \le k \le l$ in perturbation theory.
These originate from collinear singularities screened by the heavy-quark mass 
due to the constrained phase space for gluon emission from massive quark lines,
and as a prefactor of these logarithms one has the standard splitting functions.
In addition to logarithmic terms, there are also power corrections 
$(m_h^2/Q^2)^l$ in the heavy-flavor Wilson coefficients, 
usually appearing in form of higher transcendental functions. 
In the asymptotic regime of $Q^2 \gg m_h^2$ the logarithms dominate and
the kinematic dependence is measured experimentally, for instance  
in the tagged flavor case for charm-quark pairs in the structure function $F_2^{c\bar{c}}$.
Logarithms of a similar kind are also experimentally observed 
in differential distributions, e.g. due to the QED corrections proportional to
$\ln^k(Q^2/m_l^2)$ with $m_l$ being the charged lepton mass, 
cf.~\cite{Kwiatkowski:1990es,Arbuzov:1995id}. 

The resummation of the logarithms $\alpha_s^l \ln^k(Q^2/m_h^2)$ 
to all orders in perturbation theory is effectively carried out by
the transition $n_f \to n_f+1$ along with the introduction of new
heavy-quark PDFs as described in Eqs.~(\ref{eq:VFNS1})--(\ref{eq:VFNS3}).
Whether such a transition is appropriate or not depends, of course, on the detailed kinematics. 
If the hard scale is closer to threshold, $Q^2 \simeq m_h^2$, 
a description with $n_f$ light flavors is more suitable, while for $Q^2 \gg m_h^2$ 
one switches to a theory with $n_f+1$ massless flavors.
In order to achieve a unified description for hard scattering cross sections 
both at low scales $Q^2 \simeq m_h^2$ and asymptotically for $Q^2 \gg m_h^2$,  
so-called variable-flavor number schemes (VFNS) have been constructed. 
Effectively, these aim at an interpolation between the asymptotic limits 
of the quarks being very light or very heavy relative to the other hard scales of the process.
At the LHC such considerations apply to processes with bottom quarks in the initial state such as 
single top-quark production as well as bottom-quark initiated Higgs boson production
(see Ref.~\cite{Maltoni:2012pa} for more recent studies and Ref.~\cite{Harlander:2011aa} 
for the so-called {\it Santander matching}
scheme for Higgs boson production in $b\bar b$ annihilation).

Of particular interest for PDF fits is the reduced cross section to the pair-production of
heavy quarks in DIS, which is parametrized in terms of the DIS heavy-quark
structure functions $F_i^h$ for $i=2,L$ in Eq.~(\ref{eqF2}) and 
with heavy-flavor Wilson coefficients which are known exactly at NLO~\cite{Laenen:1992zk},
and to a good approximation at NNLO~\cite{Kawamura:2012cr} in QCD.
For the interpolation $n_f \to n_f+1$ of the heavy-quark structure functions $F_i^h$ 
a number of so-called general-mass VFNS (GM-VFNS) have been discussed in the literature, such as 
ACOT~\cite{Aivazis:1993pi,Kramer:2000hn,Tung:2001mv}, 
BMSN~\cite{Buza:1996wv}, 
FONLL~\cite{Forte:2010ta} or RT~\cite{Thorne:2012az}.
These keep $m_h \neq 0$ and are to be distinguished from the zero-mass VFNS (ZM-VFNS), which
describes essentially the massless case.
Note that presently the GM-VFNS are applied only 
to one single heavy flavor at the time. 
That is the sequential transition $n_f \to n_f+1$, 
so that the charm or bottom quarks are not considered simultaneously 
and charm-quark mass effects in the bottom-quark structure function $F_i^b$ are neglected 
as discussed above.

The various GM-VFNS contain a number of additional assumptions, and some come in more than one variety.
The GM-VFNS differ, for instance, in the way the low-$Q^2$ region is modeled. 
This modeling is a necessary undertaking to provide a reasonable behavior of the VFNS 
in the kinematical regime of present DIS data.
Additional assumptions in the GM-VFNS are related to the matching
scale $\mu$ for the transition $n_f \to n_f+1$ as the adopted choice $\mu = m_h$ is not unique,
see~\cite{Alekhin:2013fua} for an in-depth discussion.

Briefly, the problem can be illustrated with the heavy-quark velocity, the leading order 
formula \cite{Gluck:1980cp} being
\begin{eqnarray}
\label{eq:velocity}
v \,=\, \sqrt{ 1 - \frac{4 m_h^2}{s}} \,=\, \sqrt{ 1 - \frac{4 m_h^2 x}{Q^2(1-x)}} 
\, ,\qquad\qquad
x \le \frac{1}{1+ 4 m_h^2/Q^2}
\, .
\end{eqnarray}
The transition $n_f \to n_f+1$ when the corresponding flavor is considered as nearly massless 
requires light-like velocities $v \simeq 1$. 
That implies the absence of all power corrections $(m_h^2/Q^2)^l$ in the heavy-flavor Wilson coefficients at the matching scale $\mu^2$.
In practice, the matching is often applied at the scale $\mu^2 = m_h^2$ and
for kinematics $Q^2 \ngg m_h^2$, where this condition is not fulfilled, 
which implies restrictions on the range in $x$ in Eq.~(\ref{eq:velocity}).

Finally, the logarithmic accuracy of the resummation for large scales $Q^2 \gg m_h^2$, 
or the order of perturbation theory in current implementations of GM-VFNS, 
is often not consistent. For example, NNLO evolution~\cite{Moch:2004pa,Vogt:2004mw} 
of the massless PDFs is sometimes combined with the heavy-quark OMEs at NLO~\cite{Buza:1995ie,Bierenbaum:2007qe},
omitting NNLO results~\cite{Bierenbaum:2009mv,Ablinger:2010ty,Kawamura:2012cr,Ablinger:2014nga}.

Altogether, these facts introduce a significant model dependence in any GM-VFNS implementation.
A sensitive parameter to test this model dependence is the extraction of
the charm- or bottom-quark mass used in different versions of GM-VFNS and
subsequent comparison with the Particle Data Group (PDG) results~\cite{Agashe:2014kda}.
In addition, the quality of the various GM-VFNS can be quantified with the goodness-of-fit for
the description of HERA data on DIS charm-quark production 
obtained from the combination of individual H1 and ZEUS results~\cite{Abramowicz:1900rp}.

\begin{figure}[t!]
\begin{center}
\includegraphics[width=0.32\textwidth, angle=0]{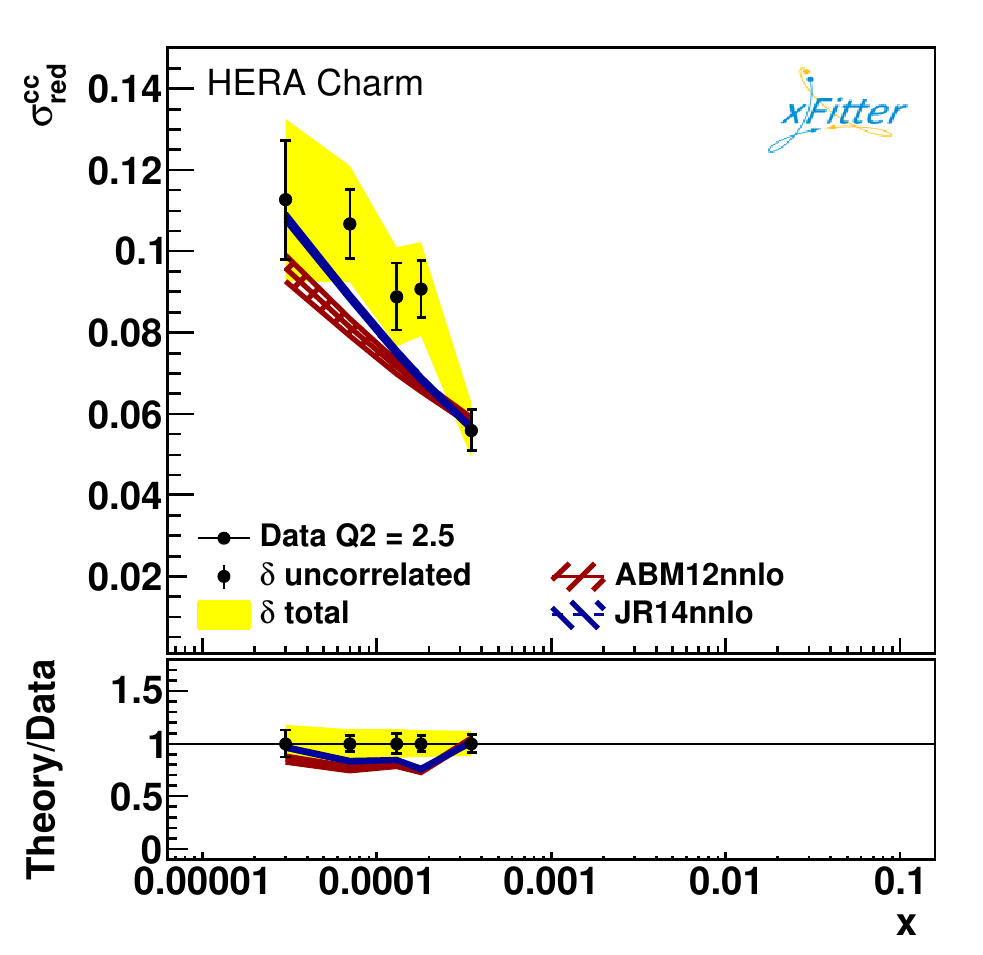}
\includegraphics[width=0.32\textwidth, angle=0]{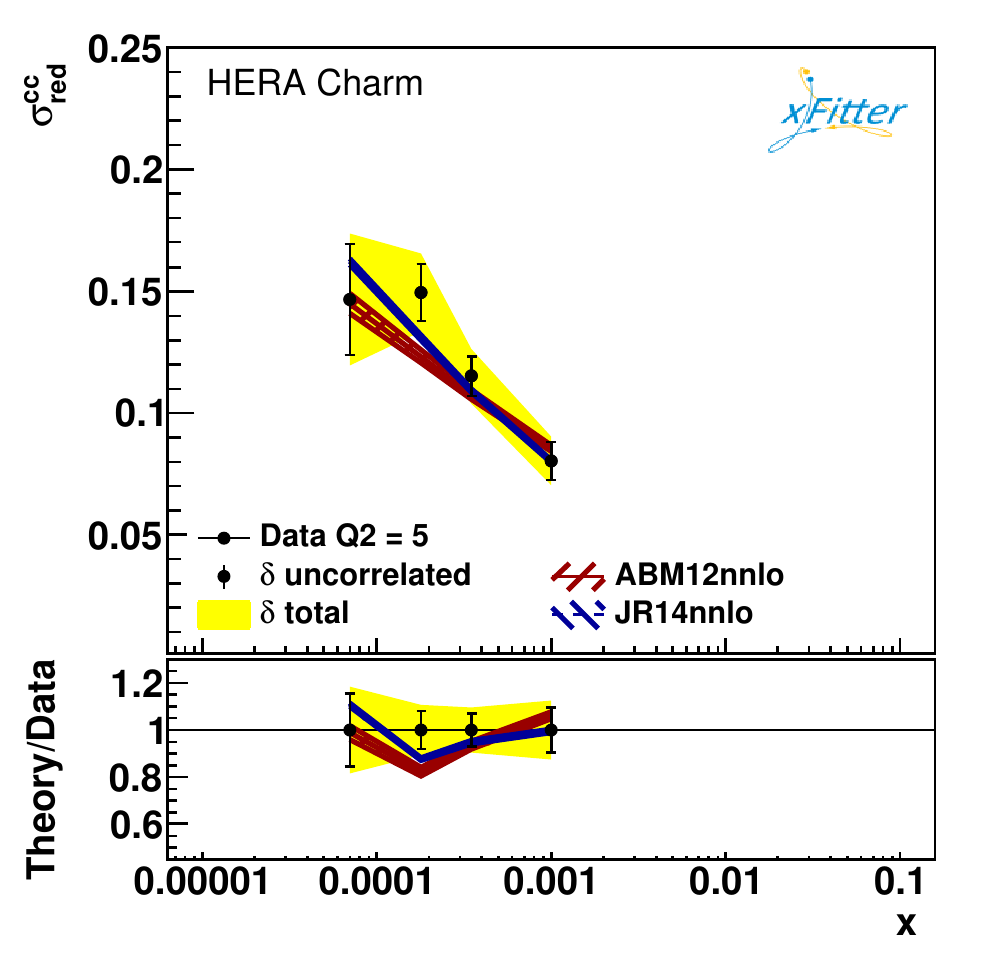}
\includegraphics[width=0.32\textwidth, angle=0]{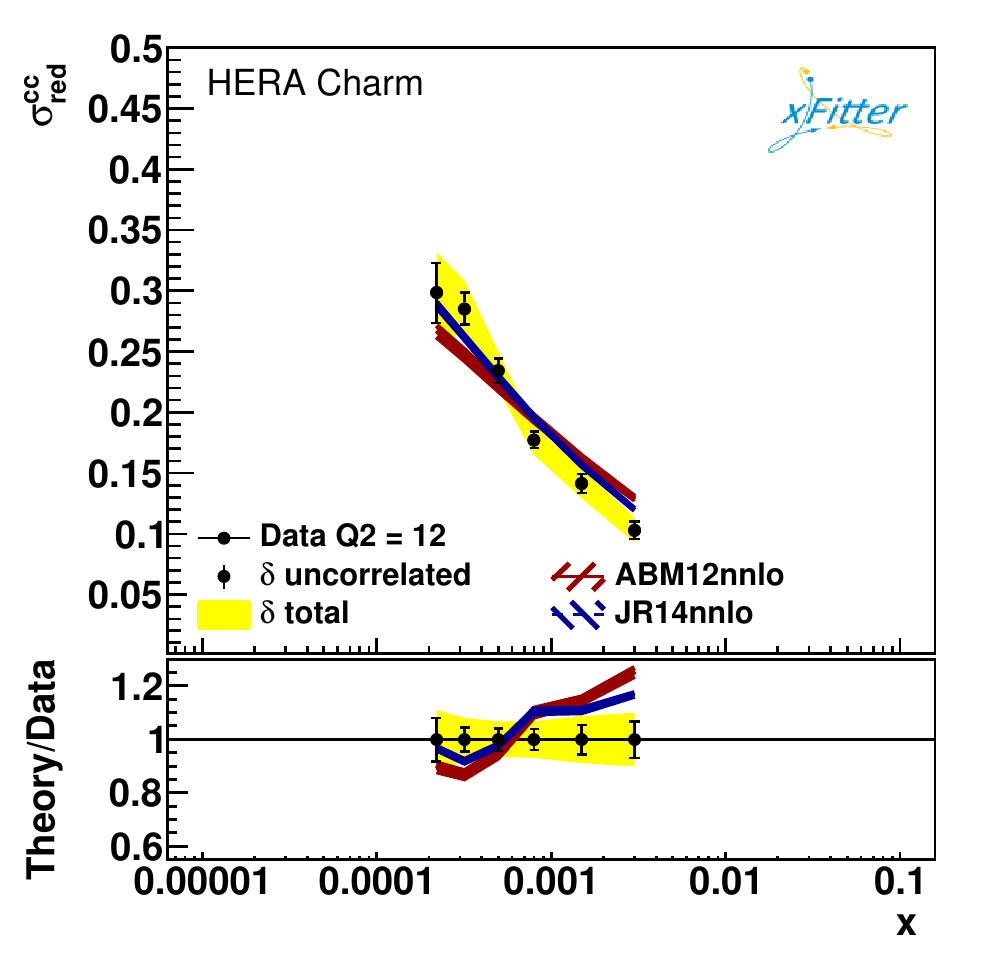}
\\[4ex]
\includegraphics[width=0.32\textwidth, angle=0]{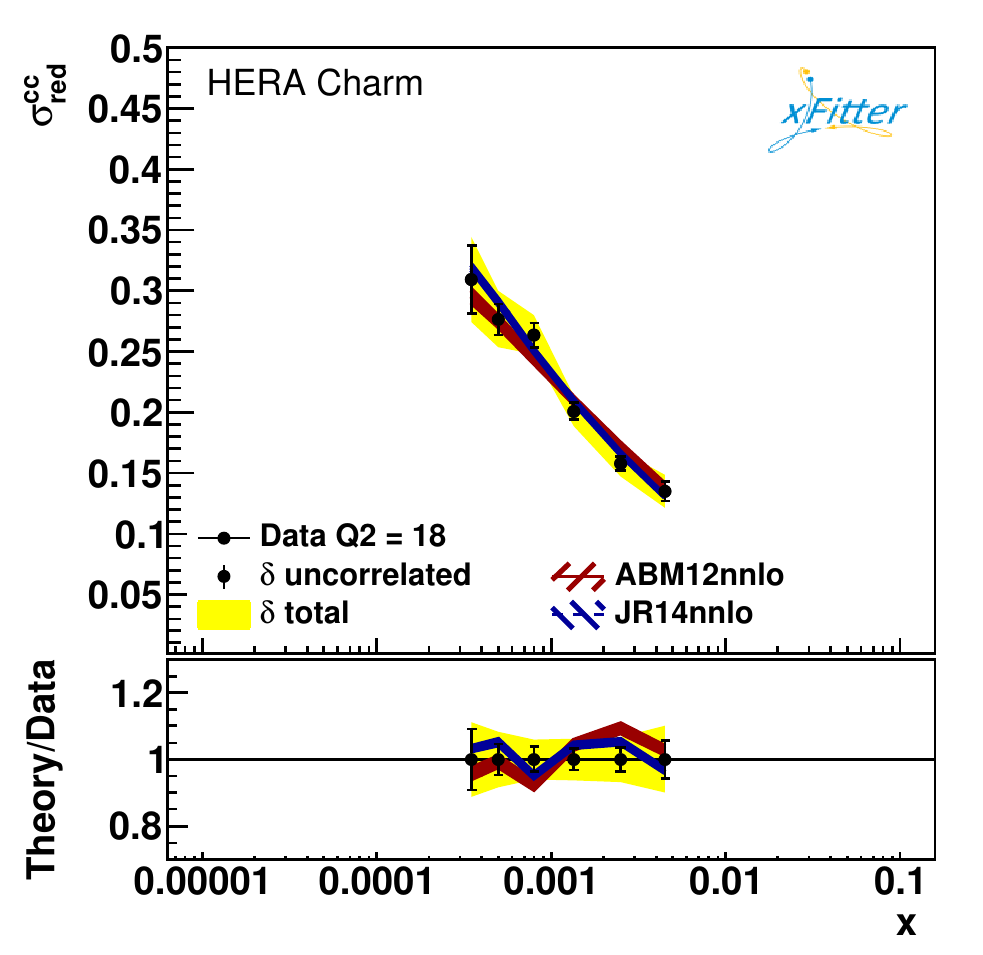}
\includegraphics[width=0.32\textwidth, angle=0]{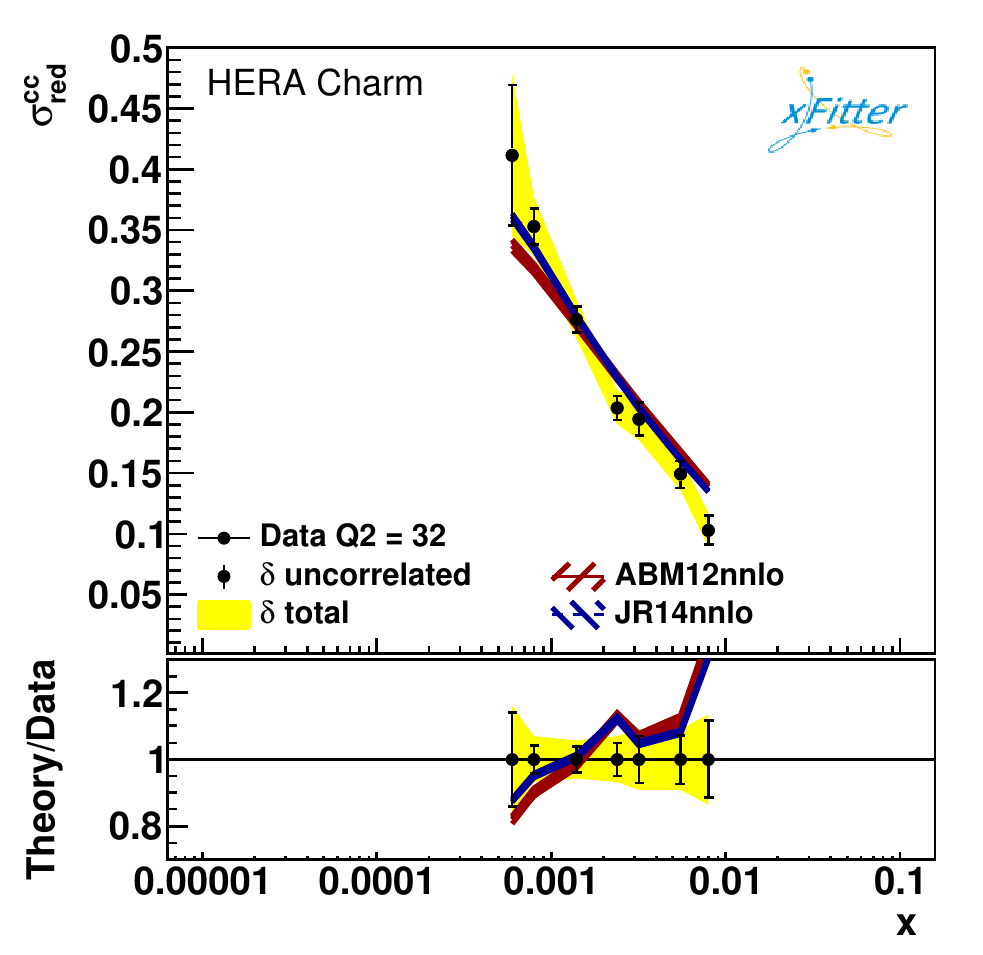}
\includegraphics[width=0.32\textwidth, angle=0]{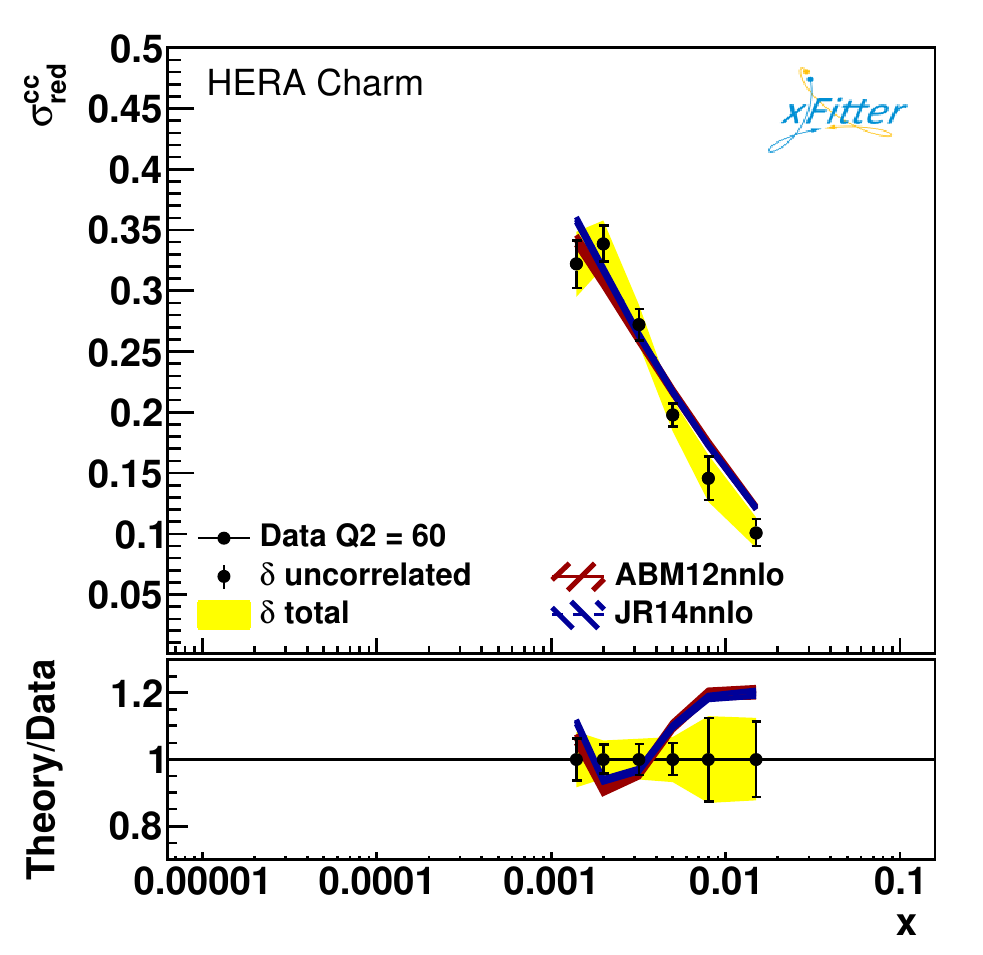}

\caption{\small 
\label{fig:hera-ccbar-dis-1} 
Comparison of HERA data for the DIS pair-production of charm quarks~\cite{Abramowicz:1900rp} to 
the QCD predictions at NNLO in the FFNS using ABM12~\cite{Alekhin:2013nda} and
JR14~\cite{Jimenez-Delgado:2014twa} PDFs with a running charm-quark mass.}
\end{center}
\end{figure}

\begin{figure}[ht!]
\begin{center}
\includegraphics[width=0.32\textwidth, angle=0]{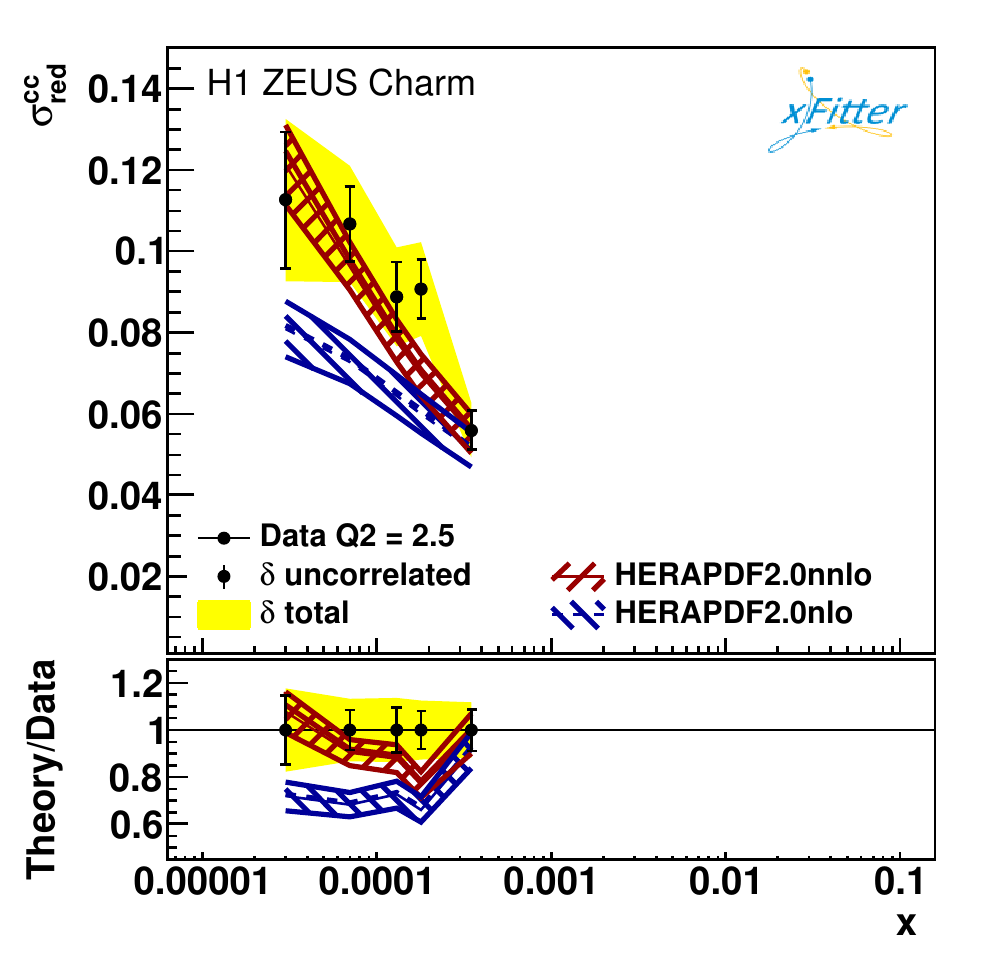}
\includegraphics[width=0.32\textwidth, angle=0]{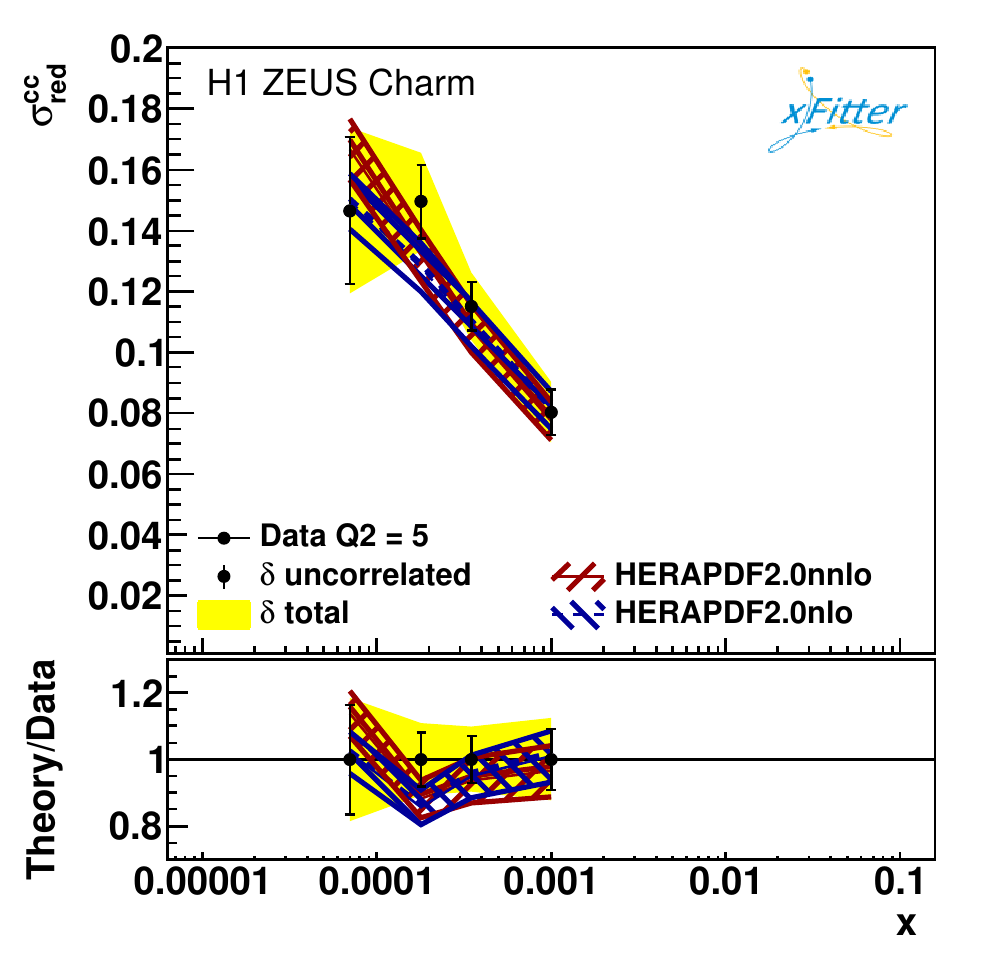}
\includegraphics[width=0.32\textwidth, angle=0]{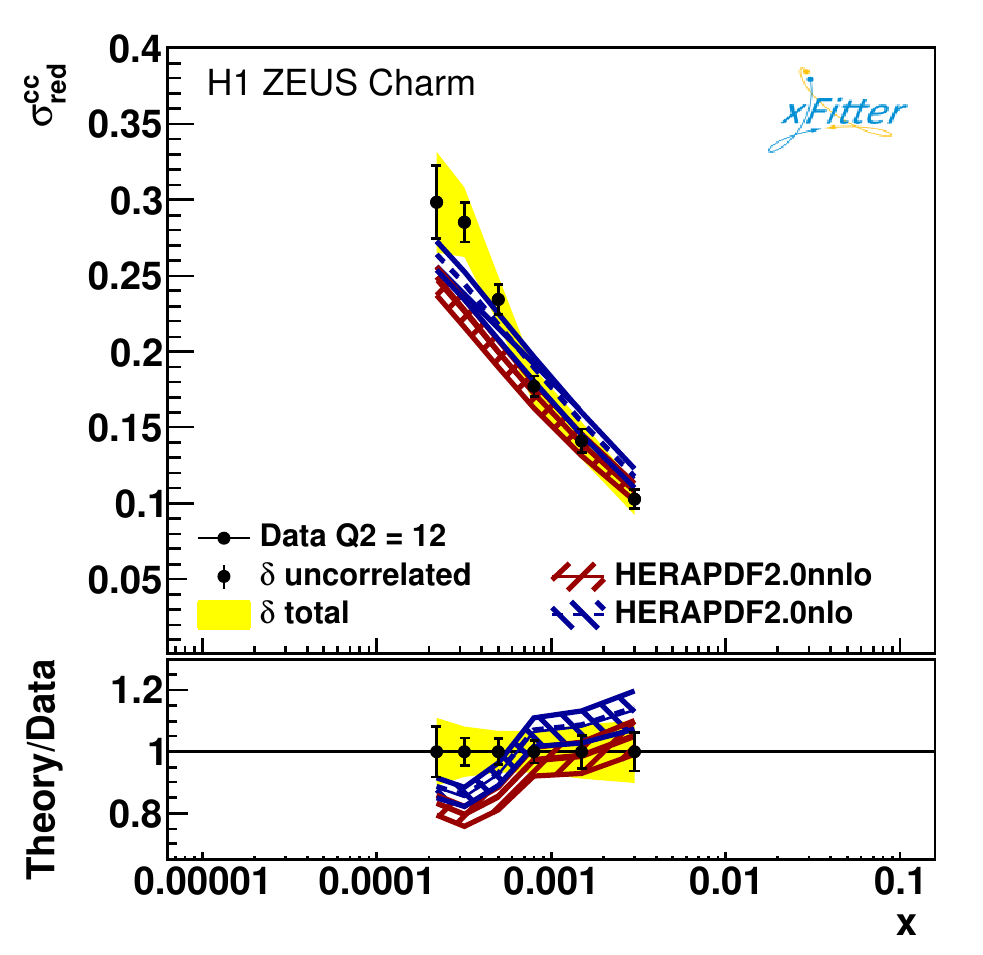}
\\[4ex]
\includegraphics[width=0.32\textwidth, angle=0]{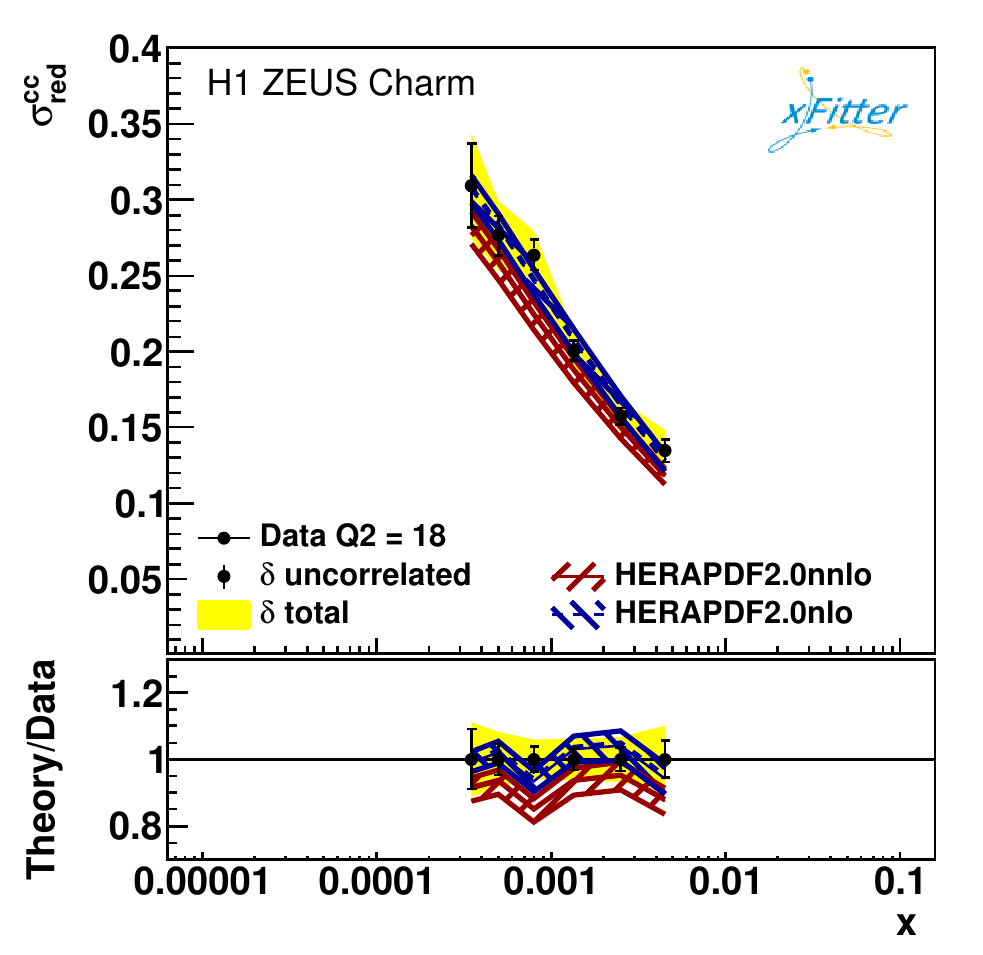}
\includegraphics[width=0.32\textwidth, angle=0]{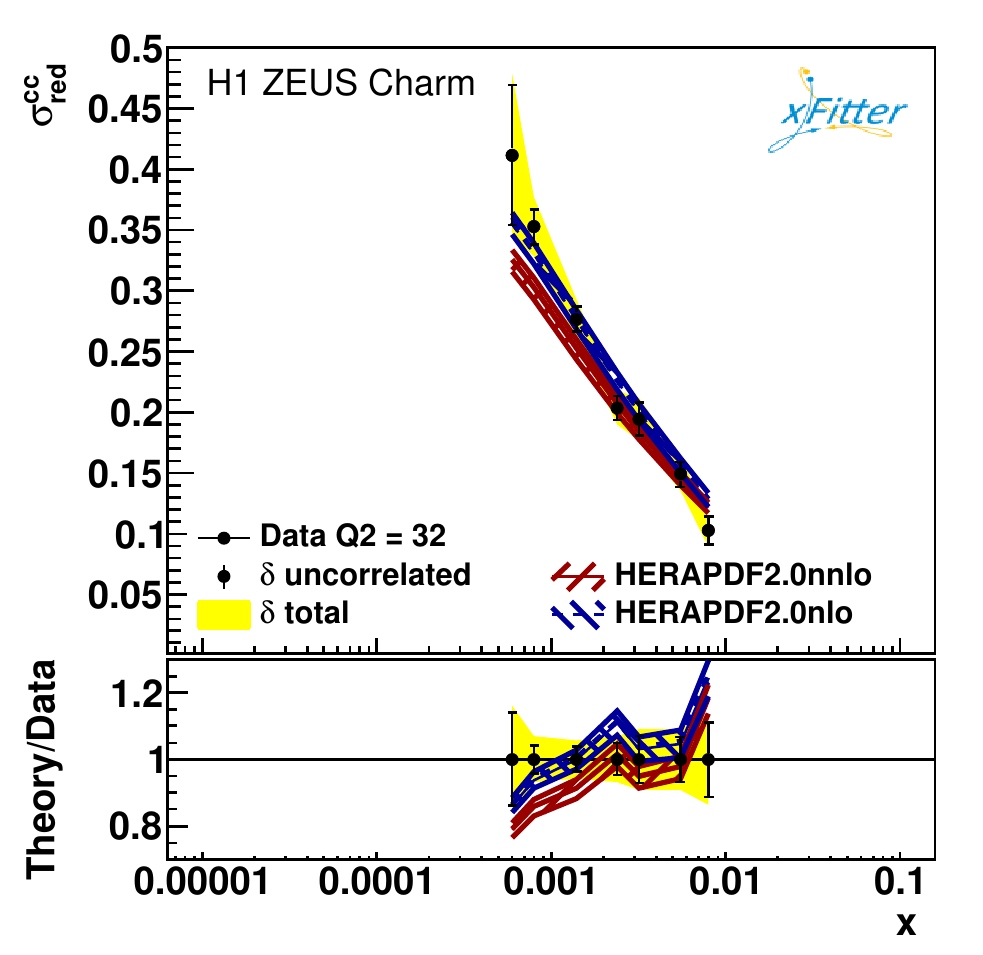}
\includegraphics[width=0.32\textwidth, angle=0]{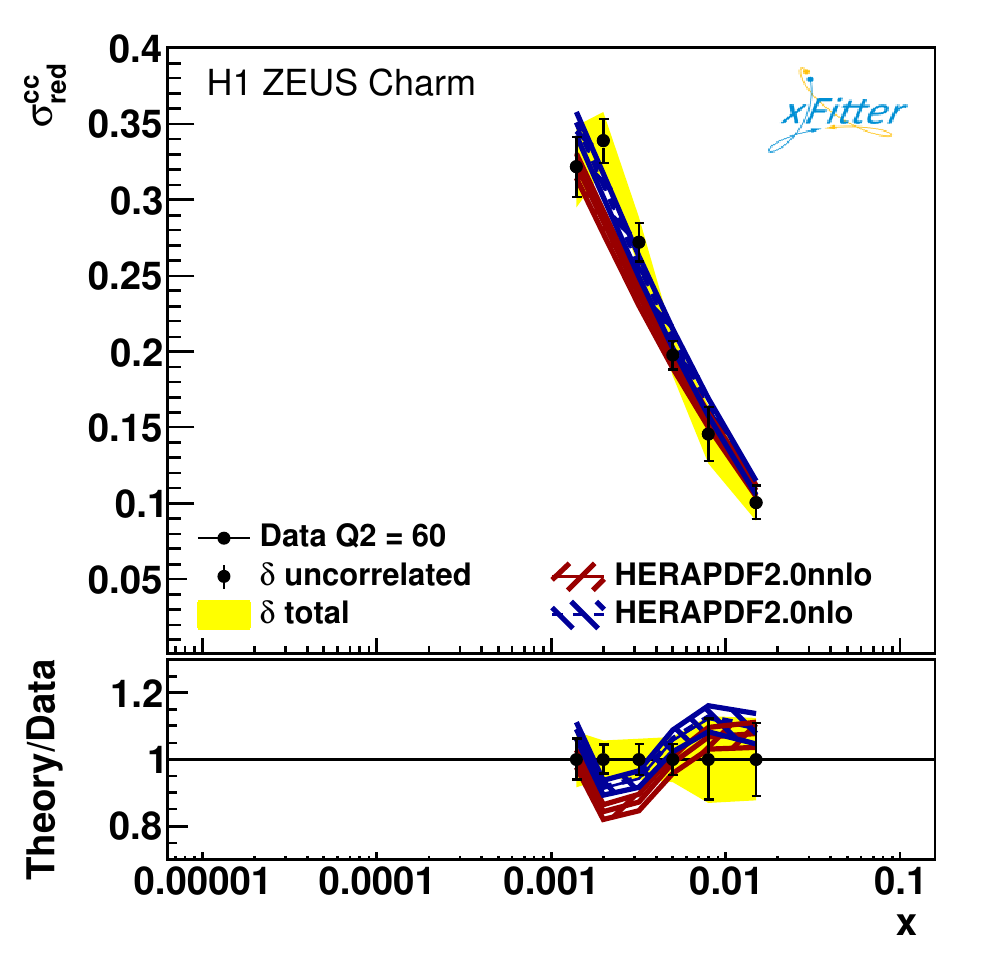}

\caption{\small 
\label{fig:hera-ccbar-dis-2} 
Same as Fig.~\ref{fig:hera-ccbar-dis-1} with QCD predictions at NLO and NNLO 
in the RT optimal~\cite{Thorne:2012az} VFNS 
using the HERAPDF2.0~\cite{Abramowicz:2015mha} PDF sets at NLO and NNLO.}
\end{center}
\end{figure}

\begin{figure}[ht!]
\begin{center}
\includegraphics[width=0.32\textwidth, angle=0]{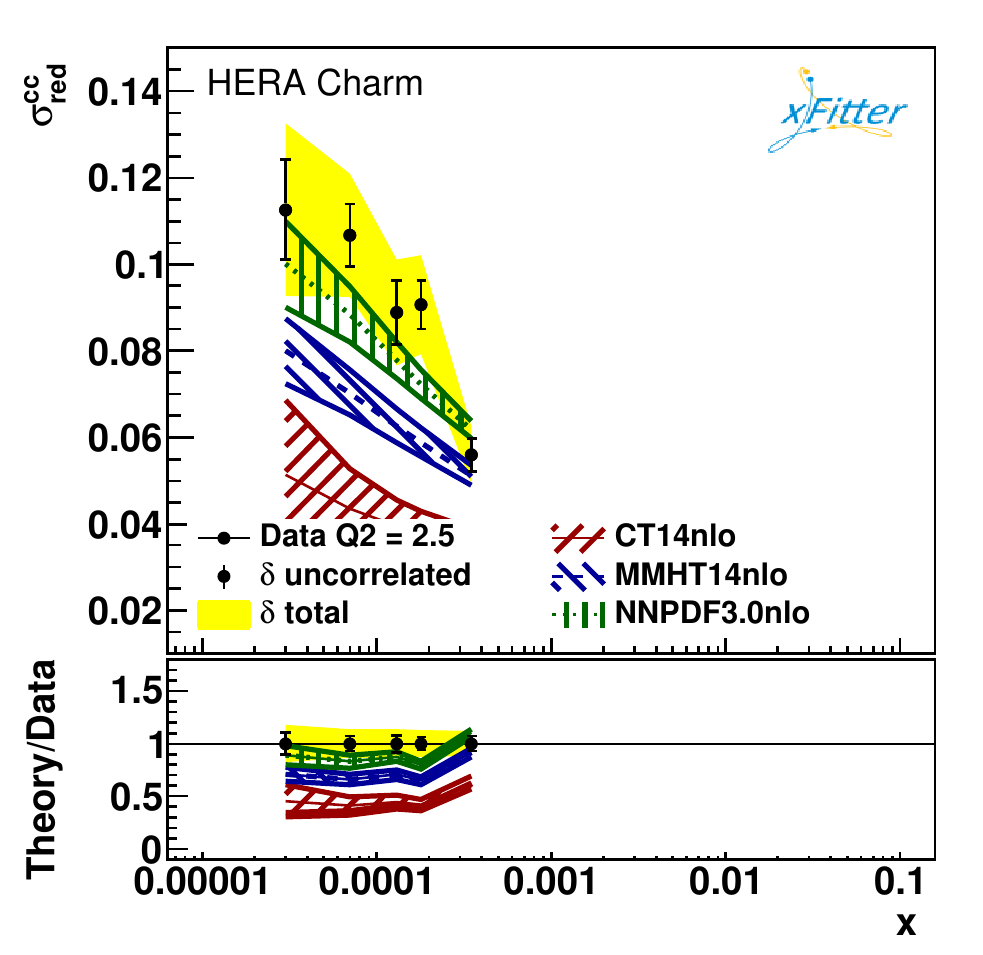}
\includegraphics[width=0.32\textwidth, angle=0]{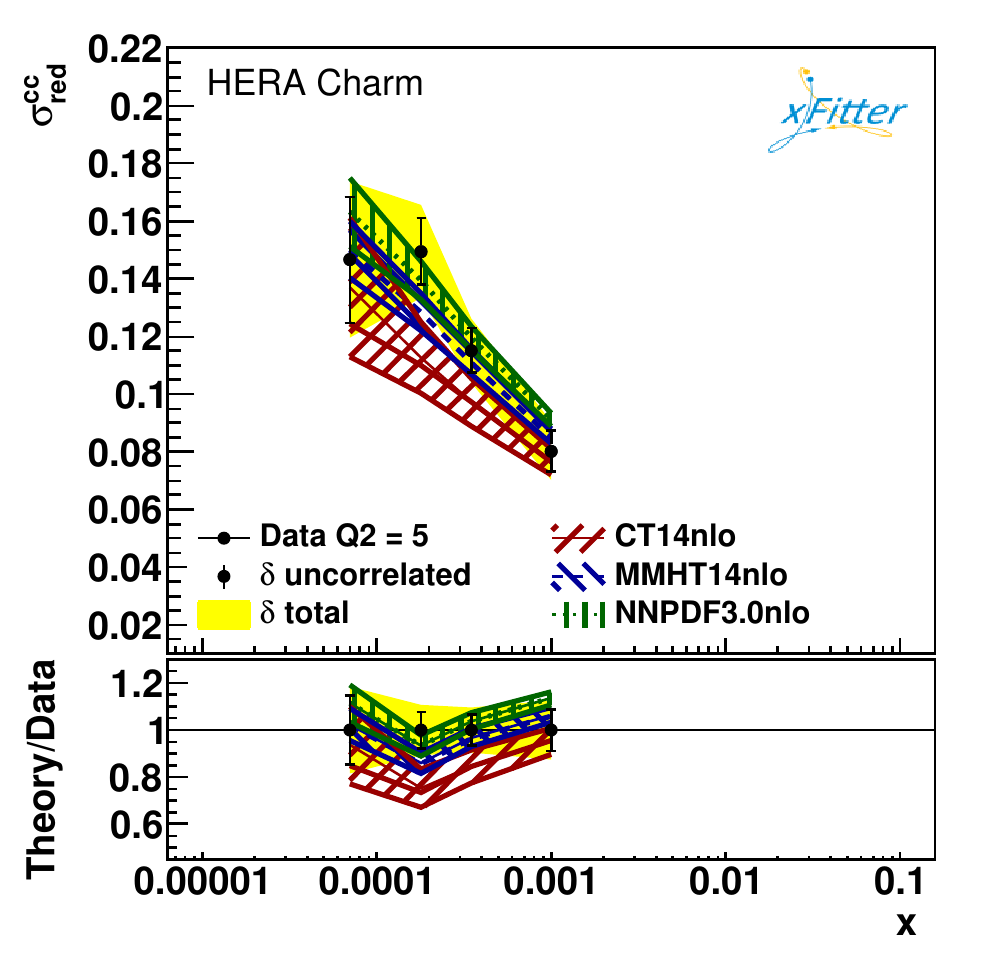}
\includegraphics[width=0.32\textwidth, angle=0]{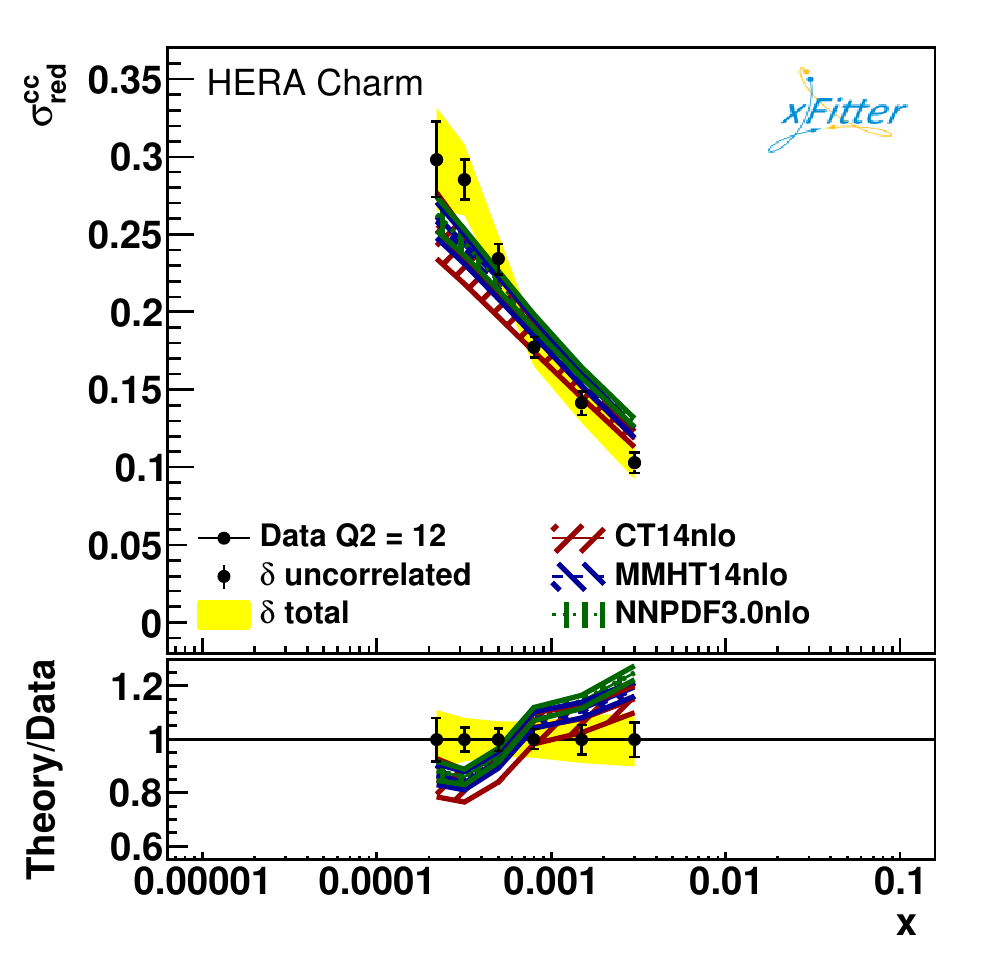}
\\[4ex]
\includegraphics[width=0.32\textwidth, angle=0]{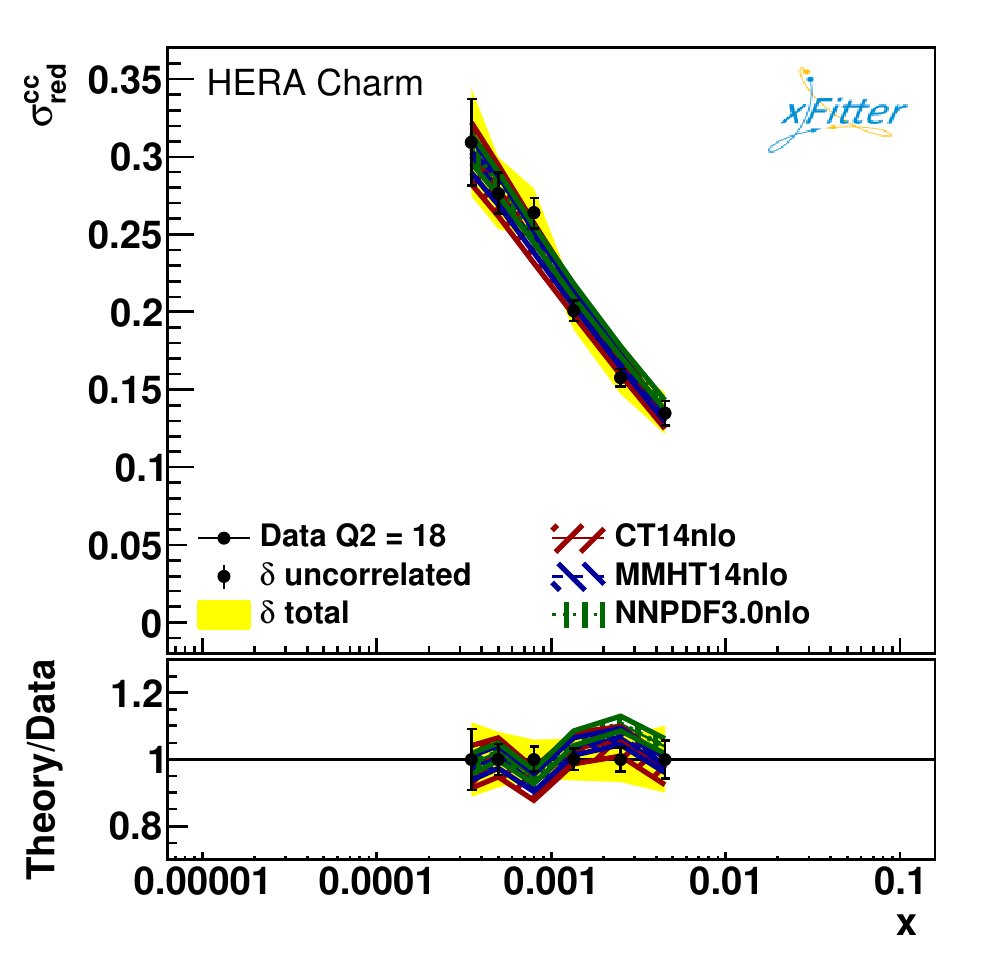}
\includegraphics[width=0.32\textwidth, angle=0]{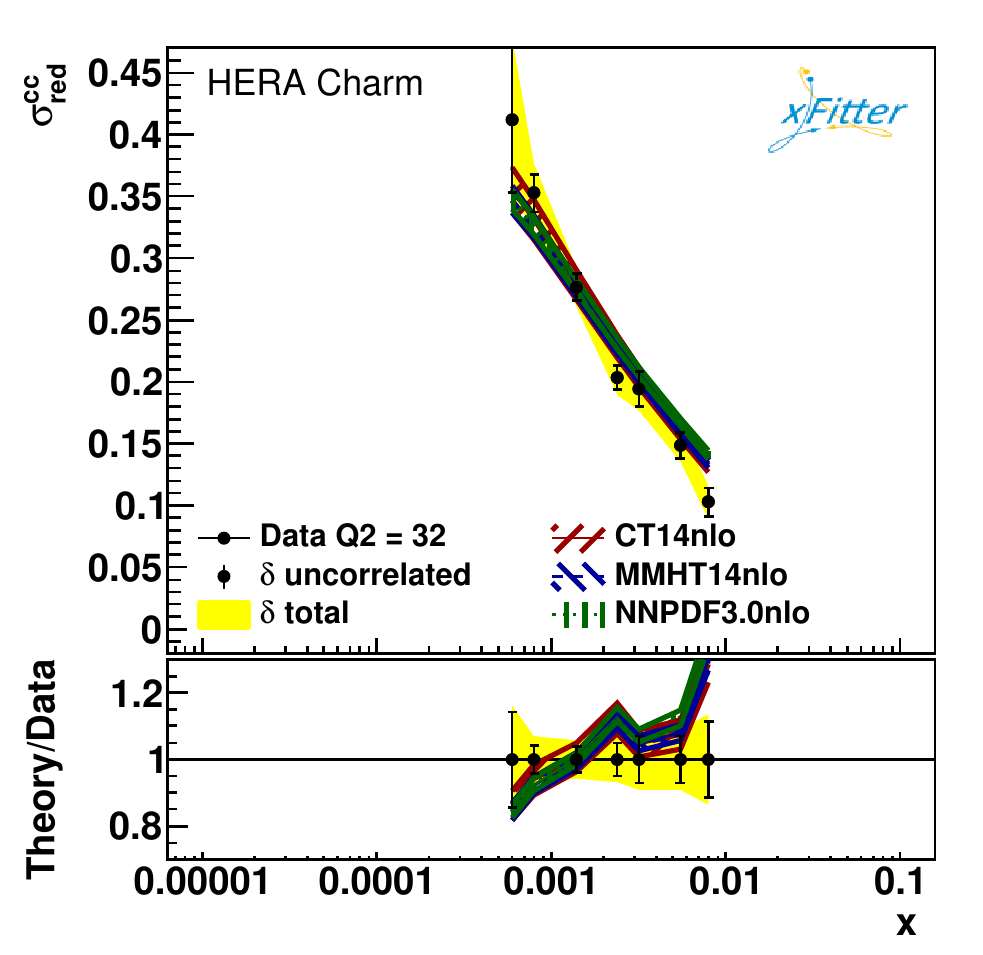}
\includegraphics[width=0.32\textwidth, angle=0]{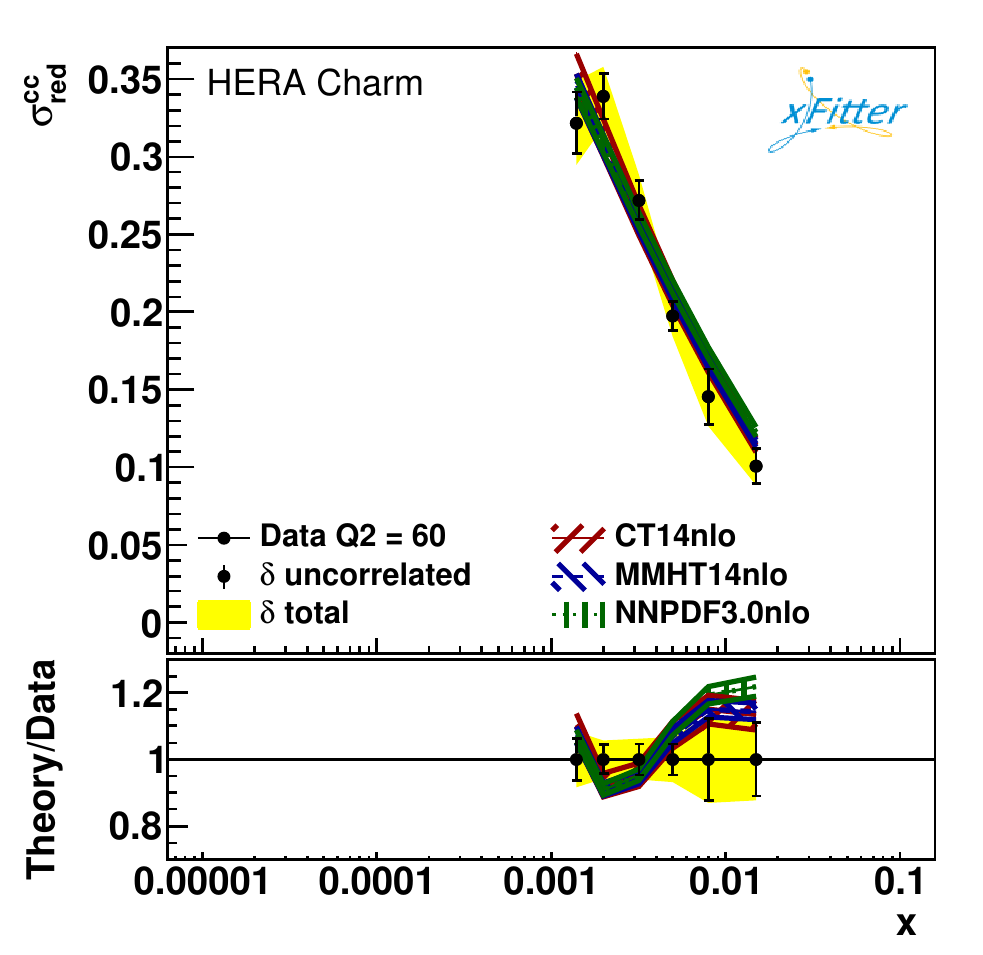}

\caption{\small 
\label{fig:hera-ccbar-dis-4} 
Same as Fig.~\ref{fig:hera-ccbar-dis-1} with QCD predictions at NLO and
different versions of VFNS using the PDFs 
CT14~\cite{Dulat:2015mca} (SACOT-$\chi$~\cite{Tung:2001mv}), 
MMHT14~\cite{Harland-Lang:2014zoa} (RT optimal~\cite{Thorne:2012az}), 
and NNPDF3.0~\cite{Ball:2014uwa} (FONLL-B~\cite{Forte:2010ta}).}
\end{center}
\end{figure}

\begin{figure}[ht!]
\begin{center}
\includegraphics[width=0.32\textwidth, angle=0]{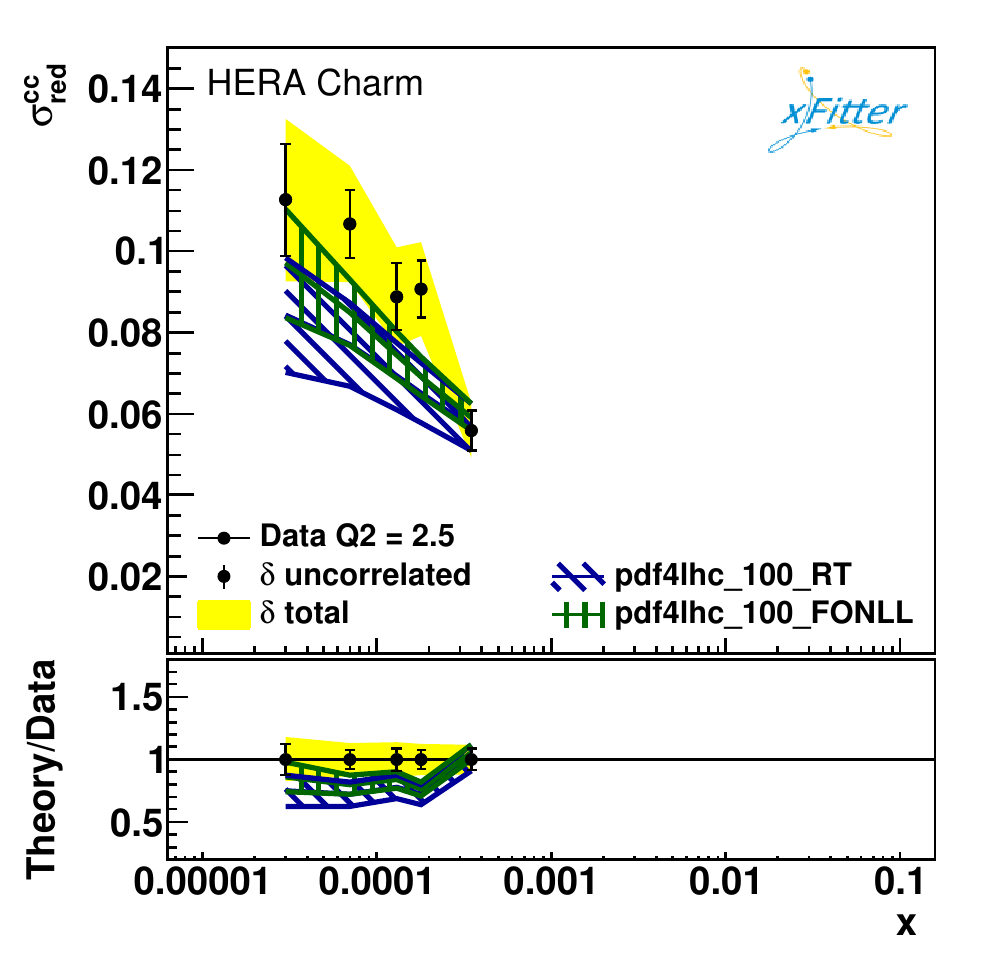}
\includegraphics[width=0.32\textwidth, angle=0]{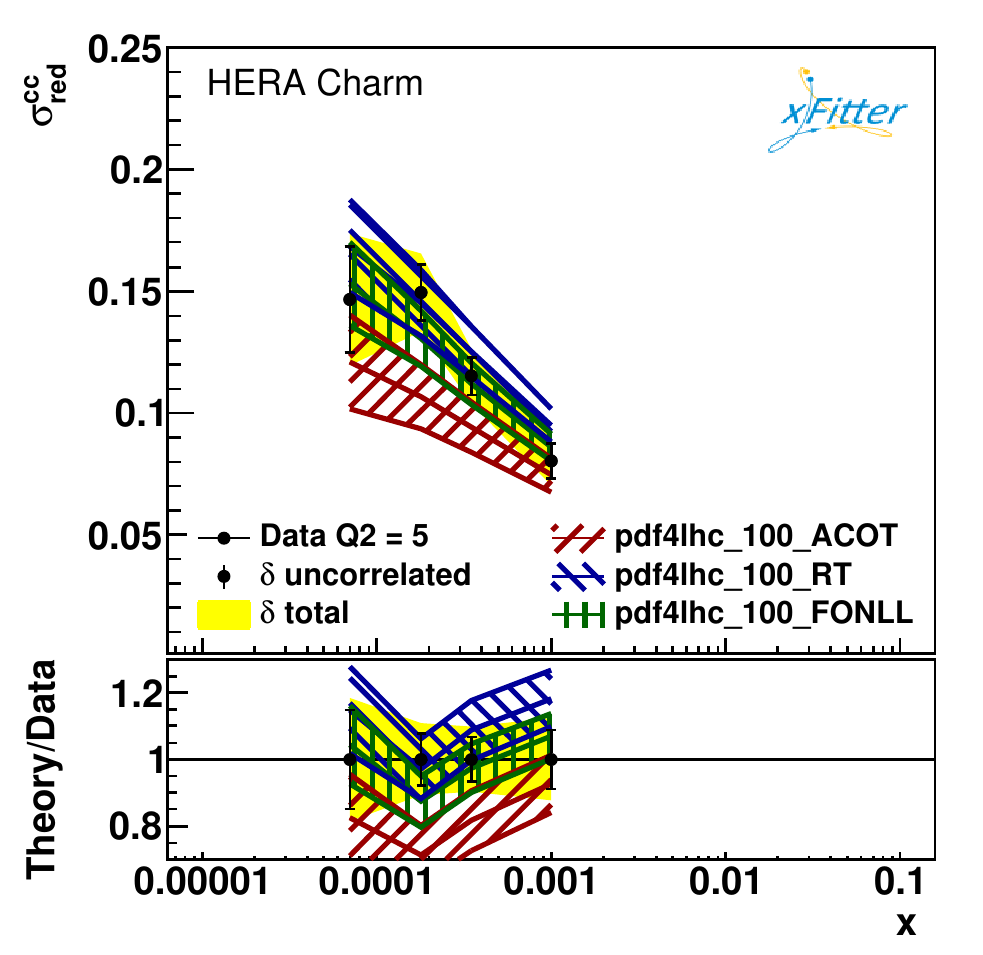}
\includegraphics[width=0.32\textwidth, angle=0]{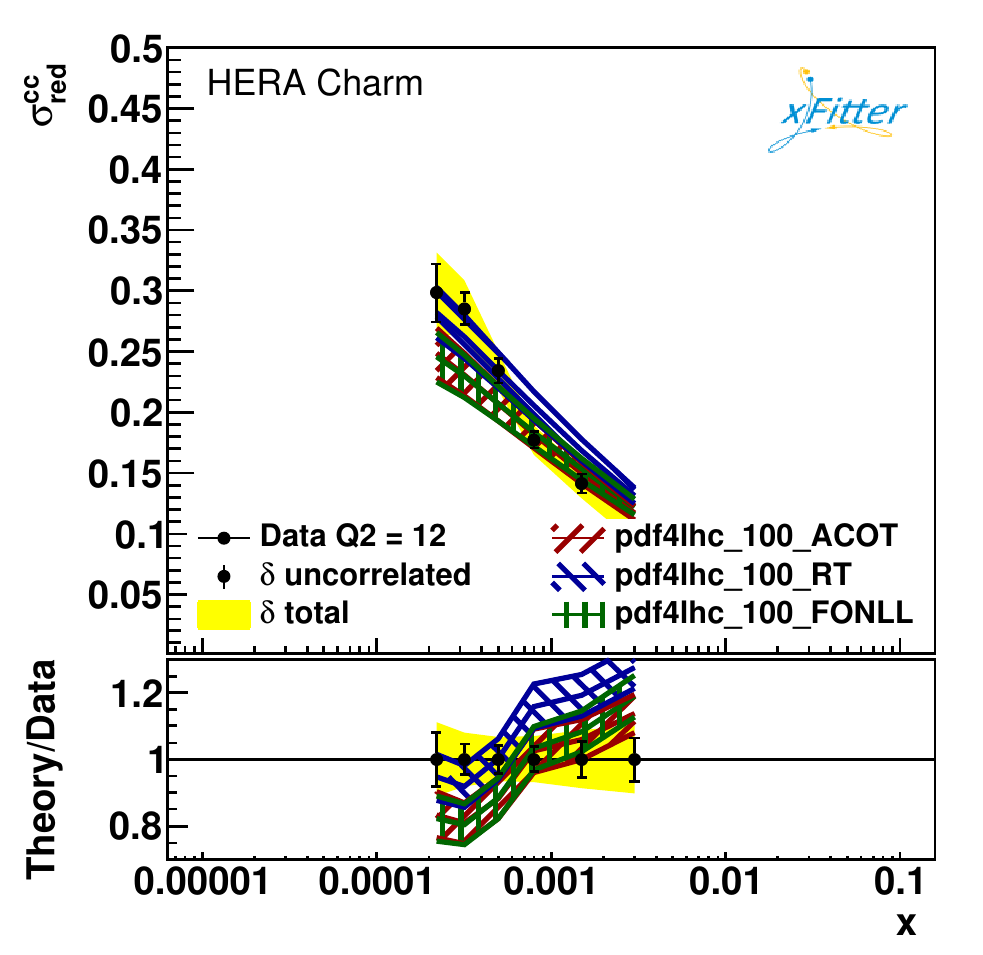}
\\[4ex]
\includegraphics[width=0.32\textwidth, angle=0]{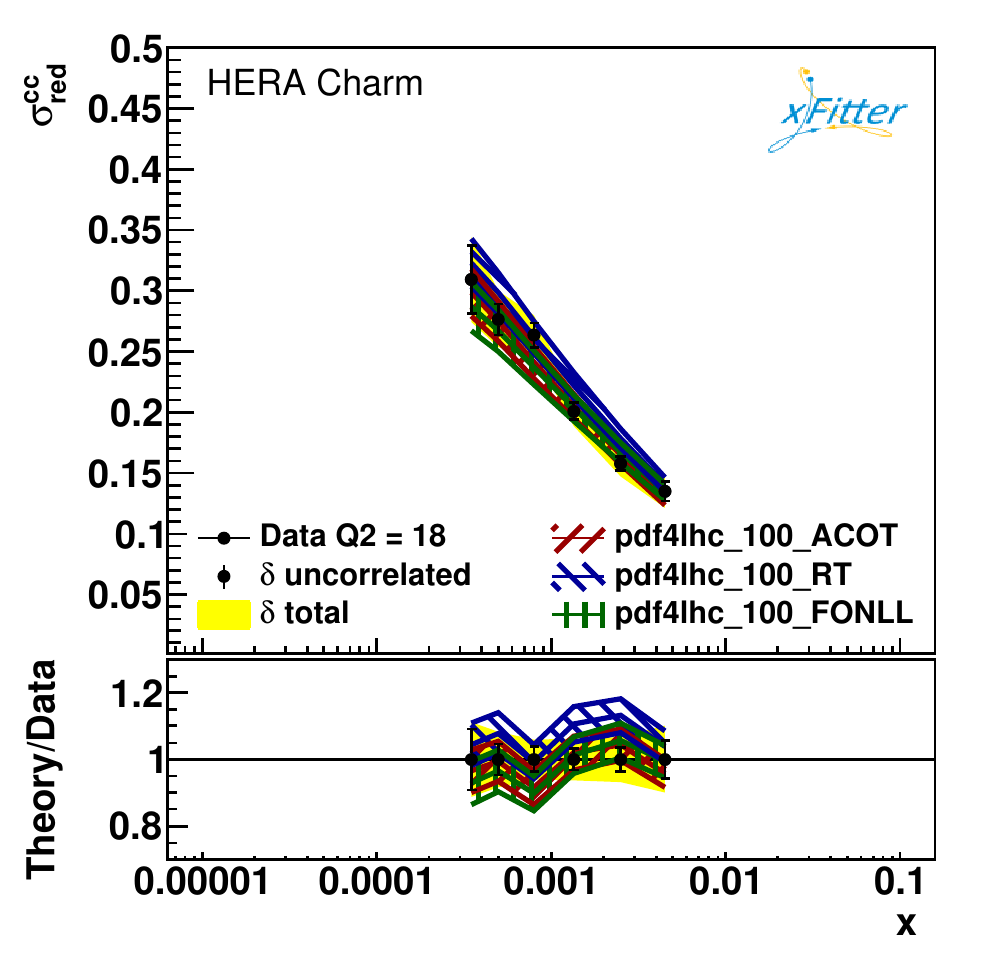}
\includegraphics[width=0.32\textwidth, angle=0]{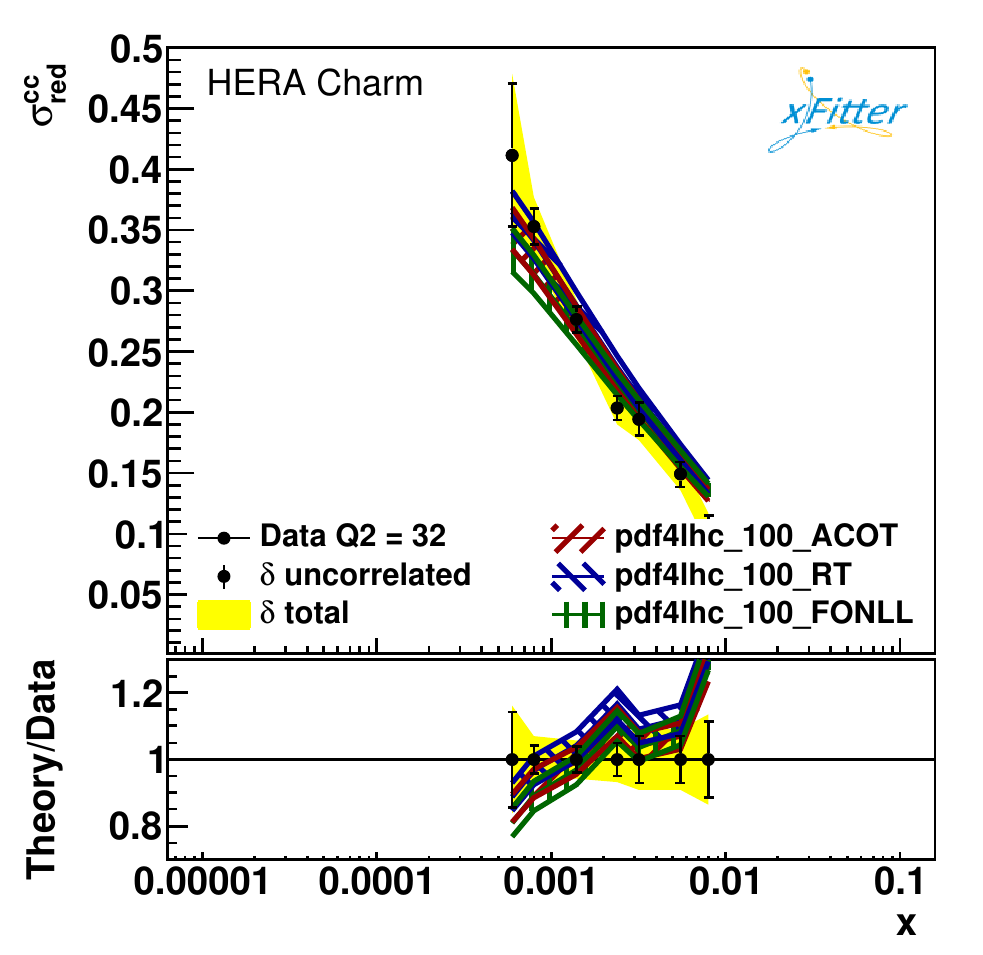}
\includegraphics[width=0.32\textwidth, angle=0]{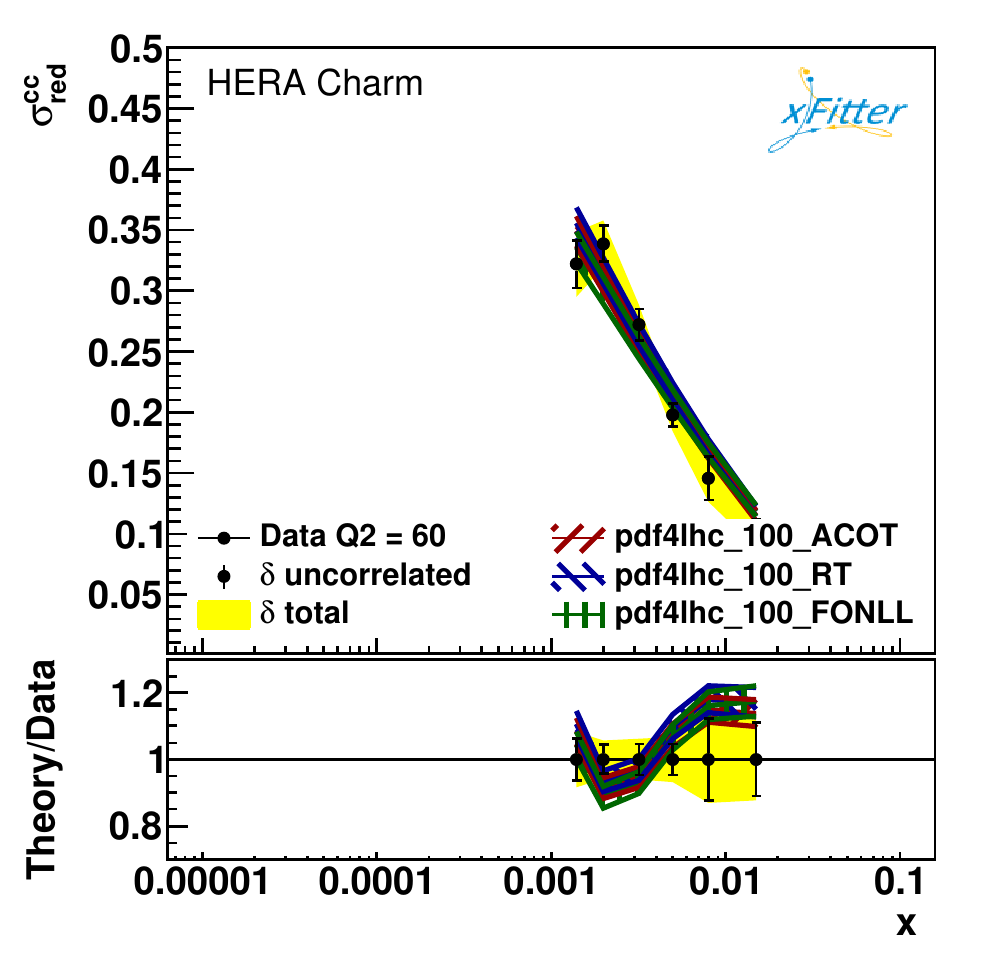}

\caption{\small 
\label{fig:hera-ccbar-dis-5} 
Same as Fig.~\ref{fig:hera-ccbar-dis-1} with QCD predictions at NLO 
using the {\tt PDF4LHC\_100} PDF set and various VFNS:
FONLL-B~\cite{Forte:2010ta}, RT optimal~\cite{Thorne:2012az} 
and SACOT-$\chi$~\cite{Tung:2001mv}.}
\end{center}
\end{figure}

\subsubsection{Validation with DIS charm-quark production}

The H1 and ZEUS combined data for the DIS charm production cross section
are unique for tests of GM-VFNS and span the region of $2.5 \le Q^2 \le 2000~\GeV^2$ and $3\times 10^{-5} \le x \le 0.05$. 
Values for the charm-quark mass and $\chi^2$/NDP 
for the individual PDF sets ABM12, CJ15, CT14, HERAPDF2.0, JR14, MMHT14, NNPDF3.0 as well
as the averaged set PDF4LHC15 are given in Tabs.~\ref{tab:mcmass} and \ref{tab:mcmass-ctd},
along with the information on the scheme choice for the heavy-quark structure functions 
and on the theoretical accuracy for the massive quark DIS Wilson coefficients. 
For reference, Tabs.~\ref{tab:mcmass} and \ref{tab:mcmass-ctd} also list
  the $\chi^2$/NDP values for the HERA inclusive cross section data~\cite{Abramowicz:2015mha}.
Comparisons to data for the DIS charm production cross section are shown in Figs.~\ref{fig:hera-ccbar-dis-1}--\ref{fig:hera-ccbar-dis-5}. 
Note that Tabs.~\ref{tab:mcmass} and \ref{tab:mcmass-ctd} adopt the standard definition of 
perturbative orders for the heavy-quark structure functions.
This is not shared by CT14, MMHT14 and NNPDF3.0 in their GM-VFNS. 
There the Born contribution to the heavy-quark Wilson coefficients
for $ep \to c{\bar c}$, which is proportional to ${\cal O}(\alpha_s)$, is referred to as being ``NLO''.
Analogously, the one-loop corrections of order ${\cal O}(\alpha_s^2)$ are denoted by ``NNLO''.

Table~\ref{tab:mcmass}, \ref{tab:mcmass-ctd} and
Fig.~\ref{fig:hera-ccbar-dis-1} show that the ABM12~\cite{Alekhin:2013nda} and 
JR14~\cite{Jimenez-Delgado:2014twa} PDFs at NNLO, using charm-quark masses in the \msbar scheme,
provide a good description of the data.
Both ABM12 and JR14 use the approximate massive three-loop Wilson
coefficients as obtained in~\cite{Kawamura:2012cr} 
by interpolating between existing ${\cal O}(\alpha_s^3)$ soft-gluon threshold resummation results 
and the ${\cal O}(\alpha_s^3)$ asymptotic $(Q^2 \gg m_c^2)$ coefficients~\cite{Bierenbaum:2009mv,Ablinger:2010ty}.
This is referred to as ${\cal O}(\alpha_s^3)_{\rm approx}$ in Tabs.~\ref{tab:mcmass} and \ref{tab:mcmass-ctd}.
The HERAPDF2.0 fit~\cite{Abramowicz:2015mha} also obtains a good description of the data, cf.~Fig.~\ref{fig:hera-ccbar-dis-2}.
This is the only set which has fitted also to the HERA inclusive cross
  section data of Ref.~\cite{Abramowicz:2015mha}.
On the other hand, the SACOT~\cite{Kramer:2000hn} GM-VFNS at NLO used by CJ15~\cite{Accardi:2016qay} does
not describe the data too well, although we should note that the HERA charm data were not included in the CJ15 fit itself.

The remarkable fact in Tabs.~\ref{tab:mcmass}, \ref{tab:mcmass-ctd} and Fig.~\ref{fig:hera-ccbar-dis-4} is, however, 
that the GM-VFNS SACOT($\chi$)~\cite{Tung:2001mv} of CT14~\cite{Dulat:2015mca} and 
RT optimal~\cite{Thorne:2012az} of MMHT14~\cite{Harland-Lang:2014zoa} have
difficulties in describing the 
DIS charm production data. 
  Note that MMHT14 models the heavy-quark Wilson coefficient functions at 
  ${\cal O}(\alpha_s^3)$ for low $Q^2$ as described in~\cite{Thorne:2012az} 
  using known leading threshold logarithms~\cite{Laenen:1998kp} and $\ln(1/x)$
  terms~\cite{Catani:1990eg}, which have been shown not to be leading.
  This is indicated as ${\cal O}(\alpha_s^2)$ in Tab.~\ref{tab:mcmass-ctd}.
Note that CT14 has applied a universal cut of $Q^2 \ge 4~\GeV^2$ on all DIS data, 
excluding the bin at $Q^2 = 2.5~\GeV^2$ in the HERA data~\cite{Abramowicz:1900rp} 
from the fit (cf.~the upper left plot in Fig.~\ref{fig:hera-ccbar-dis-4}).
We have checked that including the low $Q^2$ bin leads to a dramatic deterioration of the fit quality.

In addition, the schemes SACOT($\chi$) and RT optimal as well as FONLL-C~\cite{Forte:2010ta} of NNPDF3.0~\cite{Ball:2014uwa} 
do not improve the fit quality when comparing NLO and NNLO fits.
We note in this context that those fits do not include 
the exact~\cite{Bierenbaum:2009mv,Ablinger:2010ty,Ablinger:2014vwa,Ablinger:2014nga} and approximate \cite{Kawamura:2012cr} 
$O(\alpha_s^3)$ results
for the heavy-quark Wilson coefficients
in their theory predictions.
The averaged set PDF4LHC15~\cite{Butterworth:2015oua}, shown in Fig.~\ref{fig:hera-ccbar-dis-5}, mixes PDFs
derived with different mass schemes (ACOT, FONNL and RT) and does not describe the
data very well for virtualities up to $Q^2 \lsim 20~\GeV^2$.

\subsubsection{Charm-quark mass}

Dedicated studies of the charm-quark mass dependence 
have been performed by several groups.
In the \msbar scheme, the value of $m_c(m_c)=1.24~^{+~0.04}_{-~0.08}~$GeV
has been obtained in \cite{Alekhin:2012vu} together with $\chi^2/$NDP=61/52 
for the description of the HERA data \cite{Abramowicz:1900rp} 
as a variant of the ABM11 fit~\cite{Alekhin:2012ig}.
Other groups, which keep a fixed value of $m_c$ in the analyses,
cf.~Tabs.~\ref{tab:mcmass} and \ref{tab:mcmass-ctd},
have studied the effects of varying $m_c$ in predefined ranges. 
This has been done, for example, in the older NNPDF2.1~\cite{Ball:2011mu} and MSTW
analyses~\cite{Martin:2010db} 
as well as for the MMHT PDFs~\cite{Harland-Lang:2015qea}.
The latter yields a pole mass of $m_c^{\rm pole}=1.25~\GeV$ as the best fit 
with $\chi^2$/NDP = 75/52, while the nominal fit 
uses $m_c^{\rm pole}=1.4~\GeV$ at the price of a deterioration in the value of $\chi^2$/NDP = 
82/52.
HERAPDF2.0~\cite{Abramowicz:2015mha} has performed a scan of the values of $\chi^2$/NDP 
leading to $m_c^{\rm pole} = 1.43~\GeV$ at NNLO quoted in Tabs.~\ref{tab:mcmass} and \ref{tab:mcmass-ctd} as the best fit.
NNPDF3.0 computes heavy quark structure functions with expressions for the
pole mass definition, but adopts numerical values for the charm quark pole mass,
$m_c^{\rm pole} = 1.275~\GeV$, which corresponds to the current PDG value for the \msbar mass. 
This value is different from the one used in NNPDF2.3, 
namely $m_c^{\rm pole} = \sqrt{2}~\GeV$.
Within the framework of the CT10 PDFs~\cite{Lai:2010vv} 
the charm-quark mass in the \msbar scheme has been determined 
in Ref.~\cite{Gao:2013wwa} using the SACOT($\chi$) scheme
at order ${\cal O}(\alpha_s^2)$, although with a significant spread 
in the central values reported ($m_c(m_c)=1.12 - 1.24~$GeV) depending 
on assumption in the fit.

In this context, it is worth to point out that the running mass $m_c(\mu)$ in
the \msbar scheme is free from renormalon ambiguities and can therefore be
determined with high precision.
The PDG~\cite{Agashe:2014kda} quotes  $m_c(m_c) = 1.275 \pm 0.025~\GeV$ based
on the averaging different mass determination in various kinematics. 
DIS charm-quark production analyzed in the FFNS $(n_f=3)$ 
leads to $m_c(m_c) = 1.24 \pm 0.03~^{+~0.03}_{-~0.03}$~GeV at NNLO~\cite{Alekhin:2012vu}, 
while measurements of the \msbar mass in $e^+e^-$ annihilation give, for instance, 
$m_c(m_c) = 1.279 \pm 0.013$~GeV~\cite{Chetyrkin:2009fv} and 
$m_c(m_c) = 1.288 \pm 0.020$~GeV~\cite{Dehnadi:2015fra}.
The determination from quarkonium $1S$ energy levels yields 
$m_c(m_c) = 1.246 \pm 0.023$~GeV~\cite{Kiyo:2015ufa}.
All these values are consistent with each other within the uncertainties.

In contrast, the accuracy of the pole mass $m_c^{\rm pole}$ is limited to be of the order
of the QCD scale $\Lambda_{\rm QCD}$ and, 
moreover, the conversion from the \msbar mass $m_c(m_c)$ at low scales to the
pole mass $m_c^{\rm pole}$ does not converge.
Using $\alpha_s(M_Z)=0.1184$, for example, the conversion yields for the central value of the PDG 
$m_c^{\rm pole}=1.47~\GeV$ at one loop,  
$m_c^{\rm pole}=1.67~\GeV$ at two loops,  
$m_c^{\rm pole}=1.93~\GeV$ at three loops,
and 
$m_c^{\rm pole}=2.39~\GeV$
at four-loops~\cite{Marquard:2015qpa}.
The PDG quotes $m_c^{\rm pole}=1.67 \pm 0.07~\GeV$ for conversion at two loops. 

The low values for the pole mass of the charm quark assumed or obtained
in some PDF fits as shown in Tabs.~\ref{tab:mcmass} and \ref{tab:mcmass-ctd} are thus not compatible with
other determinations and with the world average. 
The rigorous determination of the charm-quark mass discussed, for instance, in~\cite{Alekhin:2012vu} 
provides a more controlled way of determining $m_c$ from the world DIS data, taking also into
account its correlation with $\alpha_s(M_Z)$.

\subsection{Light-flavor PDFs}

\begin{figure}[tH!]
  \centerline{
    \includegraphics[width=0.45\textwidth]{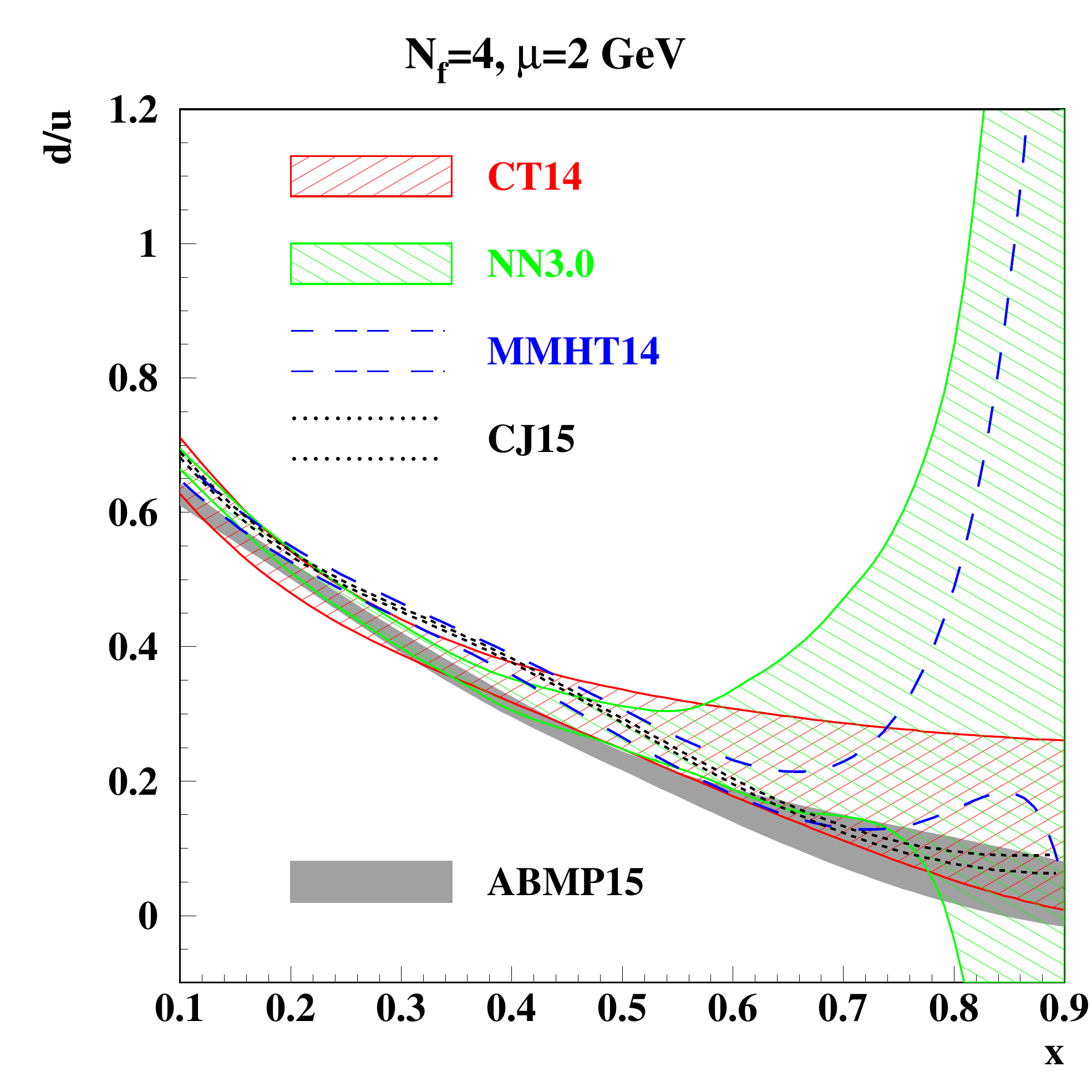}}
  \caption{\small  
    \label{fig:ducomp}
    The $1\sigma$ band for the $d/u$ ratio for the 4-flavor scheme
    and at the factorization scale $\mu=2~\GeV$
    obtained in the PDF analyses including
    forward $W^{\pm}$ data (CT14~\cite{Dulat:2015mca}: red right-tilted hatch, 
    ABMP15~\cite{Alekhin:2015cza}: gray shaded area,
    CJ15~\cite{Accardi:2016qay}: black dotted lines) 
    in comparison to those including the 
    central $W,Z$ data only and a cut of $W^2 \gtrsim 13~\GeV^2$ imposed on the 
    deuteron DIS data (MMHT14~\cite{Harland-Lang:2014zoa}: blue dashed lines, 
    NNPDF3.0~\cite{Ball:2014uwa}: green left-tilted hatch). 
  }
\end{figure}

\begin{figure}[tbh]
  \centerline{
    \includegraphics[width=0.45\textwidth]{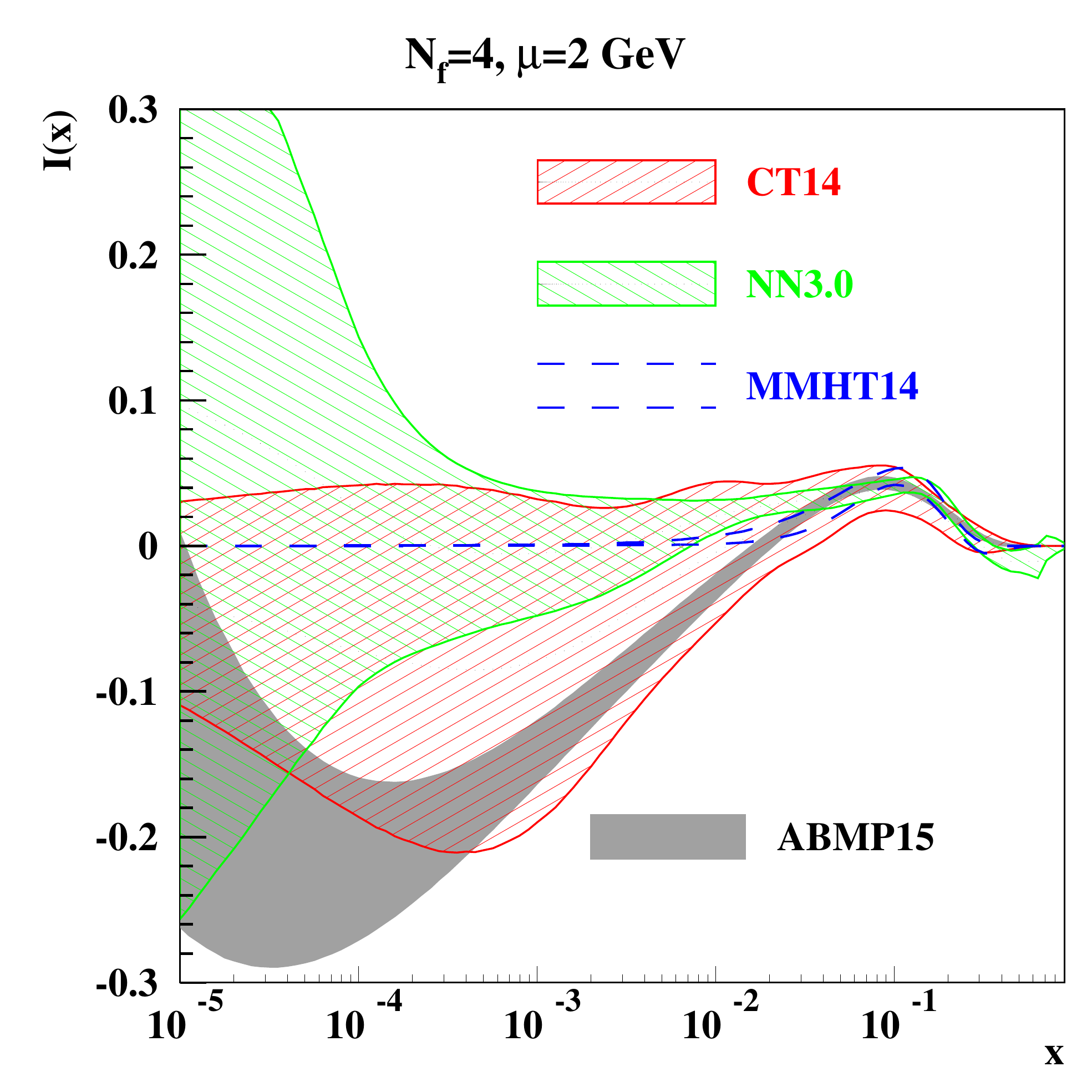}}
  \caption{\small  
    \label{fig:udmcomp}
    Same as Fig.~\ref{fig:ducomp} for the SU(2) flavor asymmetry of the
    light-quark sea, or the ``isospin'' asymmetry, $I(x)=[\bar{d}(x)-\bar{u}(x)]$.
  }
\end{figure}

%
\input{table-gauge-boson}

\begin{figure}[tbh]
  \centerline{
    \includegraphics[width=0.45\textwidth]{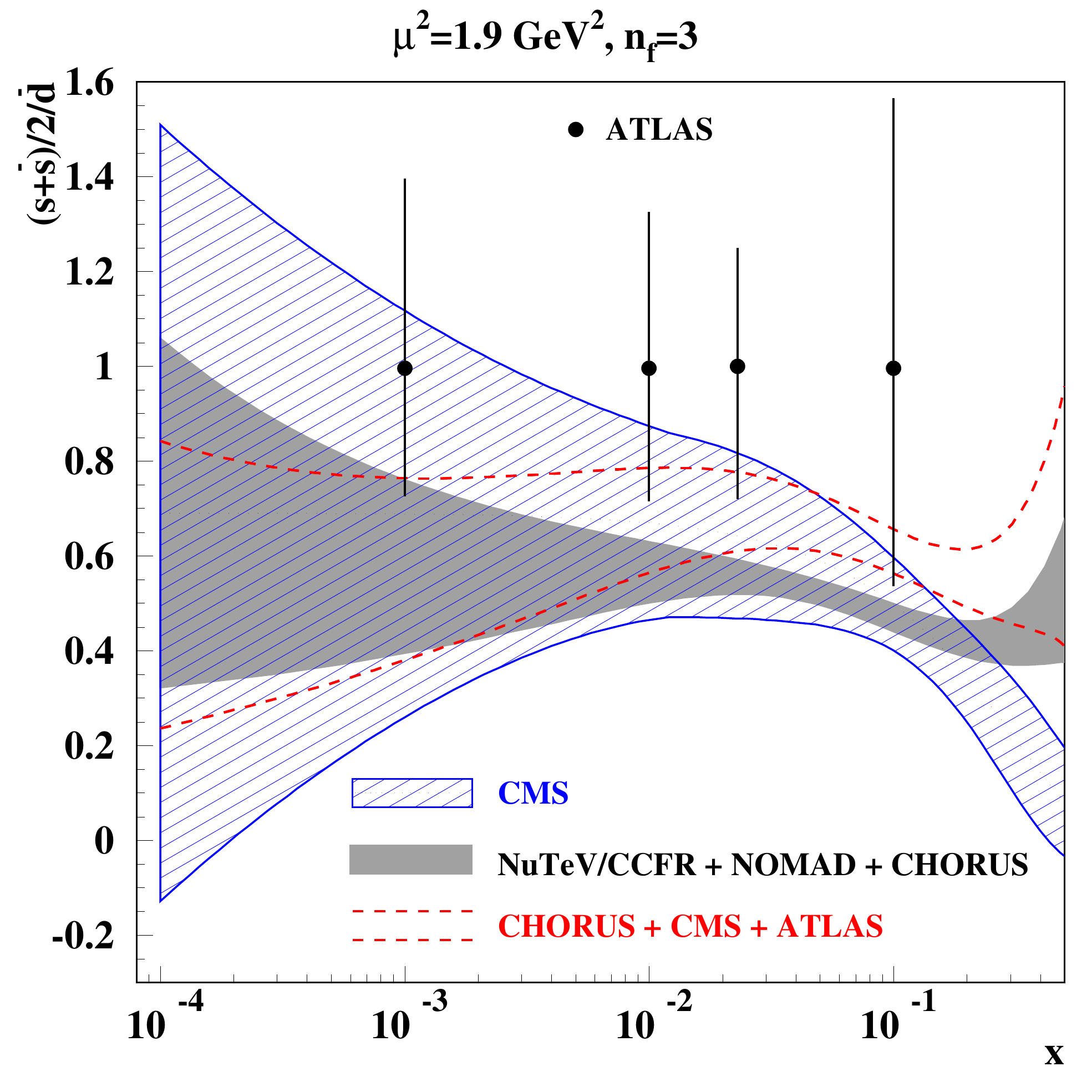}}
  \caption{\small
    \label{fig:ssup}
    The $1\sigma$ band for the 
    strange sea suppression factor $r_s=(s+\bar{s})/(2\bar{d})$ as a function of
    Bjorken $x$
    obtained in the variants of the ABM analysis~\cite{Alekhin:2014sya} 
    based on the 
    combination of the data by NuTeV/CCFR~\cite{Goncharov:2001qe},
    CHORUS~\cite{KayisTopaksu:2011mx} and NOMAD~\cite{Samoylov:2013xoa} 
    (shaded area), and CHORUS~\cite{KayisTopaksu:2011mx}, 
    CMS~\cite{Chatrchyan:2013uja} and ATLAS~\cite{Aad:2014xca} (dashed lines),  
    compared with the results obtained by the CMS 
    analysis~\cite{Chatrchyan:2013mza}
    (hatched area) and by the ATLAS $epWZ$-fit~\cite{Aad:2012sb,Aad:2014xca} at 
    different values of $x$ (full circles).
    All quantities refer to the factorization scale $\mu^2=1.9~\GeV^2$. 
  }
\end{figure}

\subsubsection{Up- and down-quark distributions}

The total quark contribution to nucleon matrix elements
is known fairly well due to constraints from the available DIS data obtained in the 
fixed-target and collider experiments in the $x$-range $10^{-4} \lsim x \lsim 0.8$. 
However, a thorough disentangling of the quark flavor structure is still a
challenging task in any PDF analysis. 
At moderate and large $x$ values this has been routinely achieved by using a combination of the 
DIS data obtained on proton and deuteron targets. 
However, uncertainties in the modeling of nuclear corrections in the deuteron introduce 
a controllable source of theoretical uncertainty on the $d$-quark PDF obtained 
in this way, especially at large $x$, as discussed below.

An alternative way to resolve the $u$- and $d$-quark contributions is 
to use data on $W$- and $Z$-boson production obtained in $pp$ and $p\bar{p}$ collisions 
at the LHC and Tevatron, respectively. 
Those experiments probe the $W$ and $Z$ rapidity distributions up to rapidities of $y=3-4$, 
depending on details of the experiments, with an integrated luminosity of $O(1)~{\rm fb}^{-1}$ 
achieved in each run. 
Such data samples are quite competitive in accuracy with the ones obtained 
in fixed-target DIS experiments, and  
provide simultaneously constraints on the quark and anti-quark PDFs at large and small $x$.  
Furthermore, the $d$-quark PDF extracted from a combination of the 
existing data on DIS off protons and $W/Z$-boson production in $pp (p\bar{p})$ collisions 
are not sensitive to nuclear corrections.
Moreover, if DIS data with small hadronic invariant masses $W^2$ are not used 
in the analyses in order to reduce the sensitivity to higher twist contributions, 
the statistical potential of the deuteron data is reduced 
and they become less competitive as compared to the collider data, cf.~Fig.~\ref{fig:ducomp}.

As mentioned above, one can further constrain the $u$- and $d$-quark flavor separated distributions 
by utilizing fixed-target deuteron DIS data. 
However, nuclear effects need to be accounted for in cross sections and structure functions 
in order to access the underlying PDFs. 
The theoretical uncertainty inherent in this nuclear correction procedure 
should be added to the statistical PDF uncertainties.
Nonetheless, the reduction of the uncertainties due to the
increased number of fitted data points is even greater, 
leading to an overall smaller $d$-quark PDF uncertainty than in fits
performed without deuterium data~\cite{Alekhin:2015cza,Owens:2012bv,Accardi:2016muk}.
Furthermore, as shown in~\cite{Accardi:2016qay}, and discussed in more detail below, it is possible
to significantly reduce the nuclear correction uncertainty by exploiting
the interplay of the deuteron DIS data and the recent high-statistics
D\O\ data on the reconstructed $W^\pm$ boson charge asymmetry at large
rapidity, which is equally sensitive to the $d/u$ ratio but is not
affected by nuclear corrections.

The $W/Z$-boson collider data also provide a valuable constraint on the small-$x$ quark PDFs. 
In particular, the charge asymmetry of leptons originating 
from the $W$ decays is sensitive to the SU(2) flavor asymmetry of the non-strange sea, 
also referred to as the ``isospin'' asymmetry $I(x)=[\bar{d}(x)-\bar{u}(x)]$ at small $x$. 
This asymmetry is constrained by the DY data from fixed-targets with protons and deuterons collected 
by the Fermilab experiment E866~\cite{Towell:2001nh}.
However, the E866 data are not sensitive to the value of $I(x)$ at small $x$ ($x \lesssim 0.2$).
Therefore, $I(x)$ is sometimes parametrized in a Regge-like form as $I(x) \sim x^{0.5}$ 
such that it vanishes at $x=0$ (cf.~the MMHT results in Fig.~\ref{fig:udmcomp}). 

The large-rapidity tail of the $W/Z$-boson production data 
allows for a model-independent check of $I(x)$ at small $x$.
The asymmetry preferred by the combination of the 
currently available LHC and Tevatron data turns out to be negative at $x<0.01$, 
while the Regge-like limit with a vanishing $I(x)$ can still be recovered at $x \lsim 10^{-5}$, 
cf.~the ABMP15 results in Fig.\ref{fig:udmcomp}. 
The CT14 analysis only includes the Tevatron forward DY data, but 
also confirms the negative trend in $I(x)$ at small $x$, with 
errors in $I(x)$ being substantially larger than those from ABMP15.

Finally, an important issue is the theoretical accuracy which is employed 
in the description of the DY data. 
There are significant differences as shown in Tab.~\ref{tab:WZ-boson} and these
cause an additional spread in the fit quality and the results for the PDFs
when comparing different NNLO PDF sets.

\subsubsection{Strange-quark distribution}

The main information on the strange sea distribution comes from charm-quark production 
in neutrino-induced charged-current DIS experiments. 
The publication of data from CHORUS and NOMAD has recently enlarged 
the statistics available for those experiments.
As a net result, the uncertainty in the strange PDF is now reduced down to a few percent at $x\sim 0.1$
(cf.~Fig.~\ref{fig:ssup}). 
However, at small $x$ the strange sea distribution is still poorly known in
view of the restricted kinematics of the production of charm quarks from fixed targets. 
Furthermore, since neutrino DIS experiments
usually involve nuclear targets, care needs to be taken when extracting
free-nucleon PDFs from the nuclear cross sections.  Nuclear effects in
neutrino DIS and possible differences between those in charged-lepton
DIS have been discussed recently in Refs.~\cite{Kulagin:2007ju,Kovarik:2010uv,Accardi:2016muk}, 
for instance.  Supplementary information on
the strange sea at small $x$, independent of nuclear effects, can be
obtained from the associated production of charm quarks and $W$ bosons
in the $pp$ collisions at the LHC.
A constraint from collider data on $W+c$ is potentially less sensitive to the 
$c$-quark fragmentation model compared to the one from semi-leptonic decays of charm, 
which plays major role in the existing fixed-target DIS experiments.
The $W+c$ data collected by ATLAS and CMS prefer a somewhat enhanced strange sea 
as compared to the fixed-target determination, cf.~Fig.~\ref{fig:ssup}.  
However, the NNLO QCD corrections to this process are still unknown. 
They are not taken into account in the analysis of $W+c$ data so far 
and may have a substantial influence on the fit.
The strange sea extracted by ATLAS from an analysis of the combined  
inclusive data on the $W$- and $Z$-boson production is even further enhanced,
which suggests a restoration of SU(3) flavor symmetry in the sea distributions. 
However, the accuracy of this determination is poor due to a limited potential 
of the inclusive data in disentangling the quark flavors. 
Therefore, the ATLAS result is in fact comparable with other determinations within 
uncertainties.  

In general, the existing experimental constraints on the strange PDF are relatively poor.
Therefore, the results of various determinations demonstrate a significant
spread, which is mainly driven by the data selection. 
An additional spread between results of earlier PDF analyses appears 
due to implementation issues. 
In particular, the strong strange-sea suppression observed in the NNPDF2.1 analysis~\cite{Ball:2011mu}
was related to an error in the DIS charm-quark production cross section being off by 
a factor of two for low scales due to an additional factor of $(1+m_c^2/Q^2)$ in Eq.~(34) of Ref.~\cite{Ball:2011mu}. 
This is now correct in NNPDF3.0~\cite{Ball:2014uwa}.
The CT10 analysis~\cite{Gao:2013xoa}, which reported an enhanced strange sea, 
may be flawed due to a wrong sign of the photon-$Z$ interference for massive quarks 
in the structure function $xF_3$~\cite{Dulat:2015mca}. 
This error also concerns the earlier results on the strange--anti-strange asymmetry~\cite{Lai:2007dq}
and has now been corrected in CT14~\cite{Dulat:2015mca}.
Finally, the MSTW~\cite{Martin:2009iq} analysis suffered from an error in the 
the NLO QCD correction for the charged-current DIS charm-quark production 
as it had omitted a part of the gluon Wilson coefficient at NLO,  
which was corrected in MMHT14~\cite{Harland-Lang:2014zoa}.

\subsection{Nuclear corrections}

\begin{figure}[t!]
\includegraphics[width=0.49\textwidth]{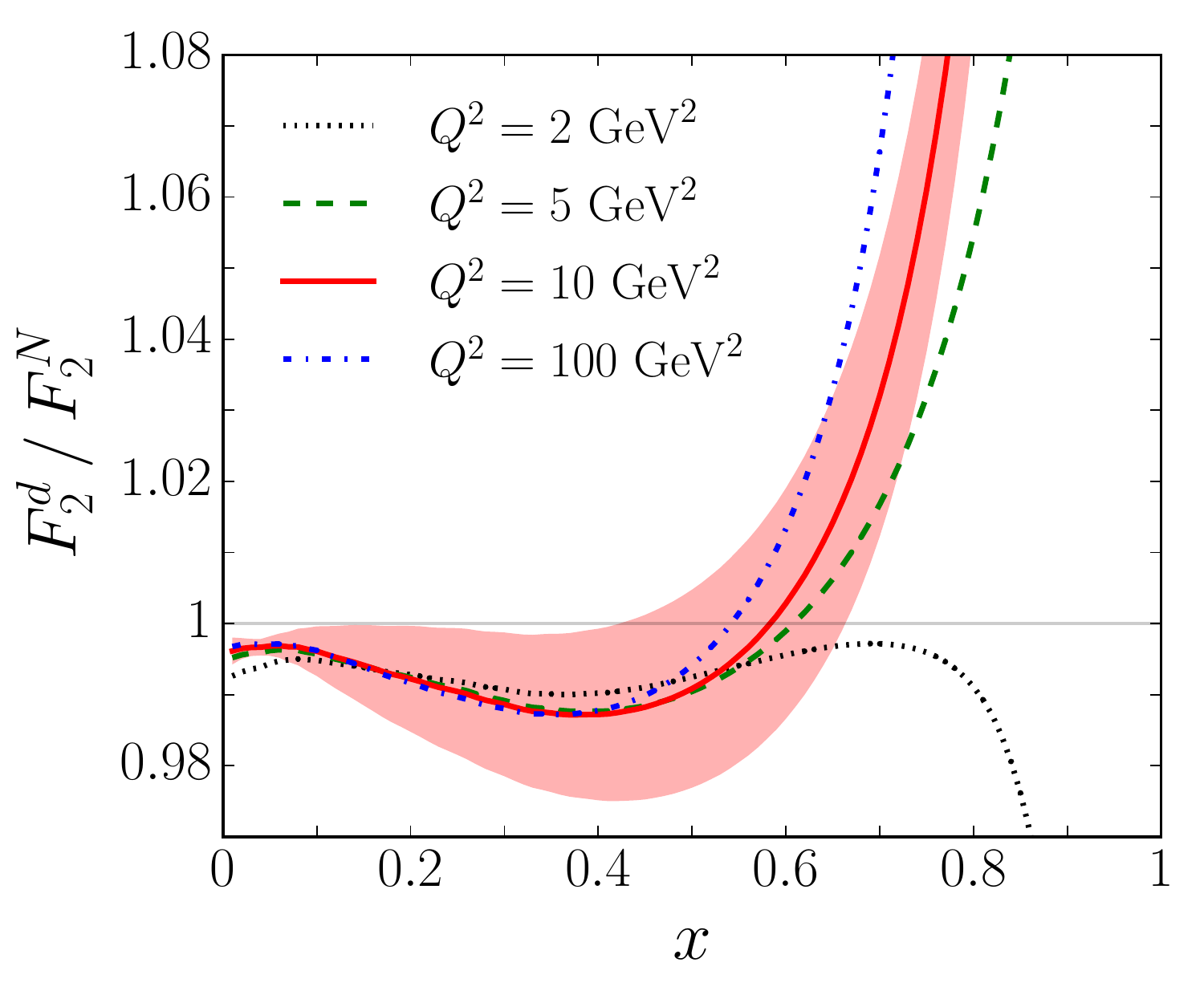}
\includegraphics[width=0.49\textwidth]{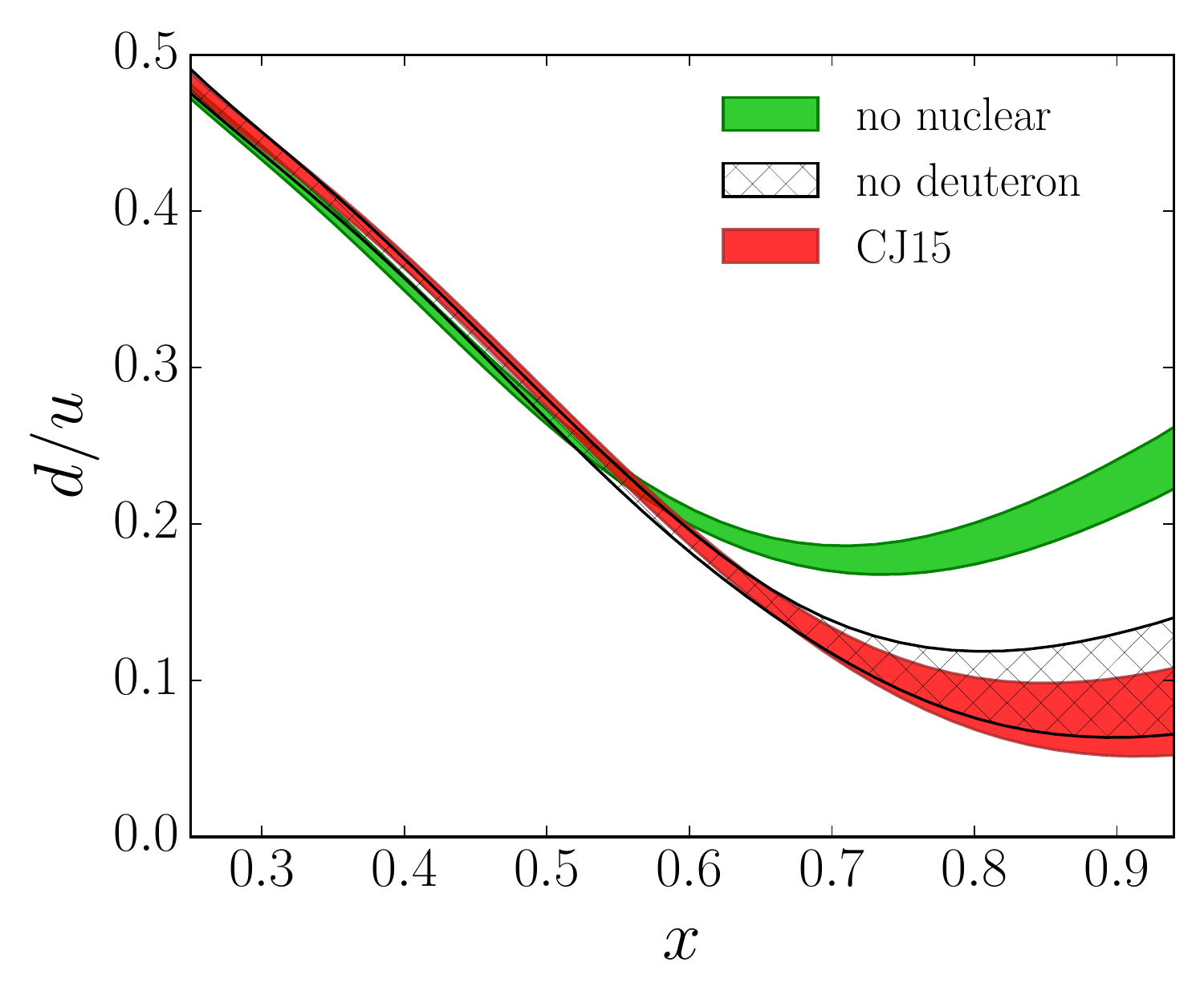}
\caption{\small 
  \label{fig:nuclear}
  (Left panel) 
  Ratio of deuteron to isoscalar nucleon structure functions
  $F_2^d/F_2^N$ computed from the CJ15 PDFs \cite{Accardi:2016qay} for 
  different values of $Q^2$. The pink envelope represents the
  fit uncertainties for $Q^2=10$~GeV$^2$.  The downturn in the
  ratio at $Q^2=2$~GeV$^2$ is due to target mass corrections.
  (Right panel)
  Impact on the $d/u$ ratio from the CJ15 fit \cite{Accardi:2016qay}
  (red band) of removing the deuterium nuclear corrections
  (green band), and omitting all deuterium data
  (cross-hatched band).}
\end{figure}

Many global PDF analyses make use of data with deuterium targets,
such as lepton-deuteron DIS and proton-deuteron DY, as a
way of obtaining stronger constraints on the flavor dependence of
PDFs that are not possible with proton data alone.  The use of
deuterium data requires that one takes into account differences  
between PDFs in the deuteron and those in the free nucleon, which
arise from effects such as nuclear Fermi motion and binding of the
nucleons in the nucleus, as well as nucleon off-shell corrections
and nuclear shadowing.  While some analyses assume that nuclear
corrections in the deuteron are negligible, a number of recent
global PDF studies have incorporated nuclear effects into their
analyses~\cite{Alekhin:2009ni,Alekhin:2012ig,Accardi:2009br,Accardi:2011fa,Owens:2012bv,Jimenez-Delgado:2014twa,Accardi:2016qay}.

Generally, the nuclear effects become increasingly important at   
large values of $x$ ($x \gtrsim 0.4$), as Fig.~\ref{fig:nuclear} 
illustrates for the ratio of the deuteron to isoscalar nucleon    
structure functions.  In this region the nuclear PDFs can be
computed through convolutions of the bound nucleon PDFs and nuclear
smearing functions describing the momentum distributions of
nucleons in the deuteron.
The latter can be expressed in terms of deuteron wave functions, 
calculated from modern potentials based on high-precision fits to    
nucleon--nucleon scattering data.  These potentials differ primarily
in their treatment of the short range $NN$ interaction, and the
different strengths of the high-momentum tails of the wave functions
translate directly to the magnitude of the nuclear corrections
at large $x$ \cite{Accardi:2009br,Arrington:2011qt}.

The nucleon off-shell corrections, on the other hand, are somewhat
more model dependent, and several model studies have been performed
to estimate their effect on nuclear 
PDFs~\cite{Melnitchouk:1993nk,Melnitchouk:1994rv,Kulagin:1994fz,Kulagin:2004ie,Ehlers:2014jpa}.  
Some earlier PDF analyses \cite{Accardi:2009br,Accardi:2011fa,Owens:2012bv} used
specific physics-motivated models for the off-shell corrections,
while more recent approaches have fitted the off-shell parameters
directly to data~\cite{Kulagin:2004ie,Accardi:2016qay}.
Other analyses \cite{Martin:2012da,Harland-Lang:2014zoa} have attempted to parametrize
the entire nuclear correction in terms of a universal,
$Q^2$-independent function, without appealing to physical constraints.
In this approach, to account for the effects of Fermi smearing a
functional form must be used that produces the steep rise in the       
$F_2^d/F_2^N$ ratio at high $x$, such as with a logarithm raised
to a high power \cite{Martin:2012da}.

The effects of the nuclear corrections are most directly visible
in the extraction of the $d$-quark PDF at large $x$, see~\cite{Accardi:2016qay}.
Figure~\ref{fig:nuclear} shows that omitting nuclear smearing
effects in the deuteron leads to an overestimated $d/u$ ratio
at $x \gtrsim 0.6$.  In fact, omitting nuclear corrections
induces a strong tension between the SLAC deuteron 
DIS data (see, e.g.,~\cite{Dasu:1993vk}) 
and the recent high precision $W$-boson asymmetry 
data from the D\O\ collaboration at the Tevatron~\cite{D0:2014kma}, which are
sensitive to the $d$-quark PDF in a similar large-$x$ range as
the SLAC data, but are not affected by nuclear corrections.
It also causes an artificial deformation of the $d$-quark distribution,
leading to essentially uncontrolled systematic errors when quark distributions
are needed beyond the $x$ range constrained by the data. This
illustrates not only the theoretical but also the phenomenological need for such
corrections when considering data at large $x$.
Of course, one can choose to avoid nuclear effects altogether
by using only proton data; however, doing so increases the
uncertainty on the $d$-quark PDF at both small and large values
of $x$, as Fig.~\ref{fig:nuclear} illustrates.
Additional details concerning the role of nuclear corrections
when using deuterium target data in global fits can be found
in Ref.~\cite{Accardi:2016qay}.
The extrapolation of nuclear effects from the deuteron to heavy
nuclei is unclear, especially in view of the differences between the
off-shell quark deformation fitted using deuteron targets~\cite{Accardi:2016qay} and using the ratio 
of heavy nuclei to deuteron structure functions~\cite{Kulagin:2004ie}.
As mentioned in the previous section, 
in general care should also be exercised when using neutrino-nucleus scattering data to obtain, for
example, constraints on strange-quark PDFs, due to the currently poor
understanding of the interaction dynamics of the final state heavy quark
propagating through the target nucleus~\cite{Accardi:2016muk}.

\subsection{Software and tools}

Data used in the PDF fits cover a wide range of kinematics 
and stem from a large number of different scattering processes.
In order to achieve an accurate theoretical description of both the PDF
evolution and the hard scattering cross sections, 
well-tested software and tools are necessary.
Benchmark numbers for the PDF evolution have long been established,
see e.g., the Les Houches report~\cite{Giele:2002hx}, 
and open-source evolution codes such as 
{\tt QCDNUM}~\cite{Botje:2010ay,Botje:2016wbq} and {\tt Hoppet}~\cite{Salam:2008qg} 
are available in Bjorken $x$-space and {\tt QCD-Pegasus}~\cite{Vogt:2004ns} in Mellin $N$-space.
This is an important development as it allows to expose 
the software used in the PDF fits to systematic validation, the need of which
can be illustrated with recent theory improvements published by various groups.
For example, MSTW~\cite{Martin:2009iq} has tested its NNLO evolution code against 
{\tt QCD-Pegasus}~\cite{Vogt:2004ns} and corrected the implementation of 
one of the heavy-quark OMEs.

For the hard scattering cross sections of the various processes, fast fitting
methods like {\tt fastNLO}~\cite{Kluge:2006xs,Britzger:2012bs} and {\tt APPLGrid}~\cite{Carli:2010rw} have been developed. 
In addition, some groups have also published open-source code for the theory
predictions of all physical cross sections employed in their analyses.
The ABM11 and ABM12 fits~\cite{Alekhin:2013nda,Alekhin:2012ig} 
use {\tt OPENQCDRAD}~\cite{OPENQCDRAD} code, which is publicly available.
The HERAPDF2.0 fit~\cite{Abramowicz:2015mha} relies on the QCD fit platform {\tt xFitter} 
(formerly known as {\tt HERAFitter})~\cite{Alekhin:2014irh,HERAFitter}, 
which is an open-source package that provides a framework for the
determination of PDFs and enables the choice of theoretical 
options for obtaining PDF-dependent cross section predictions.
In particular, {\tt xFitter} allows for a choice of different available schemes for treatment of heavy quarks in DIS. 
In Mellin $N$-space, an efficient method exists~\cite{Kosower:1997vj,Stratmann:2001pb} 
which improves on that by \cite{Graudenz:1995sk} and which has been widely
used in analyses, e.g. \cite{Stratmann:2001pb}.
However, no code has been made publicly available so far.

Given the increasing precision of PDF analyses, which is driven by the accuracy of the experimental data, 
there is ongoing demand to provide theoretical predictions that are as precise as possible.
This has stimulated recent checks of the analysis software used by various groups 
and has resulted in a number of documented improvements.
The list includes, for example, the corrections to the different parts of the DIS cross section
calculations in the NNPDF2.1, MSTW and CT10 PDF analyses as mentioned in the discussion of the PDFs for strange sea above.

This illustrates that there is a continued need for benchmarking 
of hard scattering cross sections of relevance for PDF determinations in
order to consolidate the accuracy of theory predictions for those observables. 
In this respect, open-source software may facilitate future theory
improvements and may help to establish standards for precision theory predictions.

\section{Strong coupling constant}
\label{sec:alphas}

The value of the strong coupling constant $\alpha_s(M_Z)$ has 
a direct impact on the size of a number of cross sections at the LHC, 
such as Higgs boson production, see Sec.~\ref{sec:crs-at-lhc}, 
and is therefore an important parameter.
Due to QCD factorization, $\alpha_s$ exhibits a significant correlation 
with the gluon PDF and also with the charm-quark mass,
as documented in the published correlation matrices, see
for instance \cite{Alekhin:2012ig}.
Therefore, the strong coupling constant has come to require particular attention
in the context of global PDF analyses.

Current precision determinations of $\alpha_s(M_Z)$ require NNLO accuracy in QCD 
because of the small uncertainties in the experimental data analyzed and 
the significantly reduced dependence from the variation of the renormalization
scale indicating the uncertainty due to the truncation of the perturbative series.
Extractions of $\alpha_s$ at NLO typically yield $\alpha_s(M_Z) \simeq 0.118$, however,
the NLO scale uncertainty is large, giving sizable variations $\Delta \alpha_s(M_Z) = 0.005$ 
for $\mu_r \in [Q/2,2Q]$ in DIS analyses.
Determinations of $\alpha_s$ to NNLO accuracy benefit from a significantly
reduced renormalization scale dependence, but generally result in smaller
central values for $\alpha_s(M_Z)$, with shifts downwards from NLO to NNLO of
a few percent in DIS analyses.
Beyond NNLO, the perturbative expansion converges, as illustrated in DIS 
in a valence~analysis \cite{Blumlein:2006be} at N$^3$LO which yields $\alpha_s(M_Z) = 0.1141~_{-~0.0022}^{+~0.0020}$,
in agreement with the NNLO values listed in Tab.~\ref{tab:alphas}.

%
\input{table-alphas-in-pdfs}
%
%

%
\input{table-alphas}
Of course, measurements of $\alpha_s(M_Z)$ are not limited to global fits of PDFs, but 
stem from a large number of different processes and methods at different scales, 
see, e.g., \cite{Bethke:2011tr,Moch:2014tta,d'Enterria:2015toz} for discussions and
comparisons. 
Here we restrict ourselves to issues of $\alpha_s$ arising in PDF fits.
In Tab.~\ref{tab:alphas-in-pdfs} we give an overview of the $\alpha_s$ values
currently used in the PDF analyses. There, two aspects are important.
Firstly, some PDF analyses leave  $\alpha_s$ as a free parameter in their fits,
which obviously allows one to control its correlation with other PDF parameters
and avoids potential biases.
Secondly, among the NNLO values of $\alpha_s(M_Z)$ used there exists a large
spread of $\alpha_s$ values, ranging from $\alpha_s(M_Z) = 0.1132$ to 0.1183.
Some of those fitted values of $\alpha_s(M_Z)$ are significantly smaller than,
for example, an average provided by the PDG~\cite{Agashe:2014kda} in 2014, 
which gives $\alpha_s(M_Z) = 0.1185 \pm 0.0006$ at NNLO, and is often quoted as
a motivation for fixing $\alpha_s(M_Z) = 0.118$ as in some entries in Tab.~\ref{tab:alphas-in-pdfs}.
In the recent 2015 update, the PDG~\cite{PDGupdate} reports the value $\alpha_s(M_Z) = 0.1181 \pm 0.0013$
with the uncertainty increased by a factor of two.

While the potential agreement or disagreement with the PDG average is beyond
the scope of this study, it is instructive to focus on $\alpha_s(M_Z)$ measurements 
from PDF analyses as listed in Tab.~\ref{tab:alphas} 
which have been performed since the NNLO QCD corrections in DIS first became available.
This series of measurements has led to $\alpha_s(M_Z)$ values 
which are not only mostly lower than the PDG average, 
but also exhibit a large spread in the range $\alpha_s(M_Z) = 0.1120 - 0.1175$.
This spread is significant given the small size of the experimental uncertainties in the data. 
As it turns out, the differences in the values of $\alpha_s(M_Z)$ 
can be traced back to different data sets used or to different theory
assumptions applied, as indicated in Tab.~\ref{tab:alphas}.

%
%
\input{table-th-accuracy}

For instance, the inclusion of data for the hadro-production of jets, e.g., from
the LHC, does have an impact on the value of $\alpha_s(M_Z)$ 
and can therefore provide valuable constraints.
However, it is important to note that the perturbative QCD corrections 
to the hard scattering cross section are only known completely to NLO, 
while the exact NNLO result for the $gg$ channel~\cite{Ridder:2013mf}
and approximations based on soft gluon enhancement~\cite{Kidonakis:2000gi,Kumar:2013hia,deFlorian:2013qia,Carrazza:2014hra} 
indicate corrections as large as 15\%--20\%. 
Those corrections and their magnitude depend, of course, 
on the details of the kinematics, the choice of the scale and on the jet parameters (jet radius $R$). 
For high $p_T$ they are dominated by threshold logarithms $\ln(p_T)$ accompanied by 
logarithms $\ln(R)$ for small jet radii~\cite{deFlorian:2013qia}. 

The $\alpha_s(M_Z)$ values in PDF analyses currently determined with the help of jet
data (cf.~Tab.~\ref{tab:alphas}) are, 
strictly speaking, valid to NLO accuracy only and therefore subject to significantly larger
theory uncertainties due to the variation of the renormalization scale.
The various groups employ different approaches in their NNLO analyses 
to cope with this inconsistency, 
such as using dynamical scales or applying some variant of threshold corrections, 
as detailed in Tab.~\ref{tab:jets}.
As a result of these efforts, the gluon PDF and $\alpha_s$ obtained, 
for example, in the MMHT14 and NNPDF3.0 analyses are in a good agreement.

Different modeling of important theory aspects, such as whether or not to include 
target mass corrections, higher twist contributions and nuclear corrections in
the description of DIS data, or whether or not to use a VFNS in the description of DIS heavy-quark data, 
can account for the range of $\alpha_s(M_Z)$ in Tab.~\ref{tab:alphas}. 
With largely similar model assumptions,
NNPDF2.1~\cite{Lionetti:2011pw,Ball:2011us}, MSTW~\cite{Martin:2009bu}
and MMHT~\cite{Harland-Lang:2015nxa} obtained the range $\alpha_s(M_Z) = 0.1171 - 0.1174$.
All these choices can lead to systematic shifts of the value of $\alpha_s(M_Z)$.
Let us briefly mention some of the issues in detail.

Higher twist contributions do have a big impact, because these terms are fitted within a combined analysis. 
Alternatively, the part of the DIS data significantly affected by these terms has to be removed by suitable kinematical cuts on the 
scale $Q^2$ and center-of-mass energies $W^2$.
In a variant of the ABM11 analysis~\cite{Alekhin:2012ig}, higher twist terms have been omitted
and the cuts $W^2 > 12.5~\GeV^2$ and $Q^2 > 2.5~\GeV^2$ as used by MSTW~\cite{Martin:2009bu} 
have been applied. 
This resulted in a sizable shift upwards to $\alpha_s(M_Z^2) = 0.1191 \pm 0.0016$ 
in line with earlier studies in~\cite{Vogt:1999ik}.
Yet more conservative cuts of $W^2 > 12.5~\GeV^2$ and $Q^2 > 10~\GeV^2$ 
in the ABM11 variant with higher twist terms set to zero led to $\alpha_s(M_Z^2) = 0.1134 \pm 0.0008$, 
well in agreement with the nominal value in the ABM11 analysis, cf.~Tab.~\ref{tab:alphas}.
Thus, in PDF analyses without account of higher twist contributions to DIS
data such tight cuts are essential.
  In this regard we disagree with Refs.~\cite{Thorne:2011kq,Thorne:2014toa,Ball:2013gsa} 
  which claim higher twist effects to be negligible in the framework of
  MSTW~\cite{Martin:2009iq} and NNPDF2.3~\cite{Ball:2012cx}.
We also note that NNPDF3.0 \cite{Ball:2014uwa} uses a cut of $Q^2 > 3.5~\GeV^2$
which is too low to remove the higher twist contributions.

Higher order constraints from fixed-target DIS data can also lead to shifts in $\alpha_s(M_Z)$~\cite{Alekhin:2011ey}.
For instance, NMC has measured the DIS differential cross sections 
and extracted the DIS structure functions $F_2^{\rm NMC}$~\cite{Arneodo:1996qe}.
At the time of the NMC analysis, however, the relevant DIS corrections to ${\cal O}(\alpha_s^3)$~\cite{Vermaseren:2005qc}
were not available (see discussion after Eq.~(\ref{eq:CallanGross}) above).
This information is, however, important and has to be taken into account now.
In case of fitting $F_2^{\rm NMC}$ and not describing $F_L(x,Q^2)$ at NNLO,
much larger values of $\alpha_s(M_Z^2)$ are obtained~\cite{Thorne:2011kq}.
It is therefore strongly recommended to fit the published differential scattering cross sections using
$F_L(x,Q^2)$ at ${\cal O}(\alpha_s^3)$. Presently, the MMHT \cite{Harland-Lang:2015nxa} analysis 
uses $F_L(x,Q^2)$ only at NLO. 
One should note, however, that the values of $F_L(x,Q^2)$ at NNLO are significantly different 
in the small-$x$ region (see~\cite{Thorne:2011kq}).

Finally, great care needs to be exercised when DIS data off nuclei are included in global fits, see Sec.~\ref{sec:th4pdfs}.
Details of modeling of nuclear corrections can in fact also cause systematic shifts in the value of $\alpha_s(M_Z)$. 
Therefore, Tab.~\ref{tab:alphas} indicates if scattering data on heavy nuclei have been included in the determination.
For example, MMHT~\cite{Harland-Lang:2015nxa} has reported a comparatively high value
of $\alpha_s(M_Z)$ as a consequence of fitting the NuTeV $\nu$Fe DIS data~\cite{Goncharov:2001qe}.
In general, determinations of $\alpha_s(M_Z)$ should be based upon, 
or at least cross-checked with, fits using proton and deuteron DIS data only.

\section{Cross section predictions for the LHC}
\label{sec:crs-at-lhc}

\subsection{Higgs boson production}

The dominant production mechanism for the SM Higgs boson at the LHC is the gluon-gluon fusion process. 
The large size of the QCD radiative corrections to the inclusive cross section at NLO, see, e.g. Ref.~\cite{Spira:1995rr}, 
together with the sizable scale uncertainty have motivated systematic theory improvements.
In the effective theory based on the limit of a large top-quark mass 
($m_t \to \infty$, integrating out the top-quark loop, but using the full $m_t$ dependence in the Born cross section), 
this has led to the computation of the corresponding corrections 
at NNLO~\cite{Harlander:2002wh,Anastasiou:2002yz,Ravindran:2003um} 
and even to N$^3$LO in QCD~\cite{Anastasiou:2015ema,Charalampos:2016xyz}. 
This shows an apparent, if slow, convergence of the perturbative expansion, along with
greatly reduced sensitivity to the choice for the renormalization and factorization scales $\mu_r$ and $\mu_f$.
At N$^3$LO the total scale variation amounts to 3\% 
and estimates of the four-loop corrections support these findings~\cite{deFlorian:2014vta}.

%
\input{table-higgs}
This leaves, as the largest remaining source of uncertainties in the predictions of the
physical cross section, the input for the strong coupling constant $\alpha_s$ and the PDFs.
Despite the impressive progress in theory and experiment, the situation
resembles that after the completion of the NLO QCD corrections, when it was
pointed out in Ref.~\cite{Spira:1995rr} that one of the main residual
uncertainties in the predictions was due to the gluon PDF.

In Tab.~\ref{tab:higgs} we summarize the PDF dependence of the inclusive cross section
$\sigma(H)^{\rm NNLO}$ in the effective theory (i.e., in the limit of $m_t \gg m_H$) 
at $\sqrt{s}=13$~TeV for a Higgs boson mass $m_H=125.0$~GeV,
$\mu_r = \mu_f = m_H$, and $m_t^{\rm pole}=172.5$~GeV
with uncertainties $\sigma(H)^{\rm NNLO} + \Delta \sigma(\rm{PDF}+\alpha_s)$, 
and compare the results for various PDF sets.
The PDF uncertainties are typically given at the 1$\sigma$ c.l.
We list the results for $\sigma(H)^{\rm NNLO}$ using either the values for the strong coupling constant $\alpha_s(M_Z)$
at NNLO, corresponding to the respective PDF set, or fixed values of $\alpha_s(M_Z)=0.115$ and $\alpha_s(M_Z)=0.118$.
This is done to illustrate the fact that in some PDFs the value of $\alpha_s(M_Z)$
is not obtained from a fit to data (including faithful uncertainties) 
but fixed beforehand, e.g., to the world average~\cite{Agashe:2014kda}. 
Often the same fixed value of $\alpha_s(M_Z)$ is chosen at NLO
and at NNLO independent of the order of perturbation theory, see also Sec.~\ref{sec:alphas}.
Table~\ref{tab:higgs} shows a large spread for predictions from different PDFs 
with a range $\sigma(H)^{\rm NNLO} = 38.0 - 42.6$~pb using the nominal value of $\alpha_s(M_Z)$. 
Specifically, the PDF and $\alpha_s$ differences between different sets are up to $11\%$ 
and are significantly larger than the residual scale uncertainty due to N$^3$LO QCD corrections.
In addition, the cross sections shift in the range $\sigma(H)^{\rm NNLO} = 39.0 - 44.7$~pb 
if a fixed value of $\alpha_s(M_Z)$ in the range $\alpha_s(M_Z)=0.115 - 0.118$ is used.
This amounts to a relative difference of more than $13\%$ and contradicts 
the most recent estimates of the combined PDF and $\alpha_s$ uncertainties in the inclusive cross section~\cite{Charalampos:2016xyz}, 
which quotes $3.2\%$.
In general, the findings underpin the 
importance of controlling the accuracy and the correlation of the strong coupling constant with
the PDF parameters in fits. 

%
%
\input{table-higgs-mc-mstw}

%
%

%
\input{table-higgs-mc-mmht}

%
%

%
\input{table-higgs-mc-nnpdf}

Of particular interest is the impact of additional parameters in the PDF fits,
such as the charm-quark mass, on the Higgs cross section.
The differences in the treatment of heavy quarks and the consequences for the quality of the
description of charm-quark DIS data have already been discussed in Sec.~\ref{sec:th4pdfs}.
ABM12~\cite{Alekhin:2013nda} fits the value of $m_c(m_c)$ in the \msbar scheme
and the uncertainties in the charm-quark mass are included in the uncertainties
quoted in Tab.~\ref{tab:higgs}.
Other groups keep a fixed value of the charm-quark mass in the on-shell scheme,
cf.~Tabs.~\ref{tab:mcmass} and \ref{tab:mcmass-ctd}, and vary the value of $m_c^{\rm pole}$ within some range.
Such studies have been performed in the past by NNPDF2.1~\cite{Ball:2011mu} and MSTW~\cite{Martin:2010db} 
and more recently by MMHT \cite{Harland-Lang:2015qea}.

In Tabs.~\ref{tab:higgs-mc-mstw},~\ref{tab:higgs-mc-mmht} and~\ref{tab:higgs-mc-nnpdf} 
we display the results of these fits together with the values of $\chi^2$/NDP 
for the DIS charm-quark data \cite{Abramowicz:1900rp}, mostly computed with
{\tt xFitter}~\cite{Alekhin:2014irh,HERAFitter}, as well
as the corresponding cross section for Higgs boson production to NNLO accuracy.
The MSTW analysis in Tab.~\ref{tab:higgs-mc-mstw} 
shows a linear rise of the cross section for increasing values $m_c^{\rm pole} = 1.05 - 1.75~\GeV$
in the range $\sigma(H) = 40.6 - 43.8~\mbox{pb}$, which amounts to a variation of more than $7\%$.
Even if $\alpha_s(M_Z)=0.1171$ is kept fixed, the cross section varies in the range $\sigma(H) = 41.6 - 42.6~\mbox{pb}$, which is equivalent to $2\%$.
The best fit in the MSTW analysis with $\chi^2$/NDP = 63/52 leads to $m_c^{\rm 
pole}=1.3~\GeV$ and $\alpha_s(M_Z)=0.1166$, 
both of which are lower than the ones of the nominal fit with $m_c^{\rm pole}=1.4~\GeV$ and $\alpha_s(M_Z)=0.1171$.
In Tab.~\ref{tab:higgs-mc-mmht} the same study is performed for the MMHT PDFs~\cite{Harland-Lang:2015qea}, where the reduced 
quark mass range $m_c^{\rm pole} = 1.15 - 1.55~\GeV$ still leads to cross
section variations $\sigma(H) = 40.5 - 42.1~\mbox{pb}$ (i.e., $4\%$) for the best fit $\alpha_s(M_Z)$, 
or $\sigma(H) = 42.1 - 42.6~\mbox{pb}$ (i.e., $1\%$) for a fixed $\alpha_s(M_Z)=0.118$.
The latter case leads to a best fit of $m_c^{\rm pole}=1.2~\GeV$ with $\chi^2$/NDP = 70/52, 
which is significantly smaller than the nominal fit with $m_c^{\rm pole}=1.4~\GeV$ and 
$\chi^2$/NDP = 82/52.

NNPDF has performed a study of the $m_c$ dependence in~\cite{Ball:2011mu},
which shows the same trend as for MSTW and MMHT, i.e., the smaller the chosen value
of $m_c^{\rm pole}$, the better the goodness-of-fit for the HERA data \cite{Abramowicz:1900rp}.
In addition, Tab.~\ref{tab:higgs-mc-nnpdf} displays the changes in the
charm-quark mass values from $m_c^{\rm pole} =\sqrt{2}~\GeV$ to $m_c^{\rm pole} = 1.275~\GeV$
in the evolution of the NNPDF fits from v2.1~\cite{Ball:2011mu} and v2.3~\cite{Ball:2012cx} to v3.0~\cite{Ball:2014uwa},
with the obvious correlation of smaller cross sections for Higgs boson
production with smaller chosen values of $m_c^{\rm pole}$.

As pointed out already in Sec.~\ref{sec:th4pdfs}, on-shell masses 
$m_c^{\rm pole}=1.2 - 1.3~\GeV$ as preferred by the 
goodness-of-fit analyses in Tabs.~\ref{tab:higgs-mc-mstw},~\ref{tab:higgs-mc-mmht} and~\ref{tab:higgs-mc-nnpdf} 
for the charm-quark data from HERA~\cite{Abramowicz:1900rp}, 
are not compatible with the world average of the PDG~\cite{Agashe:2014kda}.
Thus, in some PDF fits, the numerical value of the charm-quark mass effectively 
takes over the role of a ``tuning'' parameter for the Higgs cross section.
Note that the three analyses are based on partly different data sets, theory and methodology.

\subsection{Hadro-production of heavy quarks}

\subsubsection{Top-quark hadro-production: inclusive cross section}

%
\input{table-ttbar}

The cross section for the hadro-production of top-quark pairs has been measured
with unprecedented accuracy at the LHC in Run 1 with $\sqrt{s}=7~\TeV$ and $8~\TeV$.
The inclusive cross section is known to NNLO in QCD~\cite{Baernreuther:2012ws,Czakon:2012zr,Czakon:2012pz,Czakon:2013goa},
featuring good convergence of the perturbation series and 
reduced sensitivity to the renormalization and factorization scales $\mu_r$ and $\mu_f$.
These theory predictions adopt the on-shell renormalization scheme for the heavy-quark mass.
The conversion to the \msbar scheme for the heavy-quark mass has been
discussed in Refs.~\cite{Langenfeld:2009wd,Aliev:2010zk,Dowling:2013baa}.
For observables such as the inclusive cross section which are dominated by hard
scales $\mu_r \simeq \mu_f \simeq m_t$, the theory predictions in terms of 
the \msbar mass for the top quark show an even better scale stability and perturbative convergence.

In a similar study as for Higgs boson production in Tab.~\ref{tab:higgs} we
illustrate in Tab.~\ref{tab:ttbar} the PDF dependence of the
inclusive cross section $\sigma(t{\bar t})^{\rm NNLO}$ for various sets 
with uncertainties $\Delta \sigma(\rm{PDF}+\alpha_s)$.
The computation is performed in the theoretical framework as implemented in the {\texttt{HATHOR}} code~\cite{Aliev:2010zk}.
In Tab.~\ref{tab:ttbar} we choose $\sqrt{s}=13$~TeV and fix the pole mass $m_t^{\rm pole}=172.0$~GeV
and the scales at $\mu_r = \mu_f = m_t^{\rm pole}$.
For this fixed value of $m_t^{\rm pole}$, we show the impact of different
values for the strong coupling constant at NNLO. 
We choose $\alpha_s(M_Z)$ either corresponding to the respective PDF set or fixed to the values $0.115$ and $0.118$.
The results in Tab.~\ref{tab:ttbar} display a spread in a range $\sigma(t{\bar t})^{\rm NNLO} = 715 - 834~\mbox{pb}$ 
using the nominal value of $\alpha_s(M_Z)$ for each PDF set, 
which amounts to a relative range of more than $15\%$.
This decreases to about $6\%$, if the values of $\alpha_s(M_Z)$ 
are fixed to $0.115$ or $0.118$. 

The theoretical predictions at leading order depend parametrically on the strong coupling constant
and the top-quark mass to second power, as well as on the convolution of the gluon PDFs, 
$\sigma(t{\bar t})^{\rm LO} \propto (\alpha_s^2/m_t^2) \left(g \otimes g \right)$. 
Therefore, it is necessary to fully account for the correlations between the top-quark mass, 
the gluon PDF and the strong coupling when comparing to experimental data.
A number of analyses have considered $t{\bar t}$ hadro-production data. 
ABM12~\cite{Alekhin:2013nda} has included data for top-quark pair-production in a variant of the fit 
to determine the \msbar mass $m_t(m_t)$, keeping the full correlation with $\alpha_s(M_Z)$ and the gluon PDF.
On the other hand, CMS has determined the top-quark pole mass as well as the strong coupling
constant in a fit which kept all other parameters mutually fixed~\cite{Chatrchyan:2013haa},
while Ref.~\cite{Czakon:2013tha} has explored constraints on the gluon PDF from $t{\bar t}$ hadro-production data 
using fixed values for $\alpha_s(M_Z)$ and the pole mass $m_t^{\rm pole}$. 

%
\input{table-ttbar-mc-mmht}

%
%

%
\input{table-ttbar-mc-nnpdf}

In the global analyses by MMHT14~\cite{Harland-Lang:2014zoa} and NNPDF3.0~\cite{Ball:2014uwa} 
those data were also used to fit $\alpha_s(M_Z)$ and the gluon PDF.
These analyses employ a fixed value for the pole mass $m_t^{\rm pole}$, 
which is motivated by precisely measured top-quark masses from kinematic reconstructions, i.e., Monte Carlo masses, 
but does not account for the above mentioned correlation with $\alpha_s(M_Z)$ and the gluon PDF.
Moreover, the Monte Carlo mass requires additional calibration~\cite{Kieseler:2015jzh}.

For the inclusive top-quark cross section we explore in Tabs.~\ref{tab:ttbar-mc-mmht} and \ref{tab:ttbar-mc-nnpdf} 
the implicit dependence of the cross section on the charm-quark mass $m_c$ used in the GM-VFNS of the PDF fits 
and list the corresponding values of $\chi^2$/NDP for the DIS charm-quark data 
\cite{Abramowicz:1900rp}. 
This is analogous to the study for the Higgs cross section in Tabs.~\ref{tab:higgs-mc-mmht} and~\ref{tab:higgs-mc-nnpdf}.
For MMHT \cite{Harland-Lang:2015qea} the best fit with $m_c^{\rm pole} = 1.25~\GeV$ and  $\alpha_s(M_Z)=0.1167$
leads to an inclusive cross section of $\sigma(t{\bar t})^{\rm NNLO} = 814$~pb, 
which is $2\%$ lower than the value obtained for the nominal MMHT fit, cf.~Tab.~\ref{tab:ttbar}.
Likewise, the changes in the NNPDF fits from v2.1~\cite{Ball:2011mu} and v2.3~\cite{Ball:2012cx} to v3.0~\cite{Ball:2014uwa}
are documented in Tab.~\ref{tab:ttbar-mc-nnpdf}. 
The effects amount to almost $2\%$ when comparing $\sigma(t{\bar t})^{\rm NNLO}$ 
for the best fit of NNPDF2.1 with $m_c^{\rm pole} = \sqrt{2}~\GeV$ and  $\alpha_s(M_Z)=0.119$
to the cross section computed with NNPDF3.0 with $m_c^{\rm pole} = 1.275~\GeV$ and  $\alpha_s(M_Z)=0.118$.
In both Tables~\ref{tab:ttbar-mc-mmht} and \ref{tab:ttbar-mc-nnpdf} 
there is a correlation showing decreasing cross sections with decreasing values of $m_c^{\rm pole}$, although 
less pronounced than in the case of the Higgs production cross section.
The potential bias in the prediction of the inclusive top-quark pair production cross section 
due to a particular ``tuning'' of the value of the charm-quark mass for some PDFs is, 
however, of the same order of magnitude or larger than the quoted PDF uncertainties.
Therefore, this needs to be accounted for as an additional modeling uncertainty.

\begin{figure}[tH!]
\begin{center}
\includegraphics[width=0.425\textwidth]{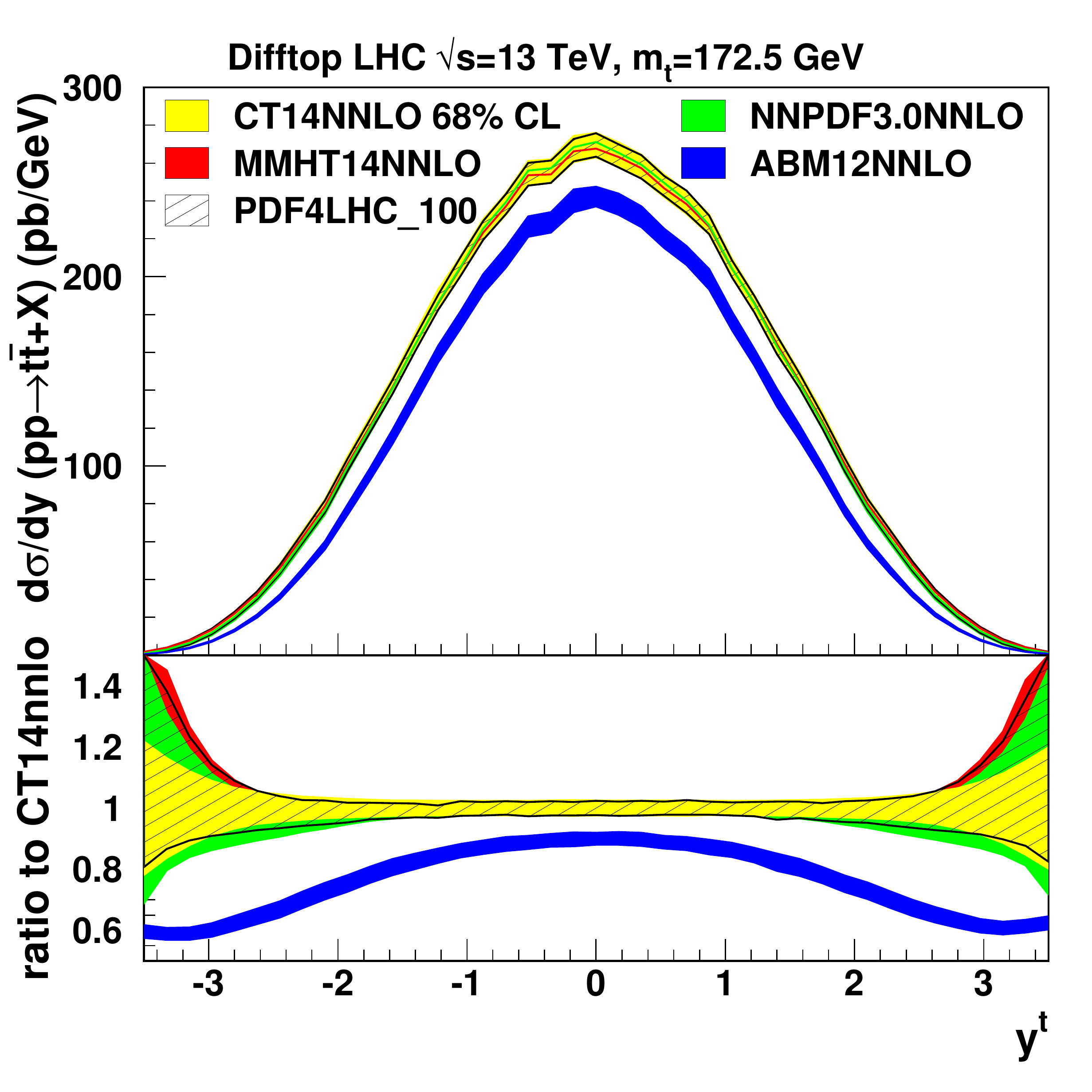}
\hspace*{7mm}
\includegraphics[width=0.425\textwidth]{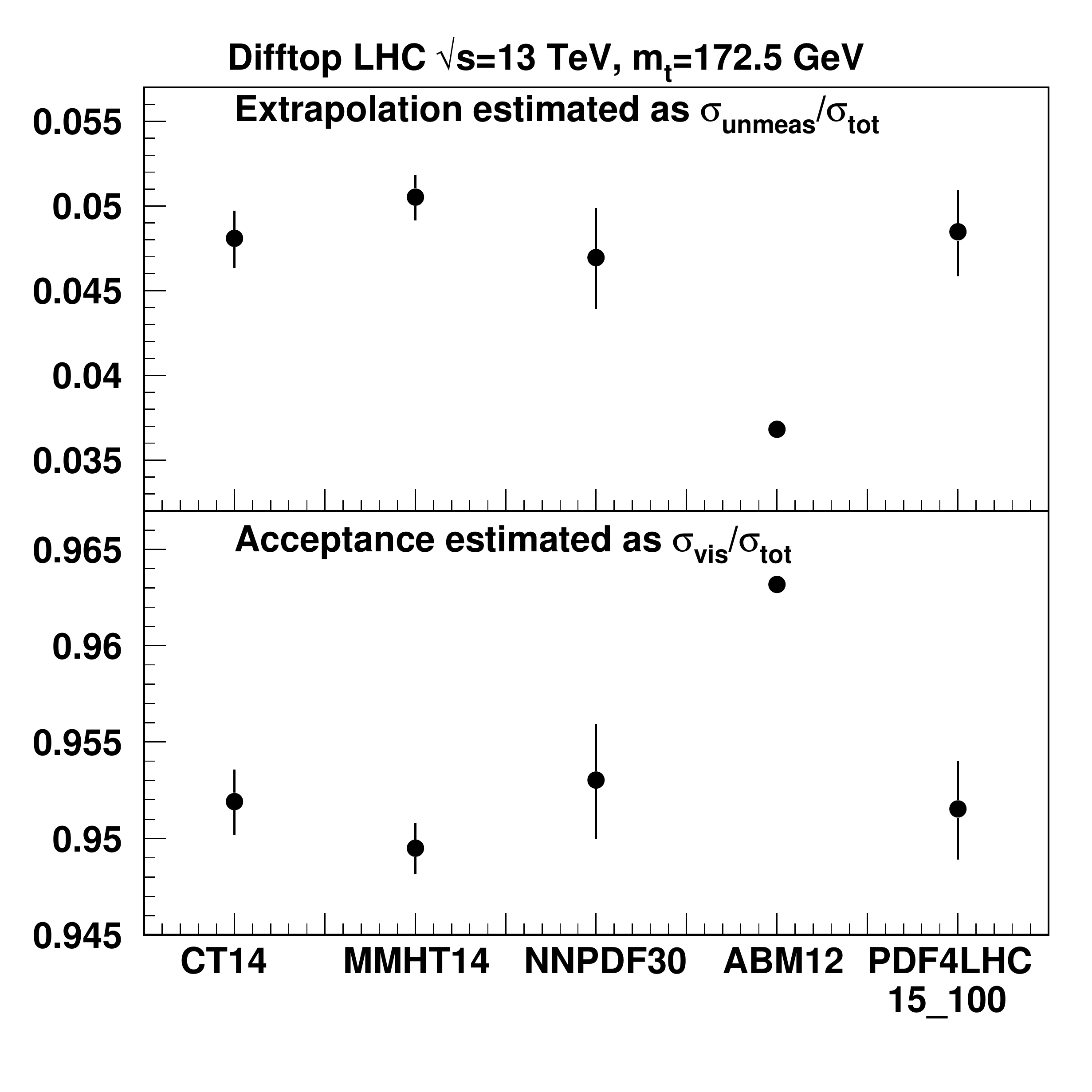}
\caption{\small 
\label{fig:difftop}
(Left panel) Predictions for top-quark pair production cross sections at approximate NNLO
as a function of the top-quark rapidity using different PDFs at NNLO with the
respective PDF uncertainty (depicted by bands of different style). 
(Right panel) The acceptance and extrapolation estimators with the respective PDF uncertainties, 
obtained by using different PDF sets.}
\end{center}
\end{figure}

\subsubsection{Top-quark hadro-production: differential distributions}

The differential cross section of the top-quark pair production is also known to NNLO in QCD~\cite{Czakon:2015owf}.
Publicly available codes such as {\tt Difftop}~\cite{Guzzi:2014wia} provide
differential distributions to approximate NNLO accuracy based on soft-gluon threshold resummation results.
We use {\tt Difftop} to calculate the distribution in the top-quark rapidity $y_t$
for proton-proton collisions at $\sqrt{s}=13$~TeV at NNLO$_{\rm approx}$
accuracy using the ABM12, CT14, MMHT14, NNPDF3.0, and the PDF4LHC15 PDF sets at NNLO 
with their respective $\alpha_s$ values. 
Here, we take the top-quark pole mass to be $m_t^{\rm pole}=172.5$~GeV, following the preferences
in the LHC analyses. The renormalization and factorization scales are set to $m_t^{\rm pole}$ 
and the choice of a dynamical scale does not change the following discussions.

By using differential cross sections, not only the sensitivity of top-quark pair production to the PDFs can be estimated, 
but also possible effects on the experimental acceptance by changing the PDF choice. 
In the experimental analysis, the PDF dependent acceptance corrections arise mostly from the
PDF dependent normalization of the production cross section and originate from
the phase space regions uncovered by the detector. 
Usually, the acceptances are determined by using Monte Carlo simulations 
as a ratio of the number of reconstructed events in the fiducial volume of the detector 
(visible phase space) to the number of events generated in the full phase space. 
In the case of top-quark pair production, the visible (full) phase space would correspond
to the top-quark rapidity range of $|y_t|<2.5$ $(|y_t|<3)$.  
Here, an acceptance estimator and a related extrapolation factor are calculated by using {\tt Difftop}
predictions for the respective cross section ratios $\sigma_{\rm vis}/\sigma_{\rm tot}$ and $\sigma_{\rm unmeasured}/\sigma_{\rm tot}$.
Such estimators are not expected to describe the true experimental efficiency,
but are helpful for drawing conclusions about PDF related effects. 

The predictions of the top-quark rapidity and the acceptance estimates obtained by
using {\tt Difftop} with different PDFs are shown in Fig.~\ref{fig:difftop}. The
largest difference in the global normalization of the predicted cross sections
is observed if the ABM12 PDFs are used instead of the CT14, NNPDF3.0 or MMHT14 sets. 
The origin of this effect is again the smaller nominal value of $\alpha_s$ in ABM12 
in combination with a smaller gluon PDF in the $x$ range relevant to top-quark
pair production at $\sqrt{s}=13~\TeV$. 
The corresponding acceptance estimators and their uncertainties, 
obtained from the error propagation of the corresponding PDF uncertainties at 68\%~c.l., 
however, demonstrate significant differences also in the expected
acceptance corrections, obtained by using ABM12 alternative to other PDFs.    

The recent PDF4LHC recommendation~\cite{Butterworth:2015oua} for calculation of the acceptance
corrections for precision observables, such as the top-quark pair-production cross section
in the LHC Run 2 data taking period, is to use the set {\tt PDF4LHC15\_100}, 
which is obtained by averaging the CT14, MMHT14 and NNPDF3.0 PDFs. 
While the central prediction obtained by using PDF4LHC15 is indeed very
close to those obtained with the CT14, MMHT14 or NNPDF3.0 PDFs, the error on the
corresponding acceptance estimator somewhat underestimates the acceptance
spread of the individual PDFs with their uncertainties. 
Furthermore, it does not cover the difference in the acceptances 
to the one using the ABM12 PDF. 
Therefore, for the conservative estimate of the acceptance correction 
and its uncertainty, as demanded in the measurement of SM precision observables,
the use of the {\tt PDF4LHC15\_100} set would lead to a significant underestimation of the
uncertainty on the resulting cross section measurement.  

A further conclusion from Fig.~\ref{fig:difftop} is that in the case of
top-quark pair production, once calculational speed is needed, it seems to
be sufficient to consider a reduced choice of PDF sets.
For instance, instead of using the averaged set {\tt PDF4LHC15\_100} one can 
take just one of the three PDFs, CT14, MMHT14 or NNPDF3.0. 
Alternative PDF choices can then always be studied to some approximation with a reweighting method.
In spite of the valiant effort in Ref.~\cite{Butterworth:2015oua} 
to provide a uniform solution, the PDF choice for measurements 
of precision observables must be decided on a case-by-case basis for each particular process.

\subsubsection{Bottom-quark hadro-production}

%
\input{table-lhcb}
Bottom-quark production in proton-proton collisions at the LHC is also dominated by the gluon-gluon fusion process. 
Therefore, the LHCb measurements of $B$-meson production in the forward region~\cite{Aaij:2013noa}
with rapidities $2.0 < y < 4.5$ at $\sqrt{s}=7$~TeV probe the gluon distributions simultaneously
at small $x$ up to $x \sim 2 \times 10^{-5}$ and at large $x \simeq 1$. 
The small-$x$ region is not accessible with HERA DIS data, for example. 
The potential improvements of PDFs near the edges of the currently covered
kinematical region, namely, at small $x$ and low scales, was first
illustrated in~\cite{Zenaiev:2015rfa, Zenaiev:2015qea} using differential LHCb
data on hadro-production of $c\bar{c}$ and $b\bar{b}$ pairs.

In the present comparison in Tab.~\ref{tab:lhcbbeauty}, 
the normalized cross sections, $({\rm d}\sigma/{\rm d}y) / ({\rm d}\sigma/{\rm d}y_0)$, 
for bottom-quark production are calculated from the absolute measurements published by LHCb, 
with ${\rm d}\sigma/{\rm d}y_0$ being the cross section in the center bin, $3 < y_0 < 3.5$, of 
the measured rapidity range in each $p_T$ bin~\cite{Zenaiev:2015rfa}. 
In the absence of NNLO QCD corrections, 
the theoretical predictions are obtained at NLO in QCD~\cite{Nason:1987xz,Beenakker:1988bq,Nason:1989zy} using a fixed number of
flavors, $n_f = 3$, for the hard scattering cross sections. 
Since data for the hadro-production of heavy quarks other than top have not been
considered for publicly available PDF fits thus far, issues of any model dependence such
as in~\cite{Harlander:2011aa} due to the use of GM-VFNS cannot be quantified.
In the calculation of the normalized cross sections, the theoretical uncertainty 
is strongly reduced, since variations of the renormalization and factorization
scales as well as of the fragmentation parameters 
do not significantly affect the shape of the $y$ distributions for heavy-flavor production, 
while this shape remains sensitive to PDFs.

The values for $\chi^2$/NDP given in Tab.~\ref{tab:lhcbbeauty} 
are computed with the QCD fit platform {\tt xFitter} 
for the individual PDF sets obtained at NLO, namely,   
  ABM11~\cite{Alekhin:2012ig}, 
  CJ15~\cite{Accardi:2016qay},
  CT14~\cite{Dulat:2015mca}, 
  HERAPDF2.0~\cite{Abramowicz:2015mha},
  JR14~\cite{Jimenez-Delgado:2014twa}, 
  MMHT14~\cite{Harland-Lang:2014zoa}, 
  NNPDF3.0~\cite{Ball:2014uwa}, 
  as well as the averaged set 
  PDF4LHC15~\cite{Butterworth:2015oua}.
All PDFs provide a good description of the data, despite the fact that none of the groups use any data 
sensitive to the gluons at very low $x$, in the region directly probed by the LHCb $B$-meson measurement. 
Remarkably, one finds that $\chi^2/\mbox{NDP} < 1$ for the vast majority of the groups
(left column in Tab.~\ref{tab:lhcbbeauty}), suggesting that the derived PDF 
uncertainties at the edges of the so far measured regions might be inflated.

\begin{figure}[t!]
\begin{center}
\includegraphics[width=0.475\textwidth]{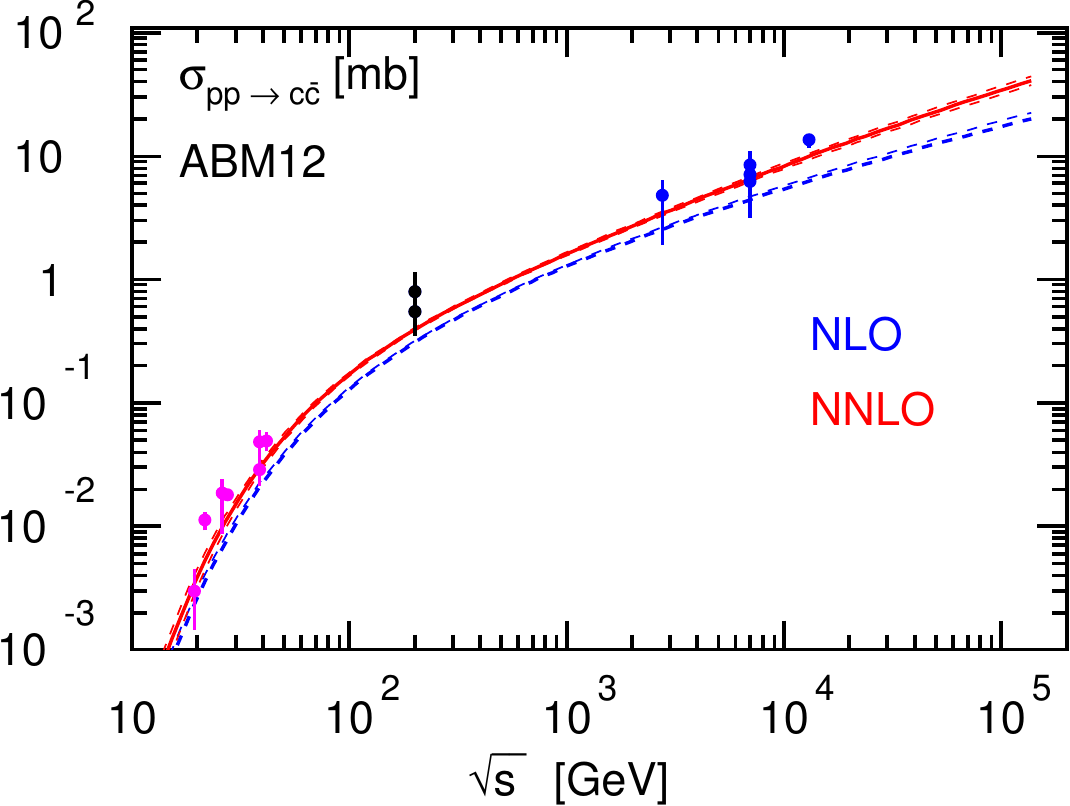}
\includegraphics[width=0.475\textwidth]{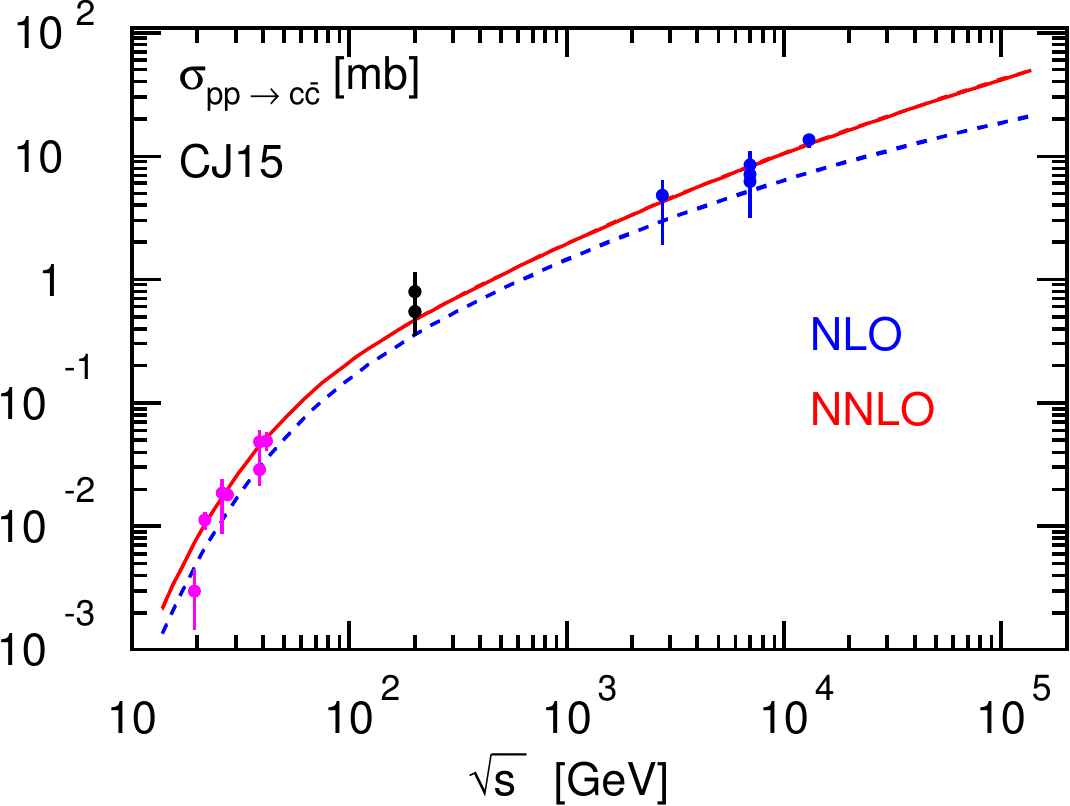} 
\\[2ex]
\includegraphics[width=0.475\textwidth]{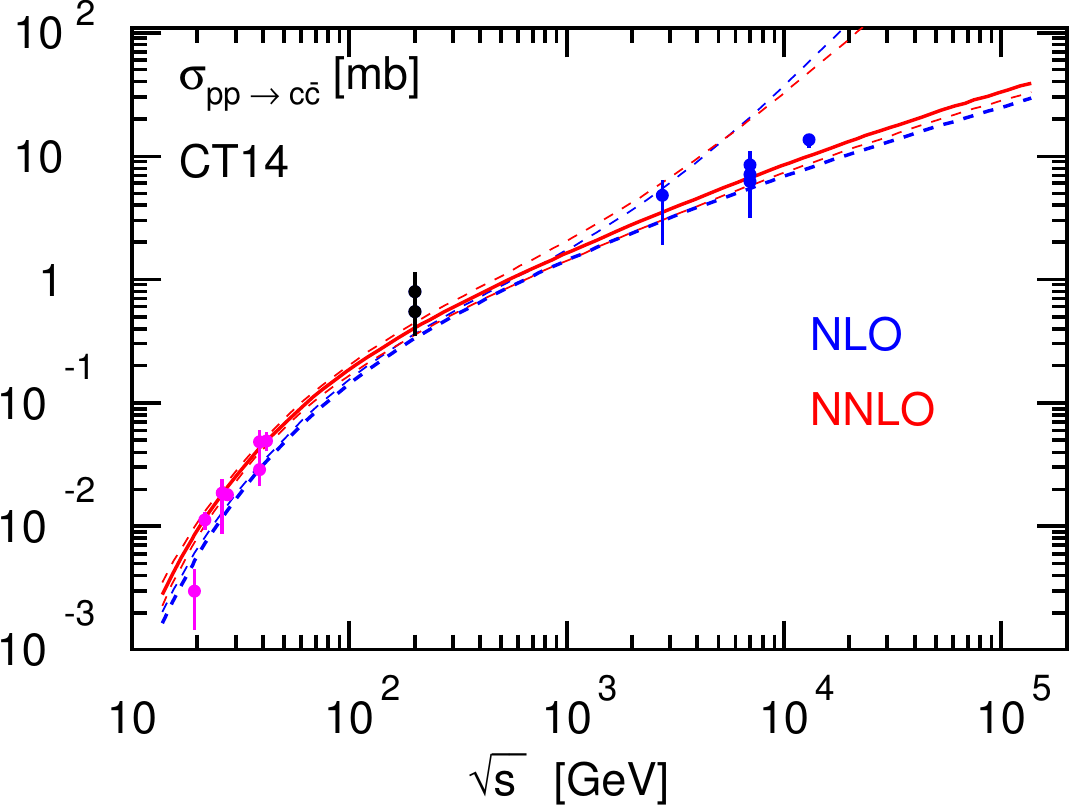}
\includegraphics[width=0.475\textwidth]{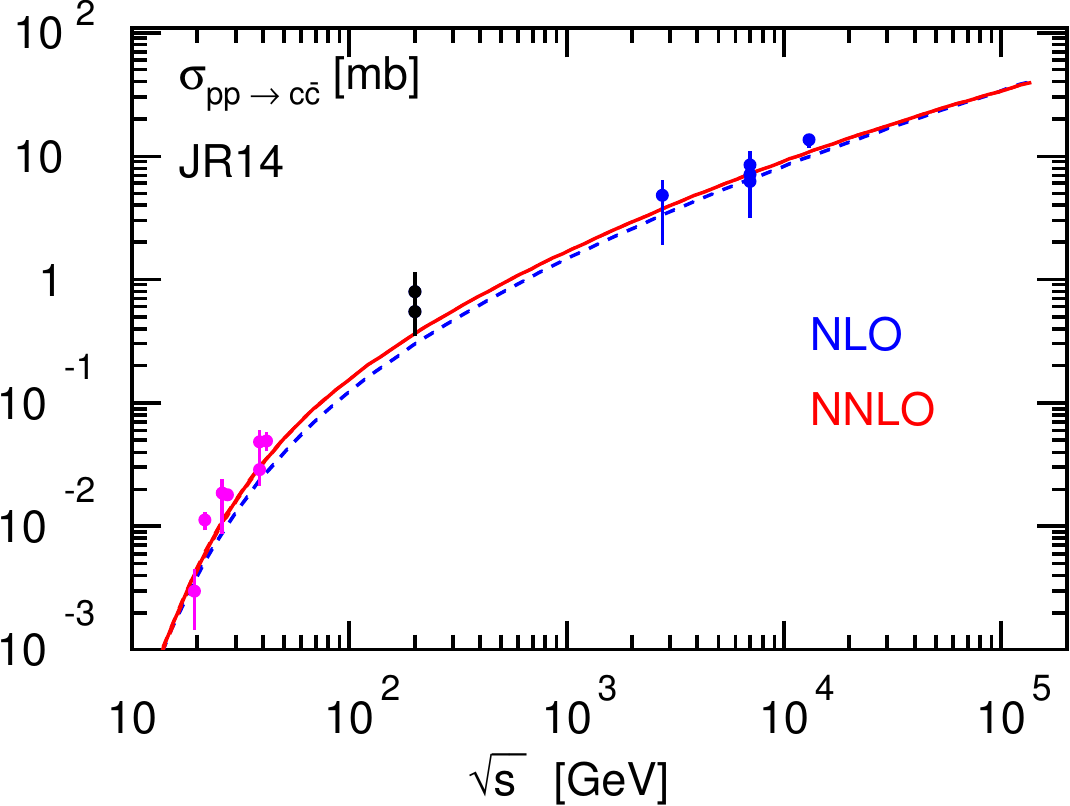}
\caption{\small 
\label{fig:sigmatot-scms-1} 
  Theoretical predictions for the total $pp \rightarrow c\bar{c}$  cross section
  as a function of the center-of-mass energy $\sqrt{s}$  
  at NLO (dashed lines) and NNLO (solid lines) QCD accuracy in the \msbar mass scheme 
  with $m_c(m_c)~=~1.27$~GeV and scale choice $\mu_R = \mu_F = 2m_c(m_c)$ 
  using the central PDF sets (solid lines) of
  ABM12~\cite{Alekhin:2013nda},
  CJ15~\cite{Accardi:2016qay}, 
  CT14~\cite{Dulat:2015mca} and 
  JR14~\cite{Jimenez-Delgado:2014twa} and the respective PDF uncertainties
  (dashed lines).
  The predictions for ABM12 (CJ15) use the NNLO (NLO) PDFs independent of the order of
  perturbation theory.
  See text for details and references on the experimental data from fixed target experiments 
  and colliders (STAR, PHENIX, ALICE, ATLAS, LHCb).
}
\end{center}
\end{figure}

\begin{figure}[t!]
\begin{center}
\includegraphics[width=0.475\textwidth]{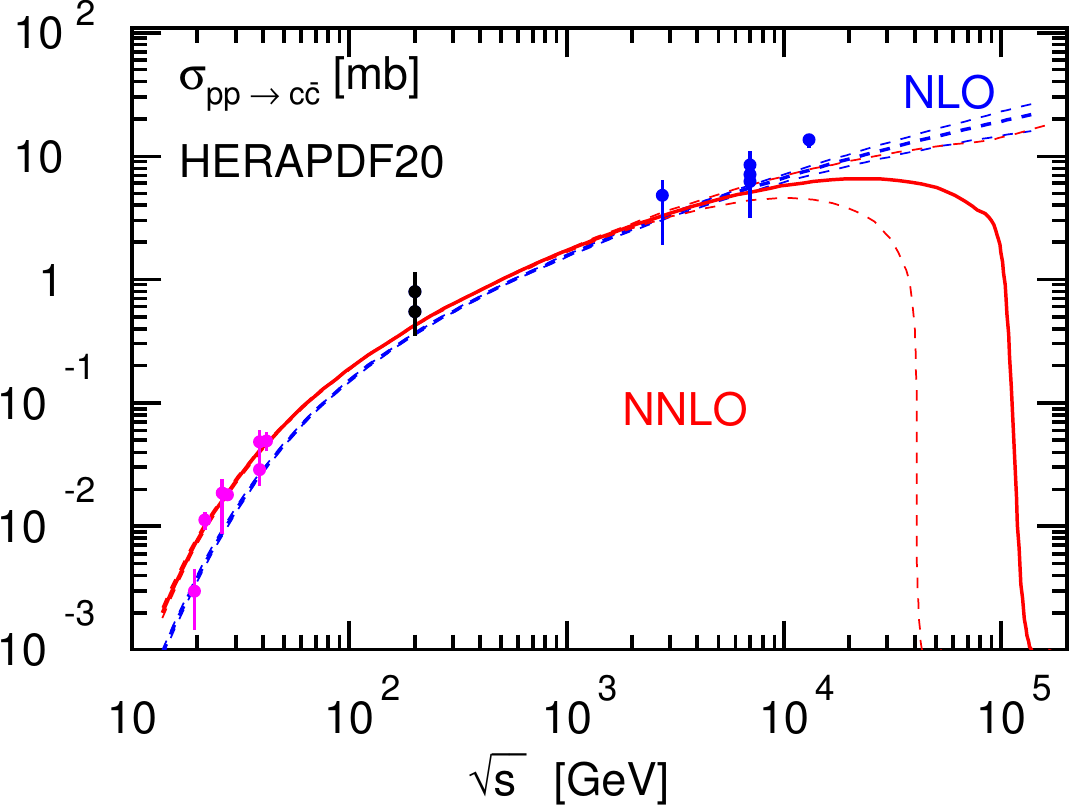}
\includegraphics[width=0.475\textwidth]{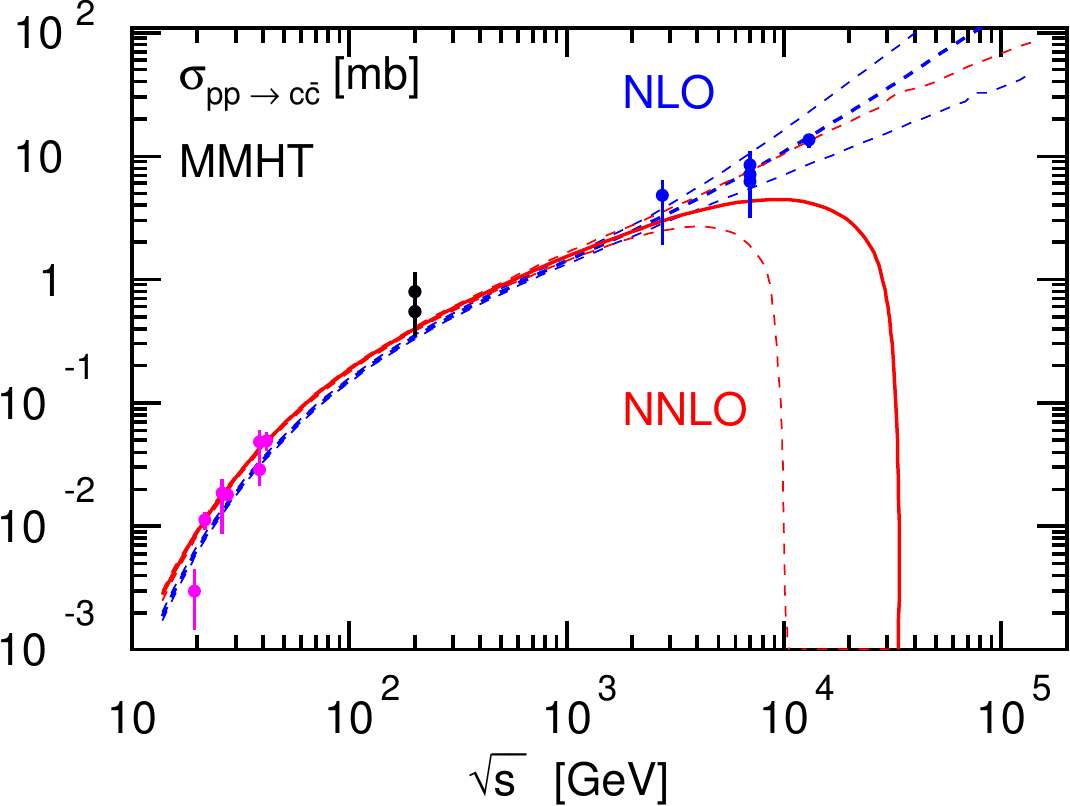} 
\\[2ex]
\includegraphics[width=0.475\textwidth]{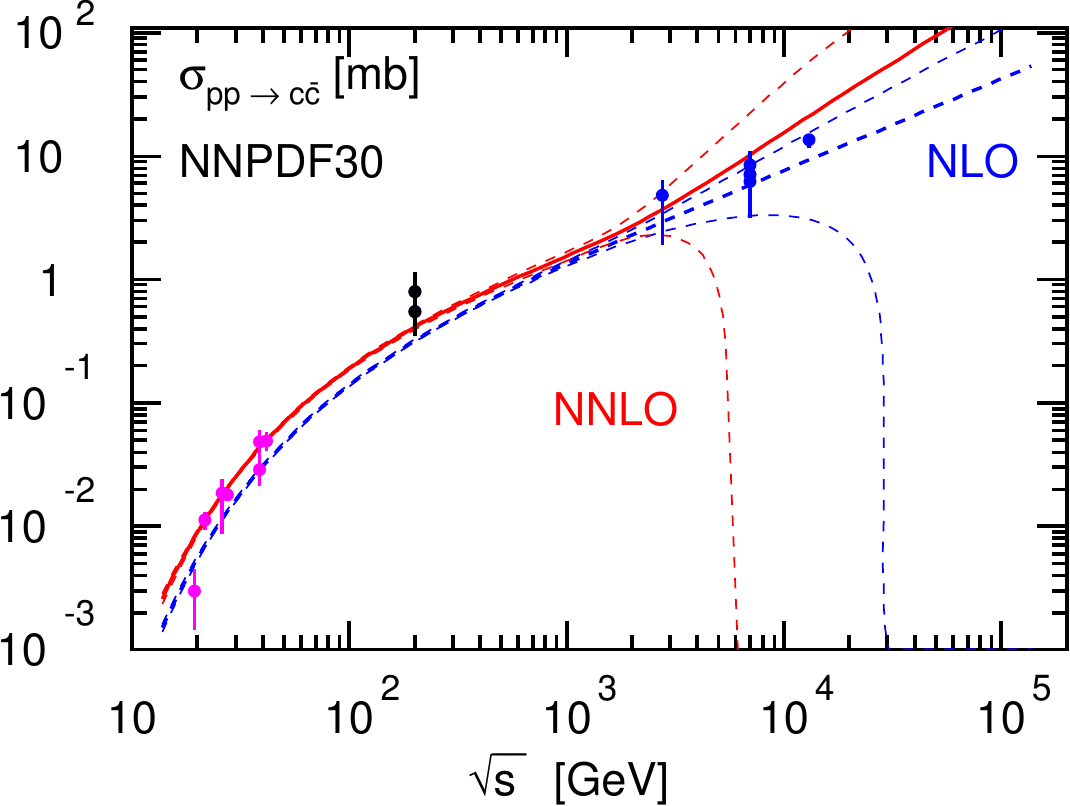}
\includegraphics[width=0.475\textwidth]{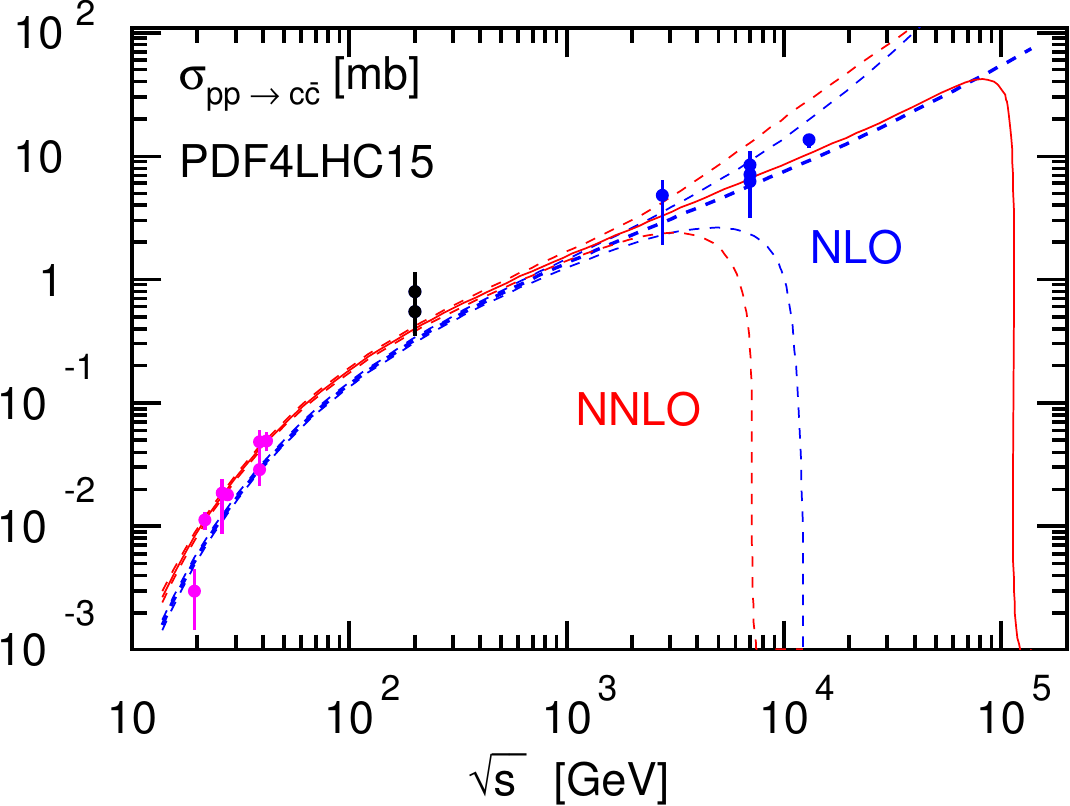}
\caption{\small 
\label{fig:sigmatot-scms-2} 
  Same as Fig.~\ref{fig:sigmatot-scms-1} 
  using the central PDF sets 
  of 
  HERAPDF2.0~\cite{Abramowicz:2015mha},
  MMHT14~\cite{Harland-Lang:2014zoa},
  NNPDF3.0~\cite{Ball:2014uwa}
  and 
  PDF4LHC15~\cite{Butterworth:2015oua}
  together with the respective PDF uncertainties.
}
\end{center}
\end{figure}

\subsubsection{Charm-quark hadro-production}

Charm-quark hadro-production offers another possibility to illustrate the consistency 
of the theory predictions for the various PDF sets.
The exclusive production of charmed mesons in the forward region at LHCb 
probe the gluon distribution down to small-$x$ values of $x \sim 5 \times 10^{-6}$ 
at $\sqrt{s}=7$~TeV, and data can be confronted with QCD predictions at NLO accuracy, 
see, e.g., \cite{Garzelli:2015psa,Gauld:2015yia}.

For the inclusive cross section of the reaction $pp$ $\rightarrow$ $c\bar{c}$ 
the QCD predictions are known up to NNLO in the \msbar scheme for the charm-quark mass and 
display good convergence of the perturbative expansion 
and stability under variation of the renormalization and factorization scales~\cite{Garzelli:2015psa}.
In Figs.~\ref{fig:sigmatot-scms-1} and \ref{fig:sigmatot-scms-2} we compare the theory predictions 
at NLO and NNLO with $m_c(m_c) = 1.27~\GeV$ in the \msbar scheme, see Sec.~\ref{sec:th4pdfs}, 
for the scale choice $\mu_r=\mu_f=2m_c(m_c)$ as a function of the center-of-mass energy $\sqrt{s}$ 
to available experimental data. 
These data span a large range in $\sqrt{s}$, 
which starts with fixed target experiments at energies up to $\sqrt{s} = 50~\GeV$
summarized in~\cite{Lourenco:2006vw} and HERA-B data~\cite{Abt:2007zg} 
(purple points in Figs.~\ref{fig:sigmatot-scms-1},~\ref{fig:sigmatot-scms-2}).
At higher energies RHIC data from PHENIX and STAR~\cite{Adare:2006hc,Adamczyk:2012af}
(black points in Figs.~\ref{fig:sigmatot-scms-1},~\ref{fig:sigmatot-scms-2})
are available and the LHC contributes measurements at energies $\sqrt{s}=2.76~\TeV$ from ALICE~\cite{Abelev:2012vra}, 
at $\sqrt{s}=7~\TeV$ from ALICE~\cite{Abelev:2012vra}, ATLAS~\cite{ATLAS:2011fea} and LHCb~\cite{Aaij:2013mga},
and at the highest available energy $\sqrt{s}=13~\TeV$ from LHCb~\cite{Aaij:2015bpa} 
(blue points in Figs.~\ref{fig:sigmatot-scms-1},~\ref{fig:sigmatot-scms-2}).
The total cross sections of LHCb have been obtained from charmed hadron production measurements 
in a limited phase space region~\cite{Aaij:2013mga,Aaij:2015bpa} 
using extrapolations based on NLO QCD predictions matched with 
parton shower Monte Carlo generators.

The theory predictions for the PDF sets ABM12, CJ15, CT14 and JR14 at NLO and NNLO are
shown in Fig.~\ref{fig:sigmatot-scms-1}, together with the respective PDF uncertainties. 
For all these PDF sets the perturbative expansion is stable, 
the theory computations agree well with the data and 
predictions, e.g., for a future collider with $\sqrt{s} \simeq 100~\TeV$,
yield positive cross sections. 
The PDF uncertainties obtained for CT14, however, do increase significantly
above energies of $\sqrt{s} \simeq 1~\TeV$.

The same information for the sets HERAPDF2.0, MMHT14, NNPDF3.0 and PDF4LHC15
is displayed in Fig.~\ref{fig:sigmatot-scms-2}.
These predictions all agree with data at low energies but 
start to behave very differently for HERAPDF2.0, MMHT14 or NNPDF3.0
at energies above $\sqrt{s} \simeq {\cal O}(10)~\TeV$ 
and for PDF4LHC15 above $\sqrt{s} \simeq {\cal O}(100)~\TeV$. 
At the same time, the associated PDF uncertainties in this region of phase space 
become very large, thereby limiting the predictive power.
Typically, the PDF uncertainties of the NNLO sets are even larger than at NLO.
In the case of MMHT14 the consistency of the NNLO predictions with LHC data 
from ALICE~\cite{Abelev:2012vra}, ATLAS~\cite{ATLAS:2011fea} and LHCb~\cite{Aaij:2013mga,Aaij:2015bpa}
at energies of $\sqrt{s}=7~\TeV$ and $13~\TeV$ deteriorates.
For NNPDF3.0 the central prediction at NNLO displays a change in slope for energies above $\sqrt{s} \simeq 3~\TeV$ 
leading to a steeply rising cross section.
The most striking feature, however, are the negative cross sections 
for HERAPDF2.0, MMHT14 and PDF4LHC15 at energies above $\sqrt{s} \simeq {\cal O}(30 - 100)~\TeV$, 
depending on the chosen set.
This is an effect of the negative gluon PDF for those sets 
at values of $x$ within the kinematic reach of current or 
future hadron colliders up to $\sqrt{s} \simeq 100~\TeV$.
This results in an instability of the perturbative expansion 
of the $\sigma_{pp \to c{\bar c}}$ cross section at high energies when the
contribution from the quark-gluon channel dominates.
The reason for a negative gluon PDF in the NNLO set of PDF4LHC15 
(being some average of the CT14, MMHT14 and NNPDF3.0 sets) is unclear. 
In contrast, other PDFs shown in Fig.~\ref{fig:sigmatot-scms-1} demonstrate stability of the perturbative expansion 
through NNLO up to very high energies and good consistency of the predictions with the experimental data.

\subsection{$W'/Z'$ production}

Cross sections sensitive to large-$x$ parton distributions typically
fall rapidly with increasing $x$ values, leading to limitations in the
quantity and precision of experimental data and the kinematic range
over which they can be obtained.
Consequently, the precision to which one can constrain large-$x$ PDFs 
decreases with $x$, and systematic uncertainties due to extrapolations
into unmeasured regions of $x$ (or those excluded by cuts) increase.
Similarly, the theoretical uncertainties due to various approximations
in the treatment of nuclear corrections for deuterium data, or target
mass and higher twist effects, also become larger.

\begin{figure}[th!]
\includegraphics[width=0.575\textwidth]{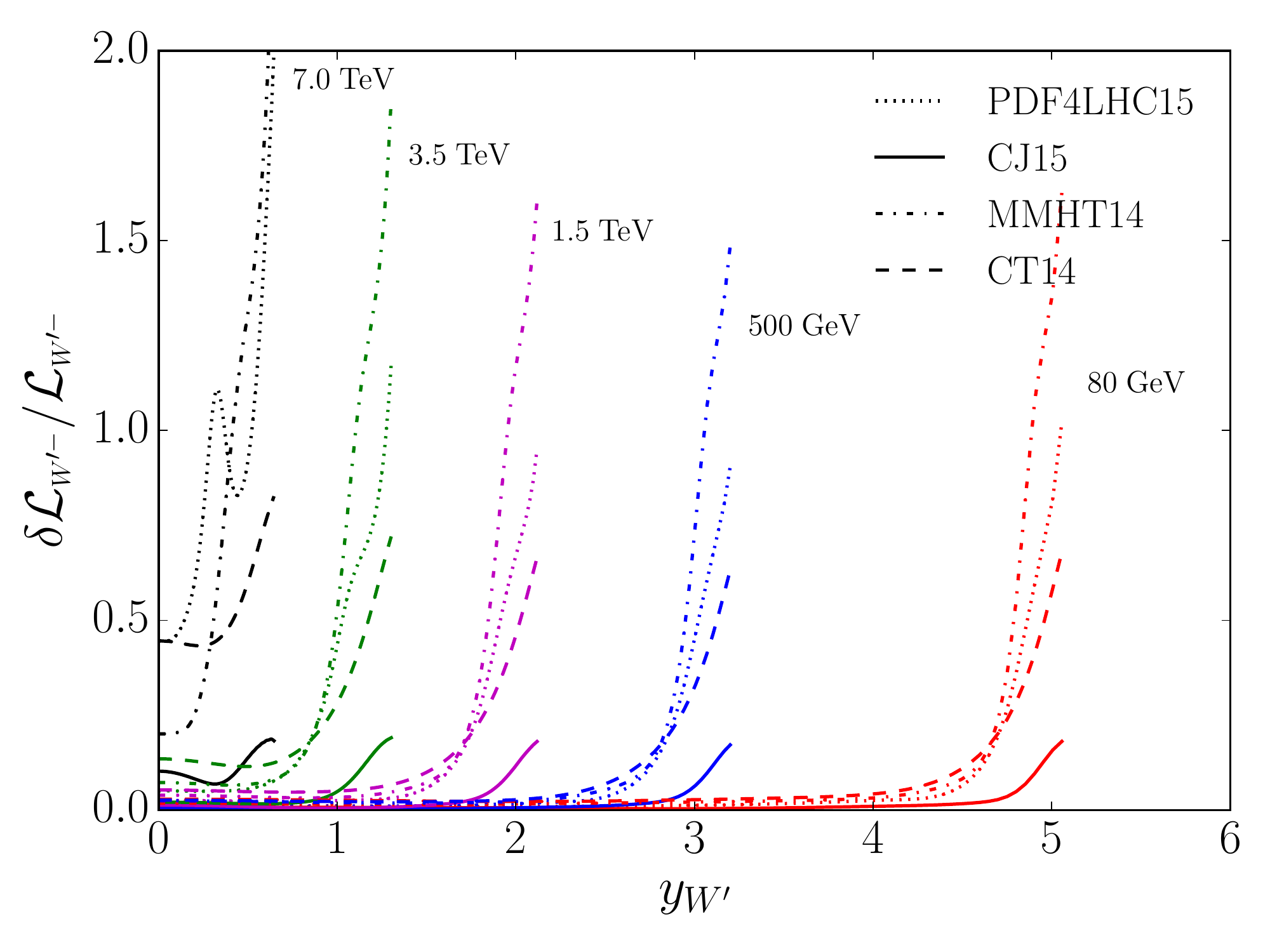}
\caption{\small 
  \label{fig:Wrap1}
  Relative uncertainty $\delta {\cal L}_{W'^-} / {\cal L}_{W'^-}$
  in the $W'^-$ luminosity as a function of rapidity $y_{W'}$ for the
  combined PDF4LHC15 set (dotted), the CJ15 (solid), MMHT14 (dot-dashed),
    and CT14 (dashed) PDFs for various $W'$ masses from 80~GeV (SM) to 7.0~TeV.  
    All PDF uncertainties have been scaled to a common 68\% c.l. as provided by the various groups.
}
\end{figure}

To illustrate this, consider the production of a heavy $W'$ boson as a
function of the $W'$ rapidity $y_{W'}$ \cite{Brady:2011hb}.  Assuming Standard
Model couplings, the parton luminosity for a produced negatively
charged $W'^-$ boson is given by
\begin{eqnarray}
{\lefteqn{ {\cal L}_{W'^-} \, = \, }} \\
&& \frac{2 \pi G_F}{3\sqrt{2}} x_1 x_2
  \bigg[ \cos^2\theta_C \big( \bar u(x_2)d(x_1) + \bar c(x_2)s(x_1) \big)
       + \sin^2\theta_C \big( \bar u(x_2)s(x_1) + \bar c(x_2)d(x_1) \big)
  \bigg]
+ (x_1 \leftrightarrow x_2)
\, ,
\nonumber
\end{eqnarray}
where $G_F$ is the Fermi constant and $\theta_C$ the Cabibbo angle.
The uncertainty $\delta {\cal L}_{W'^-}$ in the luminosity is shown
in Fig.~\ref{fig:Wrap1} for various PDF sets as a function of $y_{W'}$,
for several fixed values of the boson mass from the SM
$W$ up to $M_{W'}=7$~TeV.

Note that as the rapidity or mass of the produced boson increases,
so does the momentum fraction
  $x_{1,2} = (M_{W'}/\sqrt{s})\, e^{\pm y_{W'}}$
of one or both partons, in which case the luminosity behaves as
${\cal L}^- \sim \bar u(x_2) d(x_1)$.  Except for the highest $M_{W'}$
values, the PDF uncertainty typically remains small up to large values
of $y_{W'}$, corresponding to $x_1 \approx 0.65$, beyond which it
rises dramatically for all~$M_{W'}$.
This is precisely the region where data constraining the $d$-quark
PDF are scarce, and theoretical assumptions play an important role~\cite{Accardi:2016qay}.
This is particularly pronounced for fits that exclude DIS data at
low invariant masses, such as the three fits included in the PDF4LHC
combination \cite{Butterworth:2015oua}.  For large $W'$ masses, the $\bar u$ PDF
is evaluated at $x_2 \sim 0.2 - 0.5$, where data are either nonexistent
or have large errors, giving rise to the increased uncertainties in
some of the PDF sets at $y_{W'} \sim 0$.

 The relative uncertainties in the luminosities in
Fig.~\ref{fig:Wrap1} have been scaled to a common 68\% c.l., as in the
tables in the previous sections.  One observes a very large range of
uncertainties for the various PDF sets, which stems from different
tolerance criteria used and different methodologies employed for the
treatment of data at high values of $x$.  The smallest uncertainty
is obtained for the CJ15 PDF set, which makes use of low invariant
mass data to constrain the high-$x$ region, and does not employ
additional tolerance factors inflating the uncertainties.
The MMHT and CT14 PDF sets have larger errors, due to stronger cuts
on low-mass DIS data and larger tolerances, and consequently the
averaged PDF4LHC15 set gives similarly large uncertainties. 

   This example illustrates the problematic nature of statistically
combining PDF sets that have been determined using very different
theoretical treatments of the high-$x$ region, leading to an
overestimate of the uncertainties at these kinematics.  Using the
PDF4LHC15 set as the sole basis for background estimates, for example,
one could potentially miss signals of new physics in regions such as
at high rapidity $y_{W'}$.  A more meaningful PDF uncertainty would be
obtained when combining PDF sets obtained under similar conditions and
inputs; if large differences are found, these should be investigated
further rather than simply averaged over.

\begin{figure}[t!]
\includegraphics[width=0.575\textwidth]{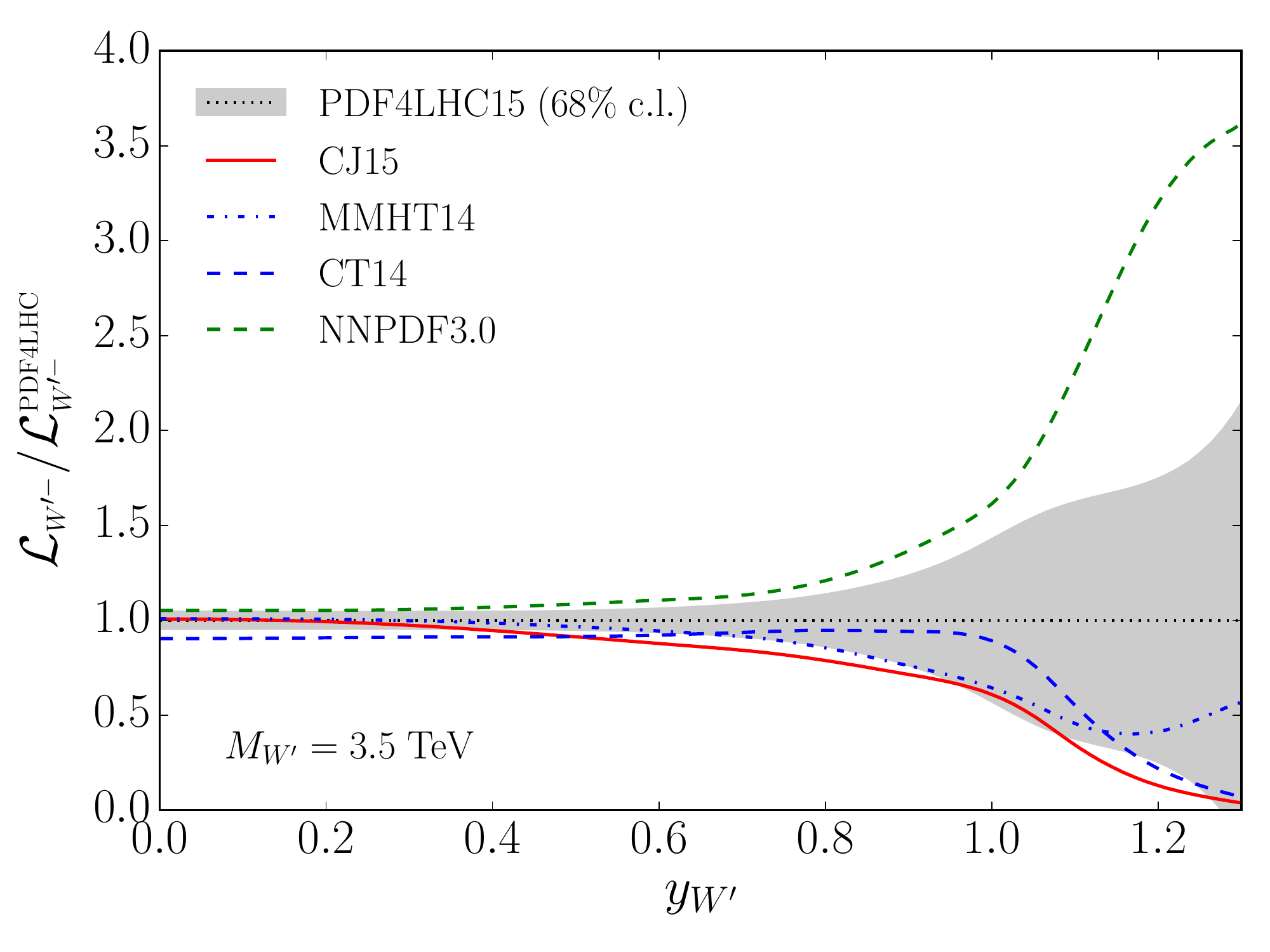}
\caption{\small 
  \label{fig:Wrap2}
  Ratio of central values of the $W'^-$ luminosity
  ${\cal L}_{W'^-}$ to the PDF4LHC value
  (dotted, 68\% c.l. shaded band)
  as a function of rapidity $y_{W'}$.
  The PDF sets CJ15 (red solid curve),
  MMHT14 (blue dot-dashed curve),
  CT14 (blue dashed curve), and
  NNPDF3.0 (green dashed curve) are compared for a
  $W'$ mass $M_{W'}=3.5$~TeV.}
\end{figure}

   This is also illustrated in Fig.~\ref{fig:Wrap2}, where the central
values for the $W'^-$ luminosity for several PDF sets are compared
relative to the luminosity computed from the central PDF4LHC15
distributions.  The different theoretical assumptions utilized in the
fits produce systematic differences in the large-$x$ PDFs, which give
rise to ratios of central values that are of the same order as the
overall PDF4LHC15 68\% c.l. uncertainty, and in the case of the
NNPDF3.0 set are about twice as large.

   The fact that the uncertainty bands of the individual sets overlap
with that of the PDF4LHC15 set is not, however, an indication that the
latter is a good estimate of the PDF uncertainties in this extrapolation
region.  Rather, the PDF4LHC15 band effectively represents a statistical
envelope of the systematic theoretical differences between the sets
included in the combination.  A comparison with the luminosity computed
using the CJ15 PDF set, which is not included in the PDF4LHC15
combination, is instructive in this respect.  The two main theoretical
assumptions affecting the $W'^-$ luminosity are the nuclear corrections
in deuterium (applied or fitted in the CJ15 and MMHT14 analyses, as well
as in JR14 and ABM12), and the parametrization of the $d$-quark PDF.

   For the latter, the traditional choice has been to assume a behavior
$\propto (1-x)^\beta$ as $x \to 1$ for both the $d$- and $u$-quark
PDFs (as, {\it e.g.}, in the MMHT14 and NNPDF3.0 analyses), in which
case the $d/u$ ratio either vanishes or becomes infinite in the
$x \to 1$ limit depending on whether the exponent $\beta$ is larger
for $d$ or $u$.  Alternatively, including an additive term in the
$d$-quark PDF proportional to $u(x)$ (as in CJ15) or constraining
$\beta_u=\beta_d$ (as in CT14) allows the $d/u$ ratio to reach
a finite, nonzero limiting value at $x \to 1$.  Furthermore,
the CJ15 distributions were also fitted to low invariant mass
(3.5~GeV$^2 < W^2 < 12.5$~GeV$^2$) DIS data, which were excluded
by kinematic cuts in the MMHT14, CT14 and NNPDF3.0 analyses.
Consequently, the following features can be observed in
Fig.~\ref{fig:Wrap2}:
\begin{itemize}
\item
The MMHT14 curve follows CJ15 closely until $y_{W'} \approx 1
(x \approx 0.65)$, after which the $d$-quark PDF turns upwards
relative to CJ15, in the region not constrained by the large-$x$
and low-$W^2$ SLAC data utilized in CJ15.
\item
The CT14 curve is lower than CJ15 at $y_{W'} \lesssim 0.6$\,
($x \lesssim 0.45$), and higher at larger $y_{W'}$, because of the
neglect of nuclear corrections.  At $y_{W'} > 1$ the $d$-quark PDF
is essentially unconstrained since neither the low-$W^2$ SLAC data
nor the reconstructed Tevatron $W$-boson production data are included
in the fit.
\item
The NNPDF3.0 fit, which excludes low-$W^2$ DIS data and does not
utilize nuclear or hadronic corrections, consistently deviates from
all others.  It is, however, compatible with those within its own
uncertainties, which at large $x$ are about four times larger than that
of the other fits.
\end{itemize}

   In summary, in extreme kinematic regions, such as at large rapidity
or for large-mass observables, caution must be exercised when utilizing
PDF error bands and nominal confidence levels provided by the various
PDF groups for precision calculations and statistically meaningful
comparisons to data.  Utilizing the PDF4LHC15 band at face value likely
overestimates the current uncertainty on large-$x$ PDFs, and could lead
to signals of new physics being missed.  Calculations performed with the
combination set should always be cross-checked with as many individual
PDF sets as possible, taking into account the amount and kind of data
included in each fit, as well as the different theoretical inputs.
The latter explore different physics issues and can vary considerably
from one PDF set to another.  When differences arise, further scrutiny
of the PDF fit results themselves may be needed before drawing any
definitive conclusions.

\section{Recommendations for PDF usage}
\label{sec:recommendations}

Recommendations for the usage of PDFs generally aim in providing guidance
for estimates of the magnitude and the uncertainties of cross sections 
in a reliable but also efficient way.
First recommendations have been provided by the 
{\it PDF4LHC Working Group} in the {\it Interim Recommendations}~\cite{Botje:2011sn}. 
There, the MSTW~\cite{Martin:2009iq} PDF was used as a central set for 
predictions at NNLO in QCD and the procedure for calculation of the PDF uncertainties,
based on an envelope of several PDF sets, was proposed.
This approach has been criticized for being impractical.
The 2015 PDF4LHC recommendations~\cite{Butterworth:2015oua} have evolved from related discussions 
and aim in improving the efficiency of cross section
computations by averaging several PDFs along with their respective uncertainties.
Here, we briefly recall these suggestions and put them into context of the findings of the previous sections.
We comment on several shortcomings of the recommendations~\cite{Butterworth:2015oua}
and propose an alternative for the PDF usage at the LHC.

\subsection{The 2015 PDF4LHC recommendations: A critical appraisal}

The 2015 PDF4LHC recommendations~\cite{Butterworth:2015oua} distinguish four cases:
(i) {\it Comparisons between data and theory for Standard Model measurements},
(ii) {\it Searches for Beyond the Standard Model phenomena},
(iii) {\it Calculation of PDF uncertainties in situations when computational speed is
  needed, or a more limited number of error PDFs may be desirable}
and (iv) {\it Calculation of PDF uncertainties in  precision observables}.

For the case (i), the recommendation is to use the individual PDF sets 
  ABM12~\cite{Alekhin:2013nda},
  CJ12~\cite{Owens:2012bv},
  CT14~\cite{Dulat:2015mca},
  JR14~\cite{Jimenez-Delgado:2014twa},
  HERAPDF2.0~\cite{Abramowicz:2015mha},
  MMHT14~\cite{Harland-Lang:2014zoa},
  and 
  NNPDF3.0~\cite{Ball:2014uwa}.
It is not clear, why the full account of the PDF dependence should be limited to SM processes only.
Deviations observed in the theory predictions obtained with the
various PDFs can often be traced back to the differences in the underlying 
theoretical asumptions and models in the PDF fits. 
With more LHC data available, tests of the compatibility of those 
data sets in the individual PDF fits will become more stringent. Studies to
quantify the constraining power of processes like hadro-production 
of $t {\bar t}$ pairs, jets or $W^\pm$ and $Z$ bosons become possible at high precision.

For the case (ii), it is recommended to employ the PDF4LHC15 sets~\cite{Butterworth:2015oua}, which 
represent the combination of the CT14~\cite{Dulat:2015mca}, MMHT14~\cite{Harland-Lang:2014zoa},
and NNPDF3.0~\cite{Ball:2014uwa}. The combination is performed using the Monte Carlo approach at different levels of precision, 
leading to the recommended sets {\tt PDF4LHC15\_30} and {\tt PDF4LHC15\_100}.
The restriction to CT14, MMHT14 and NNPDF3.0 implies a bias both for the central value 
and for the PDF uncertainties of BSM cross section predictions.
For example, a bias is introduced by fixing the central value of $\alpha_s(M_Z)$ 
to an agreed common value, currently chosen to be $\alpha_s(M_Z) = 0.118$ at both NLO and NNLO. 
This choice is in contradiction with the precision determinations of
$\alpha_s(M_Z)$ at different orders in perturbation theory, as summarized in Sec.~\ref{sec:alphas}.
Further, for searches at the highest energies, the PDFs are probed 
close to the hadronic threshold near $x \simeq 1$,  where nuclear corrections and other
hadronic effects, considered for instance in the CJ15~\cite{Accardi:2016qay} and
JR14~\cite{Jimenez-Delgado:2014twa} analyses, are important.

For the case (iii), the {\tt PDF4LHC15\_30} sets are recommended to use.
We would like to note, that here the balance between the computational speed and the precision of the 
result (in e.g. MC simulation) has to be determined by the analysers.
The problem rises from the
large deviations between data and theory predictions at low scales and also 
at the edges of the kinematical ranges of data currently used in PDF fits 
as illustrated in Secs.~\ref{sec:th4pdfs} and~\ref{sec:crs-at-lhc}. 
The average of various GM-VFNS for heavy quark production, such as
ACOT~\cite{Aivazis:1993pi}, 
FONLL~\cite{Forte:2010ta} and RT~\cite{Thorne:2012az}, 
leaves a large degree of arbitrariness in the theory predictions, cf.~Fig.~\ref{fig:hera-ccbar-dis-5}.
Note that the {\tt PDF4LHC15\_30} sets  were updated in December 2015~\cite{Rojo:private} 
to account for an extension of their validity range below the original $Q > 8~\GeV$ as 
only discussed in the later publication~\cite{Badger:2016bpw}.

For the case (iv), the set {\tt PDF4LHC15\_100} is recommended.
Recalling that this case concerns measurements of the precision observables,
it is unclear why PDFs  should be treated differently than in the case (i).
The differences
between individual PDF sets propagate  the cross section measurements
directly through the acceptance corrections or extrapolation factors, as illustrated in
Figs.~\ref{fig:difftop},~\ref{fig:sigmatot-scms-2} and~\ref{fig:Wrap2}.
Using of the {\tt PDF4LHC15\_100} is worrysome, since these differences are smeared 
out in the combination, which, in addition, is limited to only three PDF sets. 
The SM parameters, determined using the precision observables obtained in this way,
may be biased.

In summary, the recent PDF4LHC recommendations~\cite{Butterworth:2015oua}
cannot be viewed as definitive in the case of precision theory predictions, 
as the advocated averaging procedure 
introduces bias, artificially inflates the uncertainties, 
and makes it difficult to quantify potential discrepancies between the individual PDF sets.

\subsection{New recommendations for the PDF usage at the LHC}

Based on the considerations above, we propose modifications to the recommendations for PDF usage at the LHC 
in order to retain the predictive capability of the individual PDF sets. Two cases can be distinguished:

\begin{enumerate}
\item {\bf Precise theory predictions}, addressing a class of predictions, within or beyond the SM,
which encompasses any type of cross section prediction including radiative corrections of any kind,
whether at fixed-order or via resummation to some logarithmic accuracy. This class also includes the MC 
simulations used for the calculation of the acceptance corrections for precision observables, e.g. 
cross sections which might be used further for determination of SM parameters.
\begin{itemize}
\item {\it Recommendation:} 
Use the individual recent PDF sets, currently
ABM12~\cite{Alekhin:2013nda},
CJ15~\cite{Accardi:2016qay}, 
CT14~\cite{Dulat:2015mca},
JR14~\cite{Jimenez-Delgado:2014twa},
HERAPDF2.0~\cite{Abramowicz:2015mha},
MMHT14~\cite{Harland-Lang:2014zoa},
and 
NNPDF3.0~\cite{Ball:2014uwa}
(or as many as possible), together with the respective uncertainties for the chosen PDF set, 
the strong coupling $\alpha_s(M_Z)$ and the heavy quark masses $m_c$, $m_b$ and $m_t$. Once a PDF set 
is updated, the most recent version should be used.

\item {\it Rationale:} Precise theory predictions as needed for any comparisons between theory and data
for processes in the SM or beyond (such as hadro-production of jets, $W^\pm$- or $Z$-boson production,
either singly or in pairs, heavy-quark hadro-production, or generally the production of new massive
particles at the TeV scale) often depend on details of the PDF fits and the underlying theory assumptions 
and schemes used. Differences in the theory predictions based on the individual sets 
can give an indication of residual systematic
uncertainties or shed light on drawbacks and need for potential 
improvements in the physics models used in the extraction of those PDFs.
This applies in particular to measurements used for the determination of SM parameters 
such as the strong coupling $\alpha_s(M_Z)$, heavy quark masses $m_c$, $m_b$ and $m_t$
or the $W$-boson mass, because these parameters are directly 
correlated to the PDFs used in their extraction from the experimental observables.
\end{itemize}

\item {\bf Theory predictions for feasibility studies}, 
    the complementary class containing all other cross section predictions
where high precision is not required, such as those based on Born approximations and/or order of magnitude 
estimates, or in cases where precision may be sacrificed in favor of computational speed. Here, also studies 
of novel accelerators and detectors are addressed.
\begin{itemize}
\item {\it Recommendation}: 
    Use any of the recent PDF sets (listed in {\tt LHAPDFv6} or later versions).

\item {\it Rationale}: 
Often in phenomenological applications for the modern and future facilities
one is interested in a quick order of magnitude estimate for the particular cross sections. 
These are directly proportional to the parton luminosity and to the value of $\alpha_s(M_Z)$.
In these cases, one may be willing to sacrifice precision in favor of 
computational speed. Here, the usage of the sets {\tt PDF4LHC15\_30} and {\tt PDF4LHC15\_100} 
may provide an efficient estimate of PDF uncertainties, although care 
must be taken in their interpretation depending on the observable and covered kinematic range. 
Restricting the recommendation to PDFs listed in the {\tt LHAPDF(v6)}~\cite{Buckley:2014ana}
interface excludes parton luminosities with lesser precision in the interpolation of the underlying grids 
(e.g., in {\tt LHAPDF(v5)}~\cite{Whalley:2005nh}) or ``partonometers''~\cite{Ralston:1986hr} with outdated calibration.

\end{itemize}
\end{enumerate}

In the Monte Carlo generators, for example,
{\tt MadGraph5\_aMC@NLO}~\cite{Alwall:2014hca}, {\tt POWHEG-BOX (v2)}~\cite{POWHEGBOX,Alioli:2010xd} 
and {\tt SHERPA (v2)}~\cite{SHERPA,Gleisberg:2008ta}, or other recently developed generators, 
like {\tt Geneva}~\cite{Alioli:2015toa}, 
different PDF sets can be efficiently studied with reweighting methods.
This allows to generate weighted events for a given setup, and to
reweight a-posteriori each event in a fast and efficient way, by generating new
weights associated with different choices of renormalization and factorization
scales and/or PDFs. 
Please note, that at present, PDF reweighting is performed by
assuming the linear PDF weight dependence, which is not correct, since
PDFs are also present in the Sudakov form-factor.
Efforts to extend the reweighting to the entire Sudakov form-factor and to the full parton shower are ongoing.
The reweighting technique turns out to be particularly useful to compute in a
fast (although at the moment approximate) way PDF uncertainties affecting the
predictions.

\section{Conclusion}
\label{sec:conclusion}

In this report we have reviewed recent developments in the determination of PDFs in global QCD analyses.
Thanks to high precision experimental measurements and continuous theoretical improvements,
the parton content of the proton is generally well constrained and 
PDFs, along with the strong coupling constant
$\alpha_s(M_Z)$ and the heavy-quark masses $m_c$, $m_b$ and $m_t$, have been
determined with good accuracy, at least at NNLO in QCD.
This forms the foundation for precise cross section predictions at the
LHC in Run 2.

We have briefly discussed the available data used in PDF extractions and the kinematic range covered, 
and emphasized the importance of selecting mutually consistent sets of data in PDF fits 
in order to achieve acceptable $\chi^2$ values for the goodness-of-fit estimate.
The main thrust of the study has been the computation of benchmark cross
sections for a variety of processes at hadron colliders, including Higgs boson
production in gluon-gluon fusion.
We have illustrated how different choices for the theoretical description of
the hard scattering process and choices of parameters have an impact on the predicted cross sections,
and lead to systematic shifts that are often significantly larger than the 
associated PDF and $\alpha_s(M_Z)$ uncertainties.
A particular example has been the treatment of heavy quarks in DIS, where 
the quality of the various scheme choices has been quantified 
in terms of $\chi^2$/NDP values when comparing predicted cross sections to data.
We have also pointed out the inconsistently low values for the pole mass of the charm quark
used in some fits, and have stressed the correlation 
of the strong coupling constant $\alpha_s(M_Z)$ with the PDF parameters.
Ideally, $\alpha_s(M_Z)$ should be determined simultaneously with the PDFs,
and we have summarized here the state of the art in the context of PDF analyses.

Our findings expose a number of shortcomings in the recent PDF4LHC recommendations~\cite{Butterworth:2015oua}.
We have shown that these do not provide sufficient control over some theoretical uncertainties, 
and may therefore be problematic for precision predictions in Run 2 of the LHC.
Instead, we suggest new recommendations for the usage of PDFs based on a theoretically consistent procedure 
necessary to meet the precision requirements of the LHC era.

\subsection*{Acknowledgments}
We would like to thank S.~Alioli, M.~Botje, E.W.N. Glover and K.~Rabbertz for discussions, 
K.~Rabbertz also for valuable comments on the manuscript, 
and L.~Harland-Lang and R.~Thorne for providing us with the Higgs 
and $t{\bar t}$ cross sections in Tabs.~\ref{tab:higgs-mc-mmht} and \ref{tab:ttbar-mc-mmht}.

\smallskip

This work has been supported 
by Bundesministerium f\"ur Bildung und Forschung through contract (05H15GUCC1), 
by the DOE contract No.~DE-AC05-06OR23177, under which Jefferson Science Associates, LLC operates Jefferson Lab,
and by the European Commission through PITN-GA-2012-316704 ({\it HIGGSTOOLS}). 
The work of A.A. and J.F.O. was supported in part by DOE contracts No.~DE-SC0008791 and No.~DE-FG02-97ER41922,
respectively.
Two of the authors (J.B. and S.M.) would like to thank 
the Mainz Institute for Theoretical Physics (MITP) for its hospitality and support.


\providecommand{\href}[2]{#2}\begingroup\raggedright

\endgroup

\end{document}

%% file: table-data.tex
\setcounter{footnote}{0}
\renewcommand{\thefootnote}{b\arabic{footnote}}
\begin{table}[t!]
\begin{center}
\renewcommand{\arraystretch}{1.3}
\begin{tabular}{|l|c|l|}
\hline
PDF sets
  & $\Delta \chi^2$ criterion
  & data sets used in analysis
\\[1.5ex]
\hline
ABM12~\cite{Alekhin:2013nda}
  & 1
  & incl. DIS, DIS charm, DY
\\[0.5ex]
\hline
CJ15~\cite{Accardi:2016qay} 
\footnote{CJ15 use $\Delta\chi^2 = 1$ (for the 68\% c.l.) and   
the CJ15 PDF sets are provided with 90\% c.l. uncertainties ($\Delta\chi^2 = 2.71$).}
  & 1
  & incl. DIS, DY (incl. $p\bar{p}\to W^\pm X$), $p\bar{p}$ jets, $\gamma$+jet
\\[0.5ex]
\hline
CT14~\cite{Dulat:2015mca}\footnote{
The CJ14 PDFs sets are provided with 90\% c.l. uncertainties.
In addition, a two-tier tolerance test has 
been applied in case of some data sets.} 
  & 100
  & incl. DIS, DIS charm, DY,  $p\bar{p}$ jets, $pp$ jets 
\\[0.5ex]
\hline
HERAPDF2.0~\cite{Abramowicz:2015mha} 
  & 1
  & incl. DIS, DIS charm, DIS jets [only HERA data]
\\[0.5ex]
\hline
JR14~\cite{Jimenez-Delgado:2014twa} 
  & 1
  & incl. DIS, DIS charm, DY, $p\bar{p}$ jets, DIS jets 
\\[0.5ex]
\hline
MMHT14~\cite{Harland-Lang:2014zoa}
  & \multirow{2}{5em}{2.3 \dots 42.3 (dynamical)}
  & incl. DIS, DIS charm, DY, $p\bar{p}$  jets, $pp$ jets, $t\bar{t}$  
\\[3.5ex]
\hline
NNPDF3.0~\cite{Ball:2014uwa}\footnote{
A Monte Carlo method is used to estimate the errors of the PDFs. 
This method has an interpretation with respect to a level of 
tolerance only in the range in which the corresponding uncertainties are Gaussian, which 
applies to wide kinematic regions studied. In these regions the error bands correspond
to the $1~\sigma$ error obtained using the $\chi^2$ method~\cite{Forte:2016}.}
  & n.a.
  & incl. DIS, DIS charm, DY, $p\bar{p}$ jets, $pp$ jets, $t\bar{t}$, $W+\mbox{charm}$  
\\[0.5ex]
\hline
\end{tabular}
  \caption{\small 
  \label{tab:chi2}
    Summary of major hard processes used in the various PDF analyses and 
    the confidence level criteria employed. Detailed references to the  
    different specific data sets used by the various groups 
    are given 
    in Refs.~\cite{Alekhin:2013nda,Accardi:2016qay,Dulat:2015mca,Abramowicz:2015mha,Jimenez-Delgado:2014twa,Harland-Lang:2014zoa,Ball:2014uwa}
    and also the specific statistical analysis applied is 
    described in these papers.
    Note that different 
    analyses use partly different data sets for some processes.
  }
\end{center}
\end{table}

%% file: table-WandZ-data.tex
\begin{sidewaystable}[H!]
\renewcommand{\arraystretch}{1.3}
\fontsize{8.5}{9.5}\selectfont
\begin{center}
\begin{tabular}{|c|c|c|c|c|c|c|c|c|c|c|c|c|}   
\hline                           
\multicolumn{2}{|c|}{Experiment}                      
&\multicolumn{2}{c|}{ATLAS}  
&\multicolumn{4}{c|}{CMS}  

&\multicolumn{2}{c|}{D\O}
&\multicolumn{3}{c|}{LHCb}
\\
\hline                                                    
\multicolumn{2}{|c|}{$\sqrt s$~(TeV)}                      
&\multicolumn{2}{c|}{7}
&\multicolumn{2}{c|}{7}
&\multicolumn{2}{c|}{8}

&\multicolumn{2}{c|}{1.96}
&7
&\multicolumn{2}{c|}{8}
\\                                                        
\hline
\multicolumn{2}{|c|}{Final states} 
& $W^+\rightarrow l^+\nu$
& $Z\rightarrow e^+e^-$
& $W^+\rightarrow \mu^+\nu$
& $Z\rightarrow l^+l^-$

& $W^+\rightarrow \mu^+\nu$
& $Z\rightarrow l^+l^-$
&$W^+\rightarrow \mu^+\nu$
& $W^+\rightarrow e^+\nu$
&$W^+\rightarrow \mu^+\nu$
& $Z\rightarrow e^+e^-$                                                        
&$W^+\rightarrow \mu^+\nu$
\\
\multicolumn{2}{|c|}{ }                                          
& $W^-\rightarrow l^-\nu$
& $\gamma^*\rightarrow e^+e^-$
&$W^-\rightarrow \mu^-\nu$
& $\gamma^*\rightarrow l^+l^-$
&$W^-\rightarrow \mu^-\nu$
& $\gamma^*\rightarrow l^+l^-$
&$W^-\rightarrow \mu^-\nu$
&$W^-\rightarrow e^-\nu$
&$W^-\rightarrow \mu^-\nu$
&                                                         
&$W^-\rightarrow \mu^-\nu$
\\                                                        
\multicolumn{2}{|c|}{ }                                          
& $Z\rightarrow l^+l^-$
&
&
& 
&                                                         
&
&
&                                                         
& $Z\rightarrow \mu^+\mu^-$
&                                                         
& $Z\rightarrow \mu^+\mu^-$
\\
\hline                                                    
\multicolumn{2}{|c|}{Cut on the lepton $P_T$ }                      
&$P_T^l>20~{\rm GeV}$      
&$P_T^e>25~{\rm GeV}$      
&$P_T^{\mu}>25~{\rm GeV}$   
&--
&$P_T^{\mu}>25~{\rm GeV}$   
&--
&$P_T^{\mu}>25~{\rm GeV}$
&$P_T^{e}>25~{\rm GeV}$
&$P_T^{\mu}>20~{\rm GeV}$
&$P_T^{e}>20~{\rm GeV}$                       
&$P_T^{\mu}>20~{\rm GeV}$                       
\\                                                        
\hline                                                    
\multicolumn{2}{|c|}{Luminosity (1/fb)}                      
&0.035
&4.9    
&4.7
&4.5(4.8)
&18.8
&19.7
&7.3
&9.7
&1
&2                        
&2.9
\\
\hline                           
\multicolumn{2}{|c|}{Reference}                      
&\cite{Aad:2011dm}                         
&\cite{Aad:2013iua}
&\cite{Chatrchyan:2013mza}
&\cite{Chatrchyan:2013tia}
&\cite{Khachatryan:2016pev}
&\cite{CMS:2014jea}
&\cite{Abazov:2013rja}
&\cite{D0:2014kma}
&\cite{Aaij:2015gna}
&\cite{Aaij:2015vua}
&\cite{Aaij:2015zlq}
\\                                                        
\hline                                                    
\multicolumn{2}{|c|}{$NDP$}
&30
&13                      
&11  
&132
&22
&132
&10
&13
&31
&17                            
&34                            
\\                                                        
\hline
\multirow{2}{4em}{ }
& ABMP15~\cite{Alekhin:2015cza}
    \tablefootnote{
      The ABM12~\cite{Alekhin:2013nda} analysis has used 
      older data sets from CMS and LHCb
      listed in Table~\ref{tab:dydata2}.}
&29.8 
&--
 &22.5
&--
 &--
&--
 &16.9
 &18.0
&44.1
&18.2
 &--
\\
\cline{2-13}
 &CJ15~\cite{Accardi:2016qay}
 &--
 &--
&--
&--
 &--
&--
 &20
&29
 &--
&--
 &--
\\
\cline{2-13}
 &CT14~\cite{Dulat:2015mca}
&42
 &--
 &12.1~\footnote{ 
Statistically less significant data with the cut of 
$P_T^{\mu}>35~{\rm GeV}$ are used.}
&--
&--
&-- 
&--
&34.7
 &--
&--
 &--
\\
\cline{2-13}
$\chi^2$ &JR14~\cite{Jimenez-Delgado:2014twa} 
 &--
 &--
 &--
&--
&--
&--
 &--
&--
 &--
&--
 &--
\\
\cline{2-13}
&{\tt HERAFitter}~\cite{Camarda:2015zba}
 &--
 &--
 &--
&--
&--
&--
 &13
&19
 &--
&--
 &--
\\
\cline{2-13}
&MMHT14~\cite{Harland-Lang:2014zoa}
 &39
 &17
 &--
&149
&--
&--
 &21
&--
 &--
&--
 &--
\\
\cline{2-13}
&NNPDF3.0~\cite{Ball:2014uwa}
 &35.4
 &7.3\tablefootnote{
The value obtained for the sample of 5 data points.} 
 &18.9
&149.6\tablefootnote{
The value obtained for the sample of 110 data points.}
 &--
&--
 &--
&--
 &--
&--
 &--
\\
\hline                                          
\end{tabular}
\caption{\small 
\label{tab:dydata}
Compilation of precise data on $W$- and $Z$-boson production in $pp$ and $p\bar{p}$ collisions and the 
$\chi^2$ values per number of data points obtained for these data sets
in different PDF analyses 
  using their individual definitions of $\chi^2$.
  The NNLO fit results are quoted as a default, while the NLO values are 
  given for the CJ15~\cite{Accardi:2016qay} and 
  {\tt HERAFitter}~\cite{Camarda:2015zba} PDFs. 
Missing table entries indicate that the respective data sets has not been used in the analysis.
Other data sets of lower accuracy, which have become obsolete and data sets superseded are listed in Table~\ref{tab:dydata2}. 
}
\end{center}
\end{sidewaystable}

\begin{sidewaystable}[H!]
\renewcommand{\arraystretch}{1.3}
\fontsize{8.5}{9.5}\selectfont
\begin{center}
\begin{tabular}{|c|c|c|c|c|c|c|c|c|c|c|c|c|c|c|}   
\hline                           
\multicolumn{2}{|c|}{Experiment}                      
&\multicolumn{4}{c|}{CDF}
&\multicolumn{3}{c|}{CMS}
&\multicolumn{4}{c|}{D\O}
&\multicolumn{2}{c|}{LHCb}
\\
\hline                                                    
\multicolumn{2}{|c|}{$\sqrt s$~(TeV)}                      
&1.8  
&\multicolumn{3}{c|}{1.96}
&\multicolumn{3}{c|}{7}
&\multicolumn{4}{c|}{1.96}
&\multicolumn{2}{c|}{7}
\\                                                        
\hline
\multicolumn{2}{|c|}{Final states} 
& $W^+\rightarrow l^+\nu$
& $W^+\rightarrow e^+\nu$
& $Z$                                                        
& $W^+$

&$W^+\rightarrow l^+\nu$
& $Z\rightarrow e^+e^-$                                                        
&$W^+\rightarrow e^+\nu$

&$W^+\rightarrow \mu^+\nu$
&$Z$ 
&$W^+\rightarrow e^+\nu$
& $W^+$

&$W^+\rightarrow \mu^+\nu$
& $Z\rightarrow e^+e^-$                                                        
\\
\multicolumn{2}{|c|}{ }                                          
& $W^-\rightarrow l^-\nu$
&$W^-\rightarrow e^-\nu$
&
&$W^-$

&$W^-\rightarrow l^-\nu$
&
&$W^-\rightarrow e^-\nu$

&$W^-\rightarrow \mu^-\nu$
&
&$W^-\rightarrow e^-\nu$
&$W^-$

&$W^-\rightarrow \mu^-\nu$
&                                                         
\\                                                        
\multicolumn{2}{|c|}{ }                                          
&
&
&
& 
&
&
&
& 
&                                                         
&
&
& 
&                                                         
\\
\hline                                                    
\multicolumn{2}{|c|}{Cut on the lepton $P_T$ }                      
&--
&$P_T^{e}>25~{\rm GeV}$   
&--
&--

&$P_T^{l}>25~{\rm GeV}$
&--
&$P_T^{e}>35~{\rm GeV}$

&$P_T^{\mu}>20~{\rm GeV}$
&--
&$P_T^{e}>25~{\rm GeV}$   
&--

&$P_T^{\mu}>20~{\rm GeV}$   
&$P_T^{e}>20~{\rm GeV}$                       
\\                                                        
\hline                                                    
\multicolumn{2}{|c|}{Luminosity (1/fb)}                      
&0.11  
&0.17
&2.1
&1.
&0.036
&0.036
&0.84
&0.3
&0.4
&0.75
&9.7
&0.037                        
&0.94
\\\hline                           
\multicolumn{2}{|c|}{Reference}                      
&\cite{Abe:1998rv}
&\cite{Acosta:2005ud}
&\cite{Aaltonen:2010zza}
&\cite{Aaltonen:2009ta}

&\cite{Chatrchyan:2011jz}
&\cite{Chatrchyan:2011wt}
&\cite{Chatrchyan:2012xt}

&\cite{Abazov:2007pm}
&\cite{Abazov:2007jy}
&\cite{Abazov:2008qv}
&\cite{Abazov:2013dsa}

&\cite{Aaij:2012vn}
&\cite{Aaij:2012mda}
\\                                                        
\hline                                                    
\multicolumn{2}{|c|}{$NDP$}
&11  
&11
&28
&13

&12
&35
&11

&10
&28
&12
&14

&10 
&9
\\                                                        
\hline
\multirow{2}{4em}{ }
& ABMP15~\cite{Alekhin:2015cza}
&--
 &--
 &--
 &--
&--
 &--
 &--
 &--
&--
 &--
 &--
 &--
&--
\\
\cline{2-15}
 &CJ15~\cite{Accardi:2016qay}
 &12
&--
 &27
 &--
&--
&--
 &--
 &--
&16
 &--
 &14
&--
 &--
\\
\cline{2-15}
 &CT14~\cite{Dulat:2015mca}
    \tablefootnote{
 The CT14 fit includes also
the ATLAS~\cite{Aad:2011dm} and LHCb~\cite{Aaij:2012vn} data on the 
charge-lepton asymmetry not taken into account in the values of 
$NDP$ and $\chi^2$ listed.} 
 &8.9
&14
&48 
&--

&--
 &--
&10.1
&8.3\tablefootnote{
The value obtained for the sample of 9 data points.} 
&17
&--
&--

&9.9
&--
\\
\cline{2-15}
$\chi^2$ &JR14~\cite{Jimenez-Delgado:2014twa} 
&--
&--
&--
 &--
&--
&--
&--
 &--
&--
&--
&--
 &--
&--
\\
\cline{2-15}
&{\tt HERAFitter}~\cite{Camarda:2015zba}
\tablefootnote{
The values of $\chi^2$ for the charge $W$-asymmetry 
Tevatron data~\cite{Aaltonen:2010zza,Abazov:2013dsa} are obtained in the 
variant of fit with no D\O~data~\cite{D0:2014kma}
 on the electron charge asymmetry included.}
 &--
&--
 &32
 &15
&--
 &--
&--
 &--
&23
 &--
&16
 &--
&--
\\
\cline{2-15}
&MMHT14~\cite{Harland-Lang:2014zoa}
 &--
&--
 &40
 &30
&10
    \tablefootnote{
 A combination of two data samples with
 $P_T^{l}>25~{\rm GeV}$ and $P_T^{l}>30~{\rm GeV}$ was used in the fit; 
the value of $\chi^2$ correspond to the total $NDP=24$.}  
&22
&9
 &--
&16
 &27
&--
&16
 &20
\\
\cline{2-15}
&NNPDF3.0~\cite{Ball:2014uwa}
 &--
 &--
 &44.4
 &--
&--
 &--
&8.0
 &--
&17.1
 &--
&--
 &7.2
&14.3
\\
\hline                                          
\end{tabular}
\caption{\small 
\label{tab:dydata2}
Same as Tab.~\ref{tab:dydata} for the Tevatron and LHC data sets of lower
accuracy and those, which have become superseded but are still used in various
PDF analyses. 
}
\end{center}
\end{sidewaystable}

%% file: table-mc.tex
\begin{sidewaystable}[H!]
\begin{center}
%
\renewcommand{\arraystretch}{1.3}
\begin{tabular}{|l|c|c|c|c|c|c|c|c|}
\hline
PDF sets
  & $m_c$ [GeV] 
  & \multirow{2}{5em}{$m_c$ renorm. scheme}
  & \multirow{2}{6.5em}{theory method \\ ($F_2^c$ scheme)}
  & \multirow{2}{7.25em}{theory accuracy \\ for heavy quark \\ DIS Wilson coeff.} 
  & \multicolumn{2}{c|}{
  \multirow{2}{10.5em}{$\chi^2$/NDP for \\ HERA charm data \cite{Abramowicz:1900rp}
    \\
    with \\ {\tt xFitter}~\cite{Alekhin:2014irh,HERAFitter}}
  }
  & 
  \multicolumn{2}{c|}{
  \multirow{2}{10.5em}{$\chi^2$/NDP for 
   \\ inclusive HERA data \cite{Abramowicz:2015mha} ($Q^2_{\rm min}= 3.5$ GeV$^2$)
   \\ with {\tt xFitter}~\cite{Alekhin:2014irh,HERAFitter}}
  }
  \\ [8.0ex]
  &
  & 
  & 
  & 
  & (with unc.) 
  & (nominal)
  & (with unc.) 
  & (nominal)
\\[0.5ex]
\hline
ABM12~\cite{Alekhin:2013nda}
    \tablefootnote{The value of $m_c$ in ABM12 is determined from a fit to
      HERA data \cite{Abramowicz:1900rp}. 
        ABM12 uses the approximate heavy-quark Wilson coefficient functions
        of Ref.~\cite{Kawamura:2012cr}.
}
  & $1.24~^{+~0.05}_{-~0.03}$
  & \msbar\, $m_c(m_c)$ 
  & FFNS $(n_f=3)$ 
  & ${\cal O}(\alpha_s^3)_{\rm approx}$
  & \,\,\,\,\,\, 
    65/52 
    \,\,\,\,\,\,
  & 66/52
  & \,\,\,\,
    1450/1145
    \,\,\,\,
  & \,\,\,\,
    1478/1145 
    \,\,\,\,
\\[0.5ex]
\hline
CJ15~\cite{Accardi:2016qay}
  & 1.3
  & $m_c^{\rm pole}$ 
  & SACOT~\cite{Kramer:2000hn}
  & ${\cal O}(\alpha_s^2)$
  & 117/52
  & 117/52
  & 1458/1145 
  & 1465/1145
\\[0.5ex]
\hline
CT14~\cite{Dulat:2015mca}
    \tablefootnote{
    The data comparision always applies the SACOT($\chi$) scheme at NLO as
    implemented in {\tt xFitter}~\cite{Alekhin:2014irh,HERAFitter}. 
    The implementation of this scheme differs from the one used by CT14.
    Removing the $Q^2$-cut on the HERA data~\cite{Abramowicz:1900rp} 
    one obtains $\chi^2$/NDP = 158/52 (582/52) with PDF uncertainities and
    258/52 (648/52) for the central fit at NLO (NNLO).}
    & & & & & & & &\\[-0.5ex]
$\qquad$ NLO
  & $1.3$
  & $m_c^{\rm pole}$ 
  & SACOT($\chi$)~\cite{Tung:2001mv}
  & ${\cal O}(\alpha_s^2)$
  & 51/47 
  & 70/47
  & 1397/1145
  & 1455/1145
\\
$\qquad$ NNLO
  & $1.3$
  & $m_c^{\rm pole}$ 
  & SACOT($\chi$)~\cite{Tung:2001mv}
  & ${\cal O}(\alpha_s^2)$
  & 64/47 
  & 130/47
  & 1445/1145
  & 1685/1145
\\[0.5ex]
\hline
HERAPDF2.0~\cite{Abramowicz:2015mha}
& & & & & & & & \\[-0.5ex]
$\qquad$ NLO
  & $1.47$
  & $m_c^{\rm pole}$ 
  & RT optimal~\cite{Thorne:2012az}
  & ${\cal O}(\alpha_s^2)$
  & 67/52
  & 67/52
  & 1340/1145
  & 1355/1145
\\
$\qquad$ NNLO
  & $1.43$
  & $m_c^{\rm pole}$ 
  & RT optimal~\cite{Thorne:2012az}
  & ${\cal O}(\alpha_s^2)$
  & 62/52
  & 62/52
  & 1346/1145
  & 1361/1145
\\[0.5ex]
\hline
\end{tabular}
  \caption{\small 
  \label{tab:mcmass}
    Values of the charm quark mass and renormalization scheme used in the PDF
    fits together with a summary of schemes chosen for the description of the 
    charm-quark structure function $F_2^c$ and  
    the theoretical accuracy for the massive quark DIS Wilson coefficients. 
    The values of $\chi^2$/NDP 
    for the DIS charm production cross section 
    data~\cite{Abramowicz:1900rp} 
    and HERA inclusive cross section data~\cite{Abramowicz:2015mha}
    are given in two columns with the account of PDF uncertainties (with unc.,
    where CT14 PDF errors scaled from 90\% c.l. to 68\% c.l., i.e., reduced by a
    factor 1.645) and for the central prediction of each PDF set (nominal).
      In xFitter~\cite{Alekhin:2014irh,HERAFitter}, 
      the values of electroweak parameters like the Fermi constant 
      and $W$-boson mass are taken from Ref.~\cite{Agashe:2014kda}.
    The values for CT14 and for PDF4LHC with the SACOT($\chi$) scheme 
    have been determined with a cut on $Q^2 \ge 5 \GeV^2$ on the HERA
    data~\cite{Abramowicz:1900rp}.
  }
\end{center}
\end{sidewaystable}

\begin{sidewaystable}[H!]
\begin{center}
%
\renewcommand{\arraystretch}{1.3}
\begin{tabular}{|l|c|c|c|c|c|c|c|c|}
\hline
PDF sets
  & $m_c$ [GeV] 
  & \multirow{2}{5em}{$m_c$ renorm. scheme}
  & \multirow{2}{6.5em}{theory method \\ ($F_2^c$ scheme)}
  & \multirow{2}{7.25em}{theory accuracy \\ for heavy quark \\ DIS Wilson coeff.} 
  & \multicolumn{2}{c|}{
  \multirow{2}{10.5em}{$\chi^2$/NDP for \\ HERA charm data \cite{Abramowicz:1900rp}
    \\
    with \\ {\tt xFitter}~\cite{Alekhin:2014irh,HERAFitter}}
  }
  & 
  \multicolumn{2}{c|}{
  \multirow{2}{10.5em}{$\chi^2$/NDP for 
   \\ inclusive HERA data \cite{Abramowicz:2015mha} ($Q^2_{\rm min}= 3.5$ GeV$^2$)
   \\ with  {\tt xFitter}~\cite{Alekhin:2014irh,HERAFitter}}
  }
  \\ [8.0ex]
  &
  & 
  & 
  & 
  & (with unc.) 
  & (nominal)
  & (with unc.) 
  & (nominal)
\\[0.5ex]
\hline
JR14~\cite{Jimenez-Delgado:2014twa} 
    \tablefootnote{
      The $\chi^2$/NDP values are determined for the dynamical set {\tt JR14NNLO08FF}.
      JR14 uses the approximate heavy-quark Wilson coefficient functions
      of Ref.~\cite{Kawamura:2012cr}.
    }
  & $1.3$
  & \msbar\, $m_c(m_c)$ 
  & FFNS $(n_f=3)$ 
  & ${\cal O}(\alpha_s^3)_{\rm approx}$
  & 62/52
  & 62/52
  & 1561/1145
  & 1651/1145
\\[0.5ex]
\hline
MMHT14~\cite{Harland-Lang:2014zoa} 
& & & & & & & & \\[-0.5ex]
$\qquad$ NLO
  & $1.4$
  & $m_c^{\rm pole}$ 
  & RT optimal~\cite{Thorne:2012az}
  & ${\cal O}(\alpha_s^2)$
  & 72/52
  & 78/52
  & 1411/1145
  & 1526/1145
\\
$\qquad$ NNLO
  & $1.4$
  & $m_c^{\rm pole}$ 
  & RT optimal~\cite{Thorne:2012az}
  & ${\cal O}(\alpha_s^2)$
    \tablefootnote{
        MMHT14 uses the ${\cal O}(\alpha_s^2)$ 
        heavy-quark Wilson coefficient functions 
        together with some terms at ${\cal O}(\alpha_s^3)$ 
        for $Q^2 \sim m_c^2$ described in Ref.~\cite{Thorne:2012az}. 
        These terms at ${\cal O}(\alpha_s^3)$ have been shown not to be leading.
  }
  & 71/52
  & 83/52
  & 1366/1145
  & 1424/1145
\\[0.5ex]
\hline
NNPDF3.0~\cite{Ball:2014uwa}
& & & & & &  & &\\[-0.5ex]
$\qquad$ NLO
  & $1.275$
  & $m_c^{\rm pole}$ 
  & FONLL-B~\cite{Forte:2010ta}
  & ${\cal O}(\alpha_s^2)$
  & 58/52
  & 60/52
  & 1402/1145
  & 1453/1145
\\
$\qquad$ NNLO
  & $1.275$
  & $m_c^{\rm pole}$ 
  & FONLL-C~\cite{Forte:2010ta}
  & ${\cal O}(\alpha_s^2)$
  & 67/52
  & 69/52
  & 1393/1145
  & 1496/1145
\\[0.5ex]
\hline
PDF4LHC15~\cite{Butterworth:2015oua}
    \tablefootnote{
    The data comparision uses the 
    {\tt xFitter}~\cite{Alekhin:2014irh,HERAFitter} 
    implementation of the schemes  
    FONLL-B, RT optimal and SACOT($\chi$) with the set {\tt PDF4LHC\_100} at NLO.
  }
  & $-$
  & $-$
  & FONLL-B~\cite{Forte:2010ta}
  & $-$
  & 58/52 
  & 64/52
  & 1369/1145
  & 1481/1145
\\
  & $-$
  & $-$
  & RT optimal~\cite{Thorne:2012az} 
  & $-$
  & 71/52 
  & 75/52 
  & 1396/1145    
  & 1496/1145    
\\
  & $-$
  & $-$
  & SACOT($\chi$)~\cite{Tung:2001mv}
  & $-$
  & 51/47
  & 76/47
  & 1378/1145
  & 1497/1145
\\[0.5ex]
\hline
\end{tabular}
  \caption{\small 
    \label{tab:mcmass-ctd}
    Tab.~\ref{tab:mcmass} continued.
  }
\end{center}
\end{sidewaystable}

%% file: table-gauge-boson.tex
\begin{table}[ht!]
\begin{center}
\renewcommand{\arraystretch}{1.3}
\begin{tabular}{|l|c|c|l|}   
\hline    
PDF sets
&\multirow{2}{8.25em}{$\chi^2$/NDP \\ (ATLAS data~\cite{Aad:2011dm}) }
&theory accuracy
&theory method
\\[3.5ex]
\hline
ABM12~\cite{Alekhin:2013nda}
  &34.5/30
  &NNLO
  &{\tt FEWZ3.1}~\cite{Li:2012wna}, {\tt DYNNLO}~\cite{Catani:2010en}
\\[0.5ex]
\hline
ABMP15~\cite{Alekhin:2015cza}
  &32.3/30 
  &NNLO 
  &{\tt FEWZ3.1}~\cite{Li:2012wna}  
\\[0.5ex]
\hline
CT14~\cite{Dulat:2015mca} 
  &42/30
  & NNLL 
  & {\tt ResBos}~\cite{Balazs:1995nz} 
\\[0.75ex]
\hline
MMHT14~\cite{Harland-Lang:2015qea}
  &39/30
  &NNLO 
  &\multirow{2}{18.0em}{{\tt APPLGrid}~\cite{Carli:2010rw}, $C$-factors~\cite{Thorne:private}\\
  (kinematic dependence with {\tt FEWZ3.1}~\cite{Li:2012wna})}
\\[3.25ex]
\hline
NNPDF3.0~\cite{Ball:2014uwa}  
  &35.4/30
  &NNLO 
  & \multirow{2}{18.0em}{{\tt APPLGrid}~\cite{Carli:2010rw}, $C$-factors~\cite{Ball:2011uy}\\
  (kinematic dependence with {\tt FEWZ3.1}~\cite{Li:2012wna})}
\\[3.5ex]
\hline
\end{tabular}
  \caption{\small 
  \label{tab:WZ-boson}
    Description of the ATLAS data
    at $\sqrt{s} = 7$~TeV for 
    $W^\pm \rightarrow l^\pm \nu$, 
    $Z\rightarrow l^+l^-$ (Ref.~\cite{Aad:2011dm})
    used in the PDF fits. 
    The columns indicate the QCD accuracy
    of the theoretical predictions along with the tools 
    used to obtain them.
  }
\end{center}
\end{table}

%% file: table-alphas-in-pdfs.tex
\setcounter{footnote}{0}
\renewcommand{\thefootnote}{b\arabic{footnote}}
\begin{table}[t!]
\begin{center}
\renewcommand{\arraystretch}{1.3}
\begin{tabular}{|l|c|c|}
\hline
PDF sets
  & $\alpha_s(M_Z)$ 
  & \multirow{2}{8em}{method \\ of determination}
\\[3.5ex]
\hline
ABM12~\cite{Alekhin:2013nda}
  & $0.1132 \pm 0.0011$ 
  & fit at NNLO
\\[0.5ex]
\hline
CJ15~\cite{Accardi:2016qay}
  & $0.1183 \pm 0.0002$ 
  & fit at NLO
\\[0.5ex]
\hline
CT14~\cite{Dulat:2015mca} 
  & $0.118$ 
  & assumed at NNLO
\\[0.5ex]
\hline 
HERAPDF2.0Jets~\cite{Abramowicz:2015mha} 
      \tablefootnote{In detail HERAPDF2.0Jets obtains at NLO 
      $\alpha_s(M_Z) = 0.1183 \pm 0.0009 (\mbox{exp}) \pm 0.0005 (\mbox{model/parameterisation})
      \pm 0.0012 (\mbox{hadronisation})^{~+0.0037}_{~-0.0030} (\mbox{scale})$,
      which have been added in quadrature in the table entry. 
      The HERAPDF2.0 central variant uses a fixed value $\alpha_s(M_Z) = 0.118$.
    }
  & $0.1183~^{~+0.0040}_{~-0.0034}$
  & fit at NLO
\\[0.5ex]
\hline
JR14~\cite{Jimenez-Delgado:2014twa} 
  & $0.1136 \pm 0.0004$
  & dynamical fit at NNLO
\\[0.0ex]
   & $0.1162 \pm 0.0006$ 
   & standard fit at NNLO
\\[0.5ex]
\hline
MMHT14~\cite{Harland-Lang:2015qea}
    \tablefootnote{MMHT14 obtains $\alpha_s(M_Z) = 0.1172 \pm 0.0013$ at NNLO as a best fit.}
  & $0.118$ 
  & assumed at NNLO
\\[0.5ex]
\hline
NNPDF3.0~\cite{Ball:2014uwa}
  & $0.115-0.121$ 
  & assumed at NNLO; preferred value 0.118 
\\[0.5ex]
\hline
PDF4LHC15~\cite{Butterworth:2015oua}
  & $0.118$ 
  & assumed at NNLO
\\[0.5ex]
\hline
\end{tabular}
  \caption{\small 
  \label{tab:alphas-in-pdfs}
    Values of $\alpha_s(M_Z)$ obtained or used in the nominal PDF sets of the various groups.
}
\end{center}
\end{table}

%% file: table-alphas.tex
\setcounter{footnote}{0}
\renewcommand{\thefootnote}{e\arabic{footnote}}
\begin{table}[ht!]
\begin{center}
\renewcommand{\arraystretch}{1.3}
\begin{tabular}{|l|l|l|l|}
\hline
\multicolumn{1}{|c|}{ } &
\multicolumn{1}{|c|}{year} &
\multicolumn{1}{c|}{$\alpha_s({M_Z})$} &
\multicolumn{1}{c|}{method/data sets/reference} 
\\[0.5ex]
\hline
SY  & 2001 & $0.1166 \pm 0.0013$  & $F_2^{ep}$    \cite{Santiago:2001mh} \\[-1.0ex]
    & 2001 & $0.1153 \pm 0.0063$  & $xF_3^{\nu N}$ (DIS off heavy nuclei)  \cite{Santiago:2001mh} 
\\[0.5ex]
\hline
A02  & 2002 & $0.1143 \pm 0.0020$ & \cite{Alekhin:2002rk}
\\[0.5ex]
\hline
MRST03 & 2003 & $0.1153\pm 0.0020$ & \cite{Martin:2003sk}
\\[0.5ex]
\hline
BBG  & 2004(06,12) & $0.1134~_{-~0.0021}^{+~0.0019}$
         & {\rm valence~analysis, NNLO}  \cite{Blumlein:2004ip,Blumlein:2006be,Blumlein:2012se}
\\[0.5ex]
\hline
GRS  &2006    & $0.112 $ & {\rm valence~analysis, NNLO}  \cite{Gluck:2006yz}
\\[0.5ex]
\hline
AMP06 & 2006 & $0.1128 \pm 0.0015$ & \cite{Alekhin:2006zm}
\\[0.5ex]
\hline
JR   & 2008 & $0.1128 \pm 0.0010$ & {\rm dynamical~approach} \cite{JimenezDelgado:2008hf} 
\\[-1.0ex]
     & 2008 & $0.1162 \pm 0.0006$ & {\rm with jet hadroproduction at NLO} \cite{JimenezDelgado:2008hf}
\\[0.5ex]
\hline
ABKM & 2009 & $0.1135 \pm 0.0014$ & {\rm ~$n_f=3$ FFNS heavy quark scheme} \cite{Alekhin:2009ni}
\\[-1.0ex]
     & 2009 & $0.1129 \pm 0.0014$ & {\rm BMSN heavy quark scheme} \cite{Alekhin:2009ni}
\\[0.5ex]
\hline
MSTW & 2009 & $0.1171 \pm 0.0014$ & \cite{Martin:2009bu}
\\[0.5ex]
\hline
Thorne & 2013 & $0.1175$ & DIS, Drell-Yan data;
\\[-1.0ex]
 (MSTW)
     &      &          & incl. higher twist, GM-VFNS \cite{Thorne:2014toa}
\\[0.5ex]
\hline
ABM11$_J$ & 2010 & $0.1134 - 0.1149 \pm 0.0012$ & with jet production at Tevatron (NLO) \cite{Alekhin:2010iu}
\\[0.5ex]
\hline
NNPDF2.1 & 2011 & $0.1173 \pm 0.0007 \pm 0.0009$ & with data for DIS off heavy nuclei \cite{Lionetti:2011pw,Ball:2011us}
\\[0.5ex]
\hline
ABM11 & 2012 & $0.1134\pm 0.0011$ & \cite{Alekhin:2012ig}
\\[0.5ex]
\hline
ABM12 & 2013 & $0.1132\pm 0.0011$ & \cite{Alekhin:2013nda}
\\[0.5ex]
\hline
Thorne & 2013 & $0.1136$ & DIS, Drell-Yan data 
\\[-1.0ex]
(MSTW)
       &      &          & incl. higher twist, FFNS \cite{Thorne:2014toa}
\\[0.5ex]
\hline
CT10 & 2013 & $0.1140$ & with jet hadroproduction data~\cite{Gao:2013xoa} 
\\[0.5ex]
\hline
JR & 2014 & $0.1136 \pm 0.0004$ & dynamical approach \cite{Jimenez-Delgado:2014twa} 
\\[-1.0ex]
   & 2014 & $0.1162 \pm 0.0006$ & standard fit \cite{Jimenez-Delgado:2014twa} 
\\[0.5ex]
\hline
CT14 & 2015 & $0.1150~_{-~0.0040}^{+~0.0060} $ &  $\Delta \chi^2 > 1 $ and with data for DIS off heavy nuclei \cite{Dulat:2015mca}
\\[0.5ex]
\hline
MMHT &2015   & ${0.1172}  \pm 0.0013$ & with data for DIS off heavy nuclei \cite{Harland-Lang:2015nxa}
\\[0.5ex]
\hline
\end{tabular}
  \caption{\small 
  \label{tab:alphas}
    Determinations of 
    $\alpha_s(M_Z)$ values at NNLO from QCD analyses of the deep-inealstic
    world data and, partly, including
    other hard scattering data.
    For recent compilations, see \cite{Bethke:2011tr,Moch:2014tta,d'Enterria:2015toz}.
}
\end{center}
\end{table}

%% file: table-th-accuracy.tex
\setcounter{footnote}{0}
\renewcommand{\thefootnote}{b\arabic{footnote}}
\begin{table}[t!]
\begin{center}
\renewcommand{\arraystretch}{1.3}
\begin{tabular}{|l|c|c|}
\hline
PDF set
  & theory accuracy
  & data sets used 
\\[3.5ex]
\hline
CT14~\cite{Dulat:2015mca} 
  & NLO with the scale set to the individual $p_T$ of jet 
  & Tevatron + LHC 
\\[0.5ex]
\hline
MMHT14~\cite{Harland-Lang:2015qea}
  & NLO + ${\cal O}(\alpha_s^4)_{\rm approx}$ threshold corrections~\cite{Kidonakis:2000gi}
  & Tevatron
\\[3.5ex]
\hline
NNPDF3.0~\cite{Ball:2014uwa}
  &  NLO + ${\cal O}(\alpha_s^4)_{\rm approx}$ threshold corrections~\cite{Carrazza:2014hra}
  & Tevatron + LHC (``safety cuts'')
\\[0.5ex]
\hline
\end{tabular}
  \caption{\small 
    \label{tab:jets}
    The jet data sets and the theory approximations used in the NNLO PDF
    fits.
    The threshold corrections of Ref.~\cite{Kidonakis:2000gi} neglect the 
  dependence on the jet radius $R$. 
  Ref.~\cite{Carrazza:2014hra} has determined the regime of validity (``safety
  cuts'') of the threshold approximation of Ref.~\cite{deFlorian:2013qia} 
  by comparing to the exact NNLO result for the $gg$ channel~\cite{Ridder:2013mf}.
}
\end{center}
\end{table}

%% file: table-higgs.tex
\begin{table}[ht!]
\begin{center}
\renewcommand{\arraystretch}{1.3}
\begin{tabular}{|l|l|l|l|}
\hline
PDF sets
  & \multirow{2}{8em}{$\sigma(H)^{\rm NNLO}$~[pb] nominal $\alpha_s(M_Z)$}
  & \multirow{2}{8em}{$\sigma(H)^{\rm NNLO}$~[pb] $\alpha_s(M_Z)=0.115$}
  & \multirow{2}{8em}{$\sigma(H)^{\rm NNLO}$~[pb] $\alpha_s(M_Z)=0.118$}
\\[3.5ex]
\hline
ABM12~\cite{Alekhin:2013nda} 
  & $ 39.80 \pm 0.84 $
  & $ 41.62 \pm 0.46 $
  & $ 44.70 \pm 0.50 $
\\[0.5ex]
\hline
CJ15~\cite{Accardi:2016qay} \tablefootnote{
  The CJ15 PDFs have been determined at NLO accuracy in QCD.
  The PDF uncertainties quoted by CJ15 denote the 90\% c.l. and should be reduced by a factor
  of 1.645 for comparison with the 68\% c.l. uncertainties quoted by other
  groups.}
  & $ 42.45~_{-~0.18}^{+~0.43}$
  & $ 39.48~_{-~0.17}^{+~0.40}$
  & $ 42.45~_{-~0.18}^{+~0.43}$
\\[0.5ex]
\hline
CT14~\cite{Dulat:2015mca} \tablefootnote{
  The PDF uncertainties quoted by CT14 denote the 90\% c.l. and should be reduced by a factor
  of 1.645 for comparison with the 68\% c.l. uncertainties quoted by other
  groups.}
  & $ 42.33~_{-~1.68}^{+~1.43}$
  & $ 39.41~_{-~1.56}^{+~1.33}$
  & $ 42.33~_{-~1.68}^{+~1.43}$
\\[-1.0ex]
  & 
  & ($ 40.10$)
  & 
\\[0.5ex]
\hline
HERAPDF2.0~\cite{Abramowicz:2015mha} \tablefootnote{
  The model uncertainities of the {\tt HERAPDF20\_NNLO\_VAR} set are not included in the 
  uncertainty estimates.}
  & $ 42.62~_{-~0.43}^{+~0.35}$
  & $ 39.68~_{-~0.40}^{+~0.32}$
  & $ 42.62~_{-~0.43}^{+~0.35}$
\\[-1.0ex]
  & 
  & ($ 40.88$)
  &
\\[0.5ex]
\hline
JR14 (dyn)~\cite{Jimenez-Delgado:2014twa} 
  & $ 38.01 \pm 0.34 $
  & $ 39.34 \pm 0.22 $
  & $ 42.25 \pm 0.24 $
\\[0.5ex]
\hline
MMHT14~\cite{Harland-Lang:2014zoa} 
  & $ 42.36~_{-~0.78}^{+~0.56}$
  & $ 39.43~_{-~0.73}^{+~0.53}$
  & $ 42.36~_{-~0.78}^{+~0.56}$
\\[-1.0ex]
  & 
  & ($ 40.48$)
  & 
\\[0.5ex]
\hline
NNPDF3.0~\cite{Ball:2014uwa}
  & $ 42.59 \pm 0.80 $
  & $ 39.65 \pm 0.74 $
  & $ 42.59 \pm 0.80 $
\\[-1.0ex]
  & 
  & ($ 40.74 \pm 0.88 $)
  & 
\\[0.5ex]
\hline
PDF4LHC15~\cite{Butterworth:2015oua}
  & $ 42.42 \pm 0.78 $ 
  & $ 39.49 \pm 0.73 $
  & $ 42.42 \pm 0.78 $ 
\\[0.5ex]
\hline
\end{tabular}
  \caption{\small 
  \label{tab:higgs}
    The Higgs cross section at NNLO in QCD (computed in the effective theory)
    at $\sqrt{s}=13$~TeV for $m_H=125.0$~GeV
    at the nominal scale $\mu_r=\mu_f=m_H$ with the PDF (and, if available, also $\alpha_s$) uncertainties.
    The columns correspond to different choices for the central value of
    $\alpha_s(M_Z)$ using the nominal PDF set.
    The numbers in parenthesis are obtained using the PDF sets {\tt CT14nnlo\_as\_0115}, 
      {\tt HERAPDF20\_NNLO\_ALPHAS\_115}, {\tt MMHT2014nnlo\_asmzlargerange} 
      and {\tt NNPDF30\_nnlo\_as\_0115}.
  }
\end{center}
\end{table}

%% file: table-higgs-mc-mstw.tex
\begin{table}[ht!]
\begin{center}
\renewcommand{\arraystretch}{1.3}
\begin{tabular}{|l|l|c|l|l|}
\hline
    $m_c^{\rm pole}$ [GeV]
  & \multirow{2}{4em}{$\alpha_s(M_Z)$ \\ (best fit)}
  & \multirow{2}{8em}{$\chi^2$/NDP \\ (HERA data \cite{Abramowicz:1900rp})}
  & \multirow{2}{8em}{$\sigma(H)^{\rm NNLO}$~[pb] \\ best fit $\alpha_s(M_Z)$}
  & \multirow{2}{8em}{$\sigma(H)^{\rm NNLO}$~[pb] \\ $\alpha_s(M_Z)=0.1171$}
\\[3.5ex]
\hline
1.05 &  0.1157 & 73/52 & 40.65  
  &  (41.63)
 \\[0.5ex]
\hline
1.1  &  0.1159 & 69/52 & 40.85  
  &  (41.70)
 \\[0.5ex]
\hline
1.15 &  0.1160 & 66/52 & 41.04  
  &  (41.78)
 \\[0.5ex]
\hline
1.2  &  0.1162 & 64/52 & 41.25  
  &  (41.85)
 \\[0.5ex]
\hline
1.25 &  0.1164 & 64/52 & 41.47  
  &  (41.93)
 \\[0.5ex]
\hline
1.3  &  0.1166 & 63/52 & 41.69  
  &  (42.00)
 \\[0.5ex]
\hline
1.35 &  0.1168 & 63/52 & 41.93  
  &  (42.09)
 \\[0.5ex]
\hline
1.4  &  0.1171 & 65/52 & 42.16  
  &  (42.16)
 \\[0.5ex]
\hline
1.45 &  0.1173 & 68/52 & 42.42  
  &  (42.24)
 \\[0.5ex]
\hline
1.5  &  0.1175 & 73/52 & 42.64  
  &  (42.31)
 \\[0.5ex]
\hline
1.55 &  0.1177 & 80/52 & 42.88  
  &  (42.38)
 \\[0.5ex]
\hline
1.6  &  0.1180 & 88/52 & 43.16  
  &  (42.46)
 \\[0.5ex]
\hline
1.65 &  0.1182 & 99/52 & 43.34  
  &  (42.51)
 \\[0.5ex]
\hline
1.7  &  0.1184 & 112/52 & 43.59  
  &  (42.58)
 \\[0.5ex]
\hline
1.75 &  0.1186 & 127/52 & 43.81  
  &  (42.63)
 \\[0.5ex]
\hline
\end{tabular}
  \caption{\small 
  \label{tab:higgs-mc-mstw}
    The values of the charm-quark mass (on-shell scheme $m^{\rm pole}$) 
    and the strong coupling $\alpha_s(M_Z)$ 
    in the MSTW analysis~\cite{Martin:2010db}
    using the set {\tt MSTW2008nnlo\_mcrange}
    together with the value for $\chi^2$/NDP for the HERA data~\cite{Abramowicz:1900rp}
    and the Higgs cross section at NNLO in QCD (computed in the effective theory)
    at $\sqrt{s}=13$~TeV for $m_H=125.0$~GeV
    at the nominal scale $\mu_r=\mu_f=m_H$.
    The numbers in parentheses are obtained using the PDF set 
    {\tt MSTW2008nnlo\_mcrange\_fixasmz} with the value of 
    $\alpha_s(M_Z)$ fixed to $\alpha_s(M_Z)=0.1171$.
  }
\end{center}
\end{table}

%% file: table-higgs-mc-mmht.tex
 
\setcounter{footnote}{0}
\renewcommand{\thefootnote}{a\arabic{footnote}}
\begin{table}[ht!]
\begin{center}
\renewcommand{\arraystretch}{1.3}
\begin{tabular}{|l|l|c|l|l|l|}
\hline
    $m_c^{\rm pole}$ [GeV]
  & \multirow{2}{4em}{$\alpha_s(M_Z)$ \\ (best fit)}
  & \multirow{2}{8em}{$\chi^2$/NDP \\ (HERA data \cite{Abramowicz:1900rp})}
  & \multicolumn{2}{l|}{\multirow{2}{8em}{$\sigma(H)^{\rm NNLO}$~[pb]\\ best fit $\alpha_s(M_Z)$}}
  & \multirow{2}{8em}{$\sigma(H)^{\rm NNLO}$~[pb]\\ $\alpha_s(M_Z)=0.118$}
\\[3.5ex]
  &  
  &
  & this work
  & Ref.~\cite{Thorne:private}
  &
\\[0.5ex]
\hline
   1.15
   & 0.1164    
   & 78/52 
   (71/52) 
   & 40.48\,\,\,\,\,\,
&     41.01
  &  (42.05)
   \\[0.5ex]
\hline
   1.2 
   & 0.1166  
   & 76/52
   (70/52) 
   & 40.74
&     41.18
  &  (42.11)
   \\[0.5ex]
\hline
   1.25
   & 0.1167   
   & 75/52
   (76/52) 
   & 40.89
&     41.33
  &  (42.17)
   \\[0.5ex]
\hline
   1.3
   & 0.1169  
   & 76/52
   (77/52)
   & 41.16
&     41.48
  &  (42.25)
   \\[0.5ex]
\hline
   1.35
   & 0.1171  
   & 78/52
   (79/52)
   & 41.41
&     41.68
  &  (42.30)
   \\[0.5ex]
\hline
   1.4 
   & 0.1172 
   & 82/52
   (83/52)
   & 41.56
&     41.83
  &  (42.36)
   \\[0.5ex]
\hline
   1.45
   & 0.1173  
   & 88/52
   (89/52) 
   & 41.75
&     42.00
  &  (42.45)
   \\[0.5ex]
\hline
   1.5  
   & 0.1173
   & 96/52
   (96/52)
   & 41.81
&     42.14
  &  (42.51)
   \\[0.5ex]
\hline
   1.55 
   & 0.1175 
   & 105/52
   (106/52)
   & 42.08
&     42.29
  &  (42.58)
\\[0.5ex]
\hline
\end{tabular}
  \caption{\small 
  \label{tab:higgs-mc-mmht}
    Same as Tab.~\ref{tab:higgs-mc-mstw} for the MMHT14 analysis~\cite{Harland-Lang:2015qea}
    using the set {\tt MMHT2014nnlo\_mcrange\_nf5} and 
      setting $\alpha_s(M_Z)$ to the best fit value.
      The numbers of Ref.~\cite{Thorne:private} keep full 
      account of the correlation between the PDFs and $\alpha_s$.
    The values of $\chi^2$/NDP for the HERA data~\cite{Abramowicz:1900rp}
    are those quoted in \cite{Harland-Lang:2015qea} for the best fit value of $\alpha_s(M_Z)$.
    The numbers in parentheses are obtained with the value of 
    $\alpha_s(M_Z)$ fixed to $\alpha_s(M_Z)=0.118$.
  }
\end{center}
\end{table}

%% file: table-higgs-mc-nnpdf.tex
\begin{table}[ht!]
\begin{center}
\renewcommand{\arraystretch}{1.3}
\begin{tabular}{|l|l|l|c|l|l|}
\hline
PDF sets
  & $m_c^{\rm pole}$ [GeV]
  & \multirow{2}{4em}{$\alpha_s(M_Z)$ \\ (fixed)}
  & \multirow{2}{8em}{$\chi^2$/NDP \\ (HERA data \cite{Abramowicz:1900rp})}
  & \multirow{2}{7em}{$\sigma(H)^{\rm NNLO}$~[pb] fixed $\alpha_s(M_Z)$}
\\[3.5ex]
\hline
NNPDF2.1~\cite{Ball:2011mu} 
 & $\sqrt{2}$ & 0.119 & 
65/52
 & $ 44.18 \pm 0.49 $ \\[0.5ex]
\cline{2-5}
 & 1.5 & 0.119 & 
78/52
 &  $ 44.54 \pm 0.51 $ \\[0.5ex]
\cline{2-5}
 & 1.6 & 0.119 & 
92/52
 &  $ 44.74 \pm 0.50 $ \\[0.5ex]
\cline{2-5}
 & 1.7 & 0.119 & 
110/52
 & $ 44.95 \pm 0.51 $ \\[0.5ex]
\hline 
NNPDF2.3~\cite{Ball:2012cx} & $\sqrt{2}$ & 0.118 & 
71/52
 &   $ 43.77 \pm 0.41 $ \\[0.5ex]
\hline
NNPDF3.0~\cite{Ball:2014uwa} & 1.275 & 0.118 & 
67/52
 & $ 42.59 \pm 0.80 $ \\[0.5ex]
\hline
\end{tabular}
  \caption{\small 
  \label{tab:higgs-mc-nnpdf}
    Same as Tab.~\ref{tab:higgs-mc-mstw} for various NNPDF analyses.
    The values of the strong coupling $\alpha_s(M_Z)$ have been fixed in those fits. 
    The values of $\chi^2$/NDP for the description of the HERA data have been 
    determined with the FONLL-C~\cite{Forte:2010ta} scheme.
  }
\end{center}
\end{table}

%% file: table-ttbar.tex
\begin{table}[t!]
\begin{center}
\renewcommand{\arraystretch}{1.3}
\begin{tabular}{|l|l|l|l|}
\hline
PDF sets
  & \multirow{2}{8em}{$\sigma(t{\bar t})^{\rm NNLO}$~[pb] nominal $\alpha_s(M_Z)$}
  & \multirow{2}{8em}{$\sigma(t{\bar t})^{\rm NNLO}$~[pb] $\alpha_s(M_Z)=0.115$}
  & \multirow{2}{8em}{$\sigma(t{\bar t})^{\rm NNLO}$~[pb] $\alpha_s(M_Z)=0.118$}
\\[3.5ex]
\hline
ABM12~\cite{Alekhin:2013nda} 
  & $ 715.1 \pm 21.3 $
  & $ 741.7 \pm 10.3 $
  & $ 786.5 \pm 10.9 $
\\[0.5ex]
\hline
CJ15~\cite{Accardi:2016qay} \tablefootnote{
  The CJ15 PDFs have been determined at NLO accuracy in QCD.
  The PDF uncertainties quoted by CJ15 denote the 90\% c.l. and should be reduced by a factor
  of 1.645 for comparison with the 68\% c.l. uncertainties quoted by other
  groups.
}
  & $ 786.7~_{-~11.8}^{+~5.1} $
  & $ 742.0~_{-~11.1}^{+~4.8} $
  & $ 786.7~_{-~11.8}^{+~5.1} $
\\[0.5ex]
\hline
CT14~\cite{Dulat:2015mca} \tablefootnote{
  The PDF uncertainties quoted by CT14 denote the 90\% c.l. and should be reduced by a factor
  of 1.645 for comparison with the 68\% c.l. uncertainties quoted by other
  groups.}
  & $ 834.2~_{-~36.5}^{+~36.0}$
  & $ 786.7~_{-~34.4}^{+~34.0}$
  & $ 834.2~_{-~36.5}^{+~36.0}$
\\[-1.0ex]
  & 
  & ($ 791.4$)
  & 
\\[0.5ex]
\hline
HERAPDF2.0~\cite{Abramowicz:2015mha} \tablefootnote{
  The model uncertainities of the {\tt HERAPDF20\_NNLO\_VAR} set are not included in the 
  uncertainty estimates.}
  & $ 804.1~_{-~17.0}^{+~11.4}$
  & $ 757.8~_{-~16.0}^{+~10.8}$
  & $ 804.1~_{-~17.0}^{+~11.4}$
\\[-1.0ex]
  & 
  & ($ 756.8$)
  &
\\[0.5ex]
\hline
JR14 (dyn)~\cite{Jimenez-Delgado:2014twa} 
  & $ 719.3 \pm 9.1 $
  & $ 739.6 \pm 5.7 $
  & $ 784.2 \pm 6.0 $
\\[0.5ex]
\hline
MMHT14~\cite{Harland-Lang:2014zoa} 
  & $ 831.8~_{-~17.5}^{+~13.9} $
  & $ 784.5~_{-~16.5}^{+~13.1}$
  & $ 831.8~_{-~17.5}^{+~13.9} $
\\[-1.0ex]
  & 
  & ($ 794.8$)
  & 
\\[0.5ex]
\hline
NNPDF3.0~\cite{Ball:2014uwa}
  & $ 831.8 \pm 15.0 $
  & $ 784.4 \pm 14.2 $
  & $ 831.8 \pm 15.0 $
\\[-1.0ex]
  & 
  & ($ 800.9 \pm 16.5 $)
  & 
\\[0.5ex]
\hline
PDF4LHC15~\cite{Butterworth:2015oua}
  & $ 832.5 \pm 16.4 $  
  & $ 785.1 \pm 15.5 $
  & $ 832.5 \pm 16.4 $  
\\[0.5ex]
\hline
\end{tabular}
  \caption{\small 
    \label{tab:ttbar}
    The inclusive cross section for top-quark pair production at NNLO in QCD 
    at $\sqrt{s}=13$~TeV for a pole mass of $m_t^{\rm pole}=172.0$~GeV
    at the nominal scale $\mu_r=\mu_f=m_t^{\rm pole}$ with the PDF (and, if available, also $\alpha_s$) uncertainties.
    The columns correspond to different choices for the central value of
    $\alpha_s(M_Z)$ using the nominal PDF set.
    The numbers in parenthesis are obtained using PDF sets {\tt CT14nnlo\_as\_0115}, 
      {\tt HERAPDF20\_NNLO\_ALPHAS\_115}, {\tt MMHT2014nnlo\_asmzlargerange} 
      and {\tt NNPDF30\_nnlo\_as\_0115}.
    }
\end{center}
\end{table}

%% file: table-ttbar-mc-mmht.tex
 
\setcounter{footnote}{0}
\renewcommand{\thefootnote}{a\arabic{footnote}}
\begin{table}[t!]
\begin{center}
\renewcommand{\arraystretch}{1.3}
\begin{tabular}{|l|l|c|l|l|l|}
\hline
    $m_c^{\rm pole}$ [GeV]
  & \multirow{2}{4em}{$\alpha_s(M_Z)$ \\ (best fit)}
  & \multirow{2}{8em}{$\chi^2$/NDP \\ (HERA data \cite{Abramowicz:1900rp})}
  & \multicolumn{2}{l|}{\multirow{2}{8em}{$\sigma(t{\bar t})^{\rm NNLO}$~[pb]\\ best fit $\alpha_s(M_Z)$}}
  & \multirow{2}{8em}{$\sigma(t{\bar t})^{\rm NNLO}$~[pb]\\ $\alpha_s(M_Z)=0.118$}
\\[3.5ex]
  &  
  &
  & this work
  & Ref.~\cite{Thorne:private}
  &
\\[0.5ex]
\hline
   1.15
   & 0.1164    
   & 78/52 
   (71/52) 
   & 810.2
&           815.0
  &  (835.8)
   \\[0.5ex]
\hline
   1.2 
   & 0.1166  
   & 76/52
   (70/52) 
   & 813.0
&           817.3
  &  (835.4)
   \\[0.5ex]
\hline
   1.25
   & 0.1167   
   & 75/52
   (76/52) 
   & 814.0
&           818.3
  &  (834.8)
   \\[0.5ex]
\hline
   1.3
   & 0.1169  
   & 76/52
   (77/52)
   & 816.5
&           819.3
  &  (834.2)
   \\[0.5ex]
\hline
   1.35
   & 0.1171  
   & 78/52
   (79/52)
   & 819.0
&           821.8
  &  (833.4)
   \\[0.5ex]
\hline
   1.4 
   & 0.1172 
   & 82/52
   (83/52)
   & 819.0
&           822.4
  &  (831.8)
   \\[0.5ex]
\hline
   1.45
   & 0.1173  
   & 88/52
   (89/52) 
   & 820.2
&           823.1
  &  (831.5)
   \\[0.5ex]
\hline
   1.5  
   & 0.1173
   & 96/52
   (96/52)
   & 818.8
&           823.1
  &  (830.0)
   \\[0.5ex]
\hline
   1.55 
   & 0.1175 
   & 105/52
   (106/52)
   & 821.0
&           823.6
  &  (829.0)
\\[0.5ex]
\hline
\end{tabular}
  \caption{\small 
  \label{tab:ttbar-mc-mmht}
    The values of the charm-quark mass (on-shell scheme $m_c^{\rm pole}$) 
    and the strong coupling $\alpha_s(M_Z)$ 
    in the MMHT14 analysis~\cite{Harland-Lang:2015qea} 
    together the inclusive cross section for top-quark pair production at NNLO in
    QCD computed with the set {\tt MMHT2014nnlo\_mcrange\_nf5}
    at $\sqrt{s}=13$~TeV for a pole mass of $m_t^{\rm pole}=172.0$~GeV
    at the nominal scale $\mu_r=\mu_f=m_t^{\rm pole}$ and 
      setting $\alpha_s(M_Z)$ to the best fit value.
      The numbers of Ref.~\cite{Thorne:private} keep full 
      account of the correlation between the PDFs and $\alpha_s$.
    The values of $\chi^2$/NDP for the HERA data~\cite{Abramowicz:1900rp}
    are those quoted in \cite{Harland-Lang:2015qea} for the best fit value of $\alpha_s(M_Z)$.
    The numbers in parentheses for the cross section and 
    $\chi^2$/NDP are obtained using the PDF set with the value of 
    $\alpha_s(M_Z)$ fixed to $\alpha_s(M_Z)=0.118$.
  }
\end{center}
\end{table}

%% file: table-ttbar-mc-nnpdf.tex
\begin{table}[t!]
\begin{center}
\renewcommand{\arraystretch}{1.3}
\begin{tabular}{|l|l|l|c|l|l|}
\hline
PDF sets
  & $m_c^{\rm pole}$ [GeV]
  & \multirow{2}{4em}{$\alpha_s(M_Z)$ \\ (fixed)}
  & \multirow{2}{8em}{$\chi^2$/NDP \\ (HERA data \cite{Abramowicz:1900rp})}
  & \multirow{2}{7em}{$\sigma(t{\bar t})^{\rm NNLO}$~[pb] fixed $\alpha_s(M_Z)$}
\\[3.5ex]
\hline
NNPDF2.1~\cite{Ball:2011mu} 
 & $\sqrt{2}$ & 0.119 & 
65/52
 & $ 847.1 \pm 16.3 $ \\[0.5ex]
\cline{2-5}
 & 1.5 & 0.119 & 
78/52
 & $ 850.8 \pm 14.3 $ \\[0.5ex]
\cline{2-5}
 & 1.6 & 0.119 & 
92/52
 &  $ 842.9 \pm 13.6 $ \\[0.5ex]
\cline{2-5}
 & 1.7 & 0.119 & 
110/52
 & $ 840.1 \pm 14.1 $ \\[0.5ex]
\hline 
NNPDF2.3~\cite{Ball:2012cx} & $\sqrt{2}$ & 0.118 & 
71/52
 &   $ 835.7 \pm 14.9 $ \\[0.5ex]
\hline
NNPDF3.0~\cite{Ball:2014uwa} & 1.275 & 0.118 & 
67/52
 & $ 831.8 \pm 15.0 $ \\[0.5ex]
\hline
\end{tabular}
  \caption{\small 
  \label{tab:ttbar-mc-nnpdf}
    Same as Tab.~\ref{tab:ttbar-mc-mmht} for various NNPDF analyses.
    The values of the strong coupling $\alpha_s(M_Z)$ have always been fixed in those fits. 
    The values of $\chi^2$/NDP for the description of the HERA data have been 
    determined with the FONLL-C~\cite{Forte:2010ta} scheme.
  }
\end{center}
\end{table}

%% file: table-lhcb.tex
\begin{table}[t!]
\begin{center}
\renewcommand{\arraystretch}{1.3}
\begin{tabular}{|l|c|c|}
  \hline
  PDF sets         & $\chi^2$/NDP (with unc.) & $\chi^2$/NDP (nominal) \\
  \hline
  ABM11~\cite{Alekhin:2012ig}
    \tablefootnote{
      The set ABM11 fit~\cite{Alekhin:2012ig} is used here, 
      because ABM12~\cite{Alekhin:2013nda} sets are only available at NNLO.
    }
   & $222/244$ & $394/244$ 
\\[0.5ex]
CJ15~\cite{Accardi:2016qay} 
  & $241/244$ & $272/244$ 
\\[0.5ex]
  CT14~\cite{Dulat:2015mca}
  & $166/244$ & $241/244$ 
\\[0.5ex]
  HERAPDF2.0~\cite{Abramowicz:2015mha} 
  & $219/244$ & $366/244$ 
\\[0.5ex]
  JR14~\cite{Jimenez-Delgado:2014twa} 
  & $205/244$ & $217/244$ 
\\[0.5ex]
  MMHT14~\cite{Harland-Lang:2014zoa}               
  & $165/244$ & $202/244$ 
\\[0.5ex]
  NNPDF3.0~\cite{Ball:2014uwa}
  & $160/244$ & $197/244$ 
\\[0.5ex]
  PDF4LHC15~\cite{Butterworth:2015oua}
  & $173/244$ & $218/244$ 
\\
  \hline
\end{tabular}
\caption{\small 
  \label{tab:lhcbbeauty}
  The values of $\chi^2$/NDP for the normalised bottom-quark cross sections
  measured at LHCb~\cite{Aaij:2013noa} using the NLO PDFs of the individual groups.
  The left column accounts for the quoted PDF uncertainties 
  (with the CJ15 and CT14 PDF uncertainties rescaled to 68\% c.l.), 
  while the right column uses the central prediction of each PDF set.
}
\end{center}
\end{table}